%% file: main.tex
\documentclass[11pt,dvipsnames,table,xcdraw]{article}

\usepackage[utf8]{inputenc}
\usepackage[T1]{fontenc}
\usepackage[main=american,british]{babel}
\usepackage[a4paper,width=150mm,top=25mm,bottom=25mm]{geometry}

\usepackage{jheppub} 
\usepackage{settings}

\title{Algorithmic Dualization of Unitary Circular Quivers}

\preprint{\begin{flushright} 
USTC-ICTS/PCFT-26-34 \\
UWThPh  2026-6
\end{flushright}}

\author[a]{Riccardo Comi,}
\author[b,c]{Chiung Hwang,}
\author[d]{Fabio Marino}

\affiliation[a]{Abdus Salam Centre for Theoretical Physics, Imperial College London, London SW7 2AZ, UK}
\affiliation[b]{Interdisciplinary Center for Theoretical Study, University of Science and Technology of China, Hefei, Anhui 230026, China}
\affiliation[c]{Peng Huanwu Center for Fundamental Theory, Hefei, Anhui 230026, China}
\affiliation[d]{Fakultät für Physik, Universität Wien, Boltzmanngasse 5, 1090 Wien, Austria}

\emailAdd{rcomi@ic.ac.uk}
\emailAdd{chiung@ustc.edu.cn}
\emailAdd{fabio.marino@univie.ac.at}

\abstract{
We introduce a field-theoretic algorithm to find the $SL(2,\mathbb{Z})$ duality web of 3d $\mathcal{N}=4$ circular quiver theories with unitary gauge groups, extending the algorithm for linear quivers. Although circular and linear quivers share the same local structure, the circular topology requires additional ingredients, which we formulate in terms of topological and baryonic QFT blocks, together with new $SL(2,\mathbb{Z})$ duality moves acting on them. For good circular quivers, this provides a field-theoretic derivation of mirror symmetry and extends it to the full $SL(2,\mathbb{Z})$ duality web. We then study bad circular quivers, distinguishing between local badness, associated with under-balanced gauge nodes, and global badness, arising from the circular topology itself. In particular, we analyze the magnetic and electric dual frames of globally bad circular quivers and provide additional evidence for the proposed duality by matching the Higgs branch index with the dual Coulomb branch index. The latter exhibits a structure reminiscent of permutation-group gauging and reveals a refined relation to the ADHM quiver, flowing to the $\mathcal{N}=8$ infrared fixed point.
}

\begin{document} 

\maketitle
\flushbottom

\begingroup
\allowdisplaybreaks

\newpage
\section{Introduction}

Infrared dualities have long played a central role in quantum field theory by demonstrating that theories with distinct ultraviolet descriptions can nevertheless describe identical long-distance physics. A seminal example is Seiberg duality \cite{Seiberg:1994pq}, which elucidates the infrared dynamics of four-dimensional supersymmetric QCD. In three dimensions, mirror symmetry \cite{Intriligator:1996ex,Hanany:1996ie} provides another profound instance of such equivalences. It relates pairs of 3d $\mathcal{N}=4$ theories by exchanging their Higgs and Coulomb branches of vacuum moduli spaces. For systems admitting a Hanany--Witten brane realization \cite{Hanany:1996ie}, mirror symmetry admits a natural interpretation as a consequence of Type IIB S-duality. Moreover, because Higgs branches of theories with eight supercharges in dimensions $3 \le d \le 6$ are protected under circle reduction \cite{Argyres:1996eh}, mirror symmetry furnishes a powerful framework for  studying quantum moduli spaces: it translates questions about Higgs branches into problems concerning Coulomb branches of three-dimensional quiver gauge theories. This perspective has proven invaluable in the analysis of strongly coupled and even non-Lagrangian supersymmetric systems \cite{deBoer:1996mp,Aharony:1997bx,Bullimore:2015lsa}. It underlies, in particular, the magnetic-quiver program for Higgs branches at infinite coupling \cite{Ferlito:2017xdq,Bourget:2019rtl,Cabrera:2019izd}, the construction of 3d mirrors for theories of class S and of Argyres--Douglas type \cite{Beratto:2020wmn,Giacomelli:2020ryy,Carta:2021dyx}, the study of S-fold SCFTs \cite{Bourget:2020mez,Giacomelli:2020gee}, and the Hasse-diagram analysis of symplectic singularities \cite{Bourget:2019aer,Bourget:2021csg}. In addition, mirror symmetry has recently been used to uncover a new family of planar abelian dualities for 3d SQCD \cite{Benvenuti:2024seb,Benvenuti:2025huk,Benvenuti:2025qnq,Benvenuti:2026usm,Benvenuti:2026xcv}. Dimensional reductions and uplifts of mirror symmetry have also been studied \cite{Aganagic:2001uw,Hwang:2020wpd,Sacchi:2020pet,Chen:2026tzt}.

The abundance of dualities in supersymmetric quantum field theories naturally raises the question of whether they can be organized into a systematic and constructive framework. A central goal is to develop algorithmic procedures that generate dual descriptions of a given theory and clarify how seemingly distinct dualities are related, ideally reducing them to a limited set of fundamental duality moves. In this spirit, the recently proposed \emph{mirror dualization algorithm} \cite{Bottini:2021vms,Hwang:2021ulb,Comi:2022aqo} provides a constructive realization of three-dimensional mirror symmetry. Building on the local dualization method introduced in \cite{Kapustin:1999ha} for abelian mirror pairs, this framework extends the construction to non-abelian gauge theories. It was originally formulated for \emph{good} (in the sense of \cite{Gaiotto:2008ak}) linear quivers \cite{Bottini:2021vms,Hwang:2021ulb}, namely the $T_\rho^\sigma[SU(N)]$ theories \cite{Gaiotto:2008ak} (see also \cite{Cremonesi:2014uva}). Later, it has been generalized in several directions, including $SL(2,\mathbb Z)$ dualities \cite{Comi:2022aqo,Marino:2025uub}, bad linear quivers and electric dualization algorithm \cite{Giacomelli:2023zkk,Giacomelli:2024laq,Comi:2025zwu}, and $\mathcal N=2$ theories and improved bifundamentals \cite{Benvenuti:2023qtv,Benvenuti:2024mpn}.\\

In this work, we aim to generalize the algorithm to circular quivers. These theories share the same local structure as linear quivers: each gauge node is connected to two neighboring gauge nodes at most and possibly to additional flavor nodes. Since the duality moves underlying the algorithm \cite{Bottini:2021vms,Hwang:2021ulb,Comi:2022aqo} are essentially local operations, the machinery developed for linear quivers can be carried over. However, due to their different global topology, a complete description of circular quivers requires additional ingredients, which we refer to as \emph{topological and baryonic QFT blocks}, together with new $SL(2,\mathbb Z)$ duality moves acting on them. Incorporating these extra building blocks, the algorithm can be systematically extended to good circular quivers.

The data of a good circular quiver can be conveniently encoded in two partitions $\rho$ and $\sigma$ of an integer $N$, together with the rank $N_L$ of the gauge node from which the labeling starts \cite{Assel:2012cj}. The corresponding quiver can therefore be denoted by $C_\rho^\sigma[SU(N);N_L]$.
Under $\mathcal S$-dualization, namely mirror symmetry, the two partitions are exchanged:
\begin{align}
C_\rho^\sigma [SU(N);N_L] \quad \xleftrightarrow[]{\;\;\;\; \mathcal{S} \;\;\;\;} \quad C_\sigma^\rho[SU(N);N_L] \,.
\end{align}
We demonstrate this relation from a field-theoretic perspective using the dualization algorithm and further extend the analysis to the full $SL(2,\mathbb Z)$ duality group. \\

The discussion can be extended to \emph{bad} circular quivers, although these are more involved. Indeed, a circular quiver can exhibit two distinct types of badness: \emph{local} badness and \emph{global} badness, which can also occur simultaneously. Local badness is essentially identical to that in linear quivers: a gauge node is locally bad when the effective number of flavors is too small, so that some monopole operators associated with that node decouple in the IR. 
Global badness, on the other hand, arises when the circular quiver has constant gauge ranks and no flavors, and is therefore not a property associated with any particular gauge node.
In this case, there exist monopole operators whose naive R-charges violate the unitarity bound and hence decouple in the IR. In particular, these decoupling monopoles carry magnetic flux under all gauge nodes simultaneously, reflecting the intrinsically global nature of this phenomenon.

As in the linear case, the dualization of bad circular quivers leads to multiple dual descriptions associated with different operator vacuum expectation values (VEVs). The resulting patterns, however, differ for locally and globally bad quivers. 

For locally bad quivers, rather than applying the mirror algorithm, our strategy is to locally apply the results of \cite{Giacomelli:2023zkk} for bad SQCD to each under-balanced node of the theory, in the same spirit as \cite{Giacomelli:2024laq}. This procedure, known as the \emph{electric dualization algorithm}, yields a sum of \emph{frames} described by good circular quivers together with singlets, some of which are realized as Dirac delta functions in the partition function. The singlets associated with these Dirac delta functions acquire nonzero VEVs, spontaneously breaking topological symmetries and freezing the corresponding FI parameters of the theory to specific values.

On the other hand, a globally bad circular quiver has no under-balanced nodes. Thus, the electric dualization algorithm cannot be applied; instead, one must use the original mirror algorithm, as in the good case. The result is again a sum of frames weighted by Dirac delta functions, a hallmark of badness. Interestingly, each frame turns out to be an SQED with a different number of flavors, rather than a good circular quiver. These magnetic frames can then be dualized into electric dual frames via standard mirror symmetry, resulting in abelian linear quivers. Furthermore, the Dirac delta function associated with each frame can also be dualized into an additional $U(1)$ factor, which combines with the corresponding linear quiver to form an abelian circular quiver. Thus, we find that the electric dual frames of a globally bad circular quiver are described by abelian circular quivers of various lengths.

To further analyze the IR dynamics of the globally bad circular quiver, we also examine a particular limit of the superconformal index, called the Higgs branch index. We show that this index precisely agrees with the Coulomb branch index of the proposed mirror dual description, which consists of several magnetic frames weighted by Dirac delta functions. Moreover, we find that the dual Coulomb branch index has an interesting structure reminiscent of permutation-group gauging, which allows the Coulomb branch Hilbert series of the ADHM quiver to be written in terms of that of an SQED \cite{Hanany:2018cgo}. Through this structure, we identify a similarity between the indices of the globally bad circular quiver and the ADHM quiver, thereby refining the relation between the two theories, which are expected to flow to the same $\mathcal N=8$ SCFT in the IR, up to decoupled operators.\\

Our analysis shows that circular topology introduces genuinely new features into the dualization algorithm, while preserving enough locality for the linear-quiver machinery to remain effective. In particular, the appearance of topological and baryonic QFT blocks, together with the distinction between local and global badness, indicates that circular quivers provide a natural setting in which local duality moves interact nontrivially with global constraints. The framework developed here offers a unified perspective on good and bad circular quivers and opens the way to further applications of the dualization algorithm to broader classes of three-dimensional theories.\\

The rest of the paper is organized as follows.
In Section~\ref{sec:good_circular}, we introduce the 3d $\mathcal{N}=4$ circular quiver setup, together with its Type IIB brane construction, the $C_\rho^\sigma[SU(N);N_L]$ notation, and the corresponding $S^3_b$ partition function.
In Section~\ref{sec:dualization_algorithm_circular}, we develop the dualization algorithm for circular quivers. We introduce the new topological and baryonic blocks, the basic $SL(2,\mathbb Z)$ duality moves involving them, and the steps needed to implement the algorithm. We also provide several explicit examples: a mirror pair, for which the full symmetry and operator map is detailed; an $SL(2,\mathbb Z)$ triality for an ABJM-like theory; the case of $\mathcal{S}$-fold theories; and a $U(N)$ SQCD model with symmetric and antisymmetric hypermultiplets.
In Section~\ref{sec:local_badness_circular}, we explain the electric dualization algorithm applied to locally bad circular quivers. 
In contrast, Section~\ref{sec:global_badness_circular} focuses on circular theories exhibiting global badness, addressing the globally ugly and globally bad cases in turn. For the globally bad circular quiver, we also show the agreement between the Higgs branch index and the dual Coulomb branch index, which reveals the connection to the Hilbert series of the ADHM quiver and permutation-group gauging. We then discuss its electric dualization and conclude the section with brief comments on generic quivers possessing both types of badness simultaneously. Additional background material and computational details are collected in the appendices. We have also attached Mathematica code implementing both the electric and magnetic dualization algorithms.

\clearpage
\section{Good circular quivers} 
\label{sec:good_circular}

Before discussing the dualization of quivers, let us briefly review general properties of circular quivers, including their Type IIB brane engineering. We also explain a prescription for cutting open a 3d $\mathcal{N}=4$ unitary circular \emph{good} quiver into a linear quiver \cite{Assel:2012cj},\footnote{In this and the following sections, we focus on good quivers, whose gauge nodes satisfy the usual goodness condition $2 N \leq F_\text{eff}$, where $N$ denotes the gauge rank and $F_\text{eff}$ is the effective number of fundamental hypermultiplets seen by the node \cite{Gaiotto:2008ak}. There are, however, exceptions called \emph{globally bad/ugly} quivers, which satisfy this local condition yet exhibit badness or ugliness as a global property of the quiver. This class of theories will be treated separately in Section~\ref{sec:global_badness_circular}.} as this operation is necessary for implementing the dualization algorithm described in Section~\ref{sec:dualization_algorithm_circular}. \\
We also specify our choice of parameterization for circular quivers in the $S^3_b$ partition function language.

\subsection{Circular quivers from branes and cutting prescription}
\label{subsec:cutting}

The data specifying a 3d $\mathcal{N}=4$ unitary circular quiver are the rank of the gauge groups $U(N_i)$ ($i=1,\dots, L$) and the number of flavors $F_i$ at each node. Such a theory can be engineered in Type IIB string theory with a D3-D5-NS5-brane system compactified on a circle \cite{Hanany:1996ie,Assel:2012cj}, where the 5-branes are located at points along the circle, while D3-branes are extended along the circular direction. The D3-branes can go all the way around the circle, or they might end on the NS5's. We can assume without loss of generality that none of the D3's end on the D5-branes (see \cite{Witten:2009xu}). We have in total $L$ NS5-branes, $N_i$ D3-branes  and $F_i$ D5-branes in each interval between consecutive NS5-branes, as in Figure~\ref{fig:circular_brane_setup}.

\begin{figure}[!ht]
    \centering
    \includegraphics[width=0.7\linewidth]{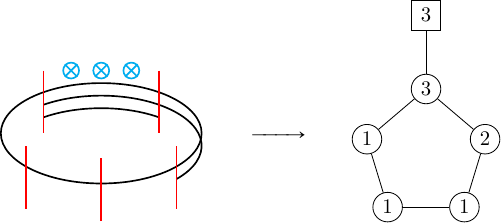}
    \caption{Type IIB brane realization (left) of a 3d unitary circular quiver (right). D3-branes stretched along a circle are depicted by black lines. NS5-branes are depicted by vertical red lines, whose number equals the number of gauge nodes in the quiver. D5-branes are depicted by cyan crosses and provide fundamental matter for the gauge groups.}
    \label{fig:circular_brane_setup}
\end{figure}

To conveniently describe a circular quiver and encode it in partitions, it is useful to begin by cutting it open \cite{Assel:2012cj}. This transforms the circular quiver into a linear quiver, which can then be described by a linear brane system in Type IIB string theory \cite{Hanany:1996ie}. The process of linearizing a circular quiver also plays a key role in implementing the dualization algorithm, which we introduce in Section~\ref{sec:dualization_algorithm_circular}. In this subsection, we expand upon the discussion in \cite{Assel:2012cj}.

To convert a circular quiver into a linear one, we first need to choose a node at which to cut the quiver open, which we designate as the special node. This node will correspond to both the zeroth and the last (i.e., $L$-th) nodes of the linearized quiver. The question then arises: how do we select this node, and in which direction -- clockwise or counterclockwise -- should we number the remaining nodes? In principle, any such choice is purely conventional, and all possibilities are equally valid.\footnote{In the string-theoretic construction of the field theory, the choice of the cut location has been interpreted as a gauge fixing for the B-field \cite{Witten:2009xu}.}

However, a particular choice of convention can simplify the description of the quiver in terms of partitions.\footnote{Selecting an arbitrary node as the $L$-th would necessitate generalizing the notion of a partition, allowing, for instance, some entries to be zero. The prescription we adopt avoids such complications. Nevertheless, it does not single out a unique preferred node for the cut, meaning multiple valid partitions may describe the same circular quiver. See Appendix~\ref{app:partitions} for further details.} This choice also proves advantageous for the dualization algorithm.
In general, the special node is chosen such that its rank $N_0=N_L$ is the smallest among all nodes. However, if multiple nodes share this minimal rank, additional criteria must be applied to determine the special node. Here we assume the total number of flavors is nonzero. The case of quivers with no flavors possess so-called global badness, which will be discussed separately in Section~\ref{sec:global_badness_circular}.
\begin{itemize}  
    \item If all nodes have the same rank, we define this common rank as $N_L$ and select one that has a nonzero number of flavors as the $L$-th node, since we assume that the total number of flavors is strictly positive. Any such choice is valid, and we may number the nodes either clockwise or counterclockwise.  
    \item If the nodes do not all have the same rank, there must exist a node with minimal gauge rank, denoted by $N_L$, such that at least one of its adjacent nodes has a higher rank; that is, a node located at the end of a chain of nodes---possibly consisting of a single node---sharing the same minimal rank. We take this node to be the $L$-th node and the adjacent node with higher rank to be the $(L-1)$-th node. The other adjacent one is then automatically identified as the first node. Again, if multiple choices satisfy these rules, any such choice is valid.
    
    An example is illustrated in Figure~\ref{fig:example_cutopen}, where $N_L=N_5=1$ and there exists a chain of three nodes, each with rank $1$. Either endpoint of the chain can be chosen as the fifth node. If we choose the left endpoint, the node with rank 3 is taken to be the fourth node and the nodes are labeled counterclockwise; if we choose the right endpoint, the node with rank 2 is taken to be the fourth node and the nodes are labeled clockwise. 
\end{itemize}  

\begin{figure}[!ht]
    \centering
    \includegraphics[width=.8\linewidth]{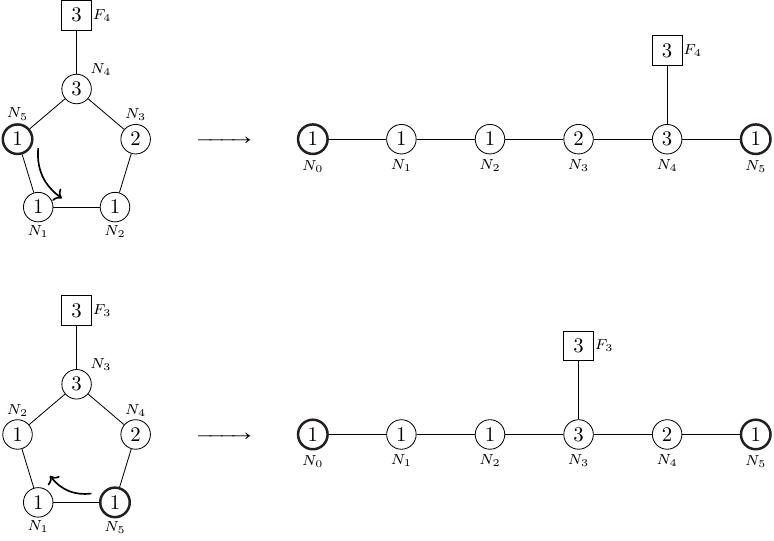}
    \caption{The quiver in Figure~\ref{fig:circular_brane_setup} provides an example of a circular theory for which the choice of the node at which to cut open the theory is non-unique. Such a node, highlighted in bold, must lie at the edge of the chain of nodes with minimal rank. The two possible choices lead to different linear quivers. Next to the nodes, we indicate the associated numbering, which follows the orientation specified by the arrow. Notice that the first and last nodes in the linearized quivers on the right are identified, namely $N_0 \equiv N_5$, since they arise from cutting open the bold node in the circular quivers on the left. In other words, they together contribute a single $\mathcal{N}=4$ vector multiplet of rank $N_0 = N_5$.
    }
    \label{fig:example_cutopen}
\end{figure}

Once the $L$-th node and the ordering are chosen following the rules described above, it is possible to encode the information of the gauge ranks $N_i$ and the number of flavors $F_i$ in partitions of a positive integer number $N$ that is defined by
\begin{align}
    N = \sum_{i=1}^L i F_i \,.
\end{align}
In analogy with the usual notation for $T_\rho^\sigma[SU(N)]$ theories \cite{Gaiotto:2008ak}, for circular theories \cite{Assel:2012cj} one can define $C_{\rho}^{\sigma}[SU(N);N_L]$, where $N_L$ denotes the rank of the node at which the quiver is cut open, and the two partitions $\rho$ and $\sigma$ of $N$ are related to $N_i$ and $F_i$ by
\begin{align}
    F_i&=(\sigma^T)_i-(\sigma^T)_{i+1} \,,\qquad
    N_i=N_L+\sum_{j\leq i}(\sigma^T)_{j}-\sum_{j\leq i}\rho_j \,,
\end{align}
for $i=1,\dots,L$. Conventionally, we also take $N_0 = N_L$ and $F_0 = 0$. Note that the data of $N_i$ and $F_i$ are entirely encoded in the integer $N_L$ and the two partitions $\rho$ and $\sigma$.
The reasoning behind assigning a partition to a circular quiver, as well as the convenience of the rules for selecting the $L$-th node, is explained in more detail in Appendix~\ref{app:partitions}. For the purpose of the dualization algorithm, it suffices to know that the data $(N_i, F_i)$ can be encoded in two partitions and that there exists a convenient choice for the $L$-th node.  

Indeed, the mirror dual quiver of a circular theory can be derived by performing S-duality on the Type IIB brane setup, which swaps NS5s and D5s. This procedure reveals that the mirror dual quiver can be simply obtained by swapping the partitions $\rho$ and $\sigma$, while the parameter $N_L$ is kept fixed:
\begin{equation}
    C^\sigma_\rho [SU(N);N_L] 
    \quad\xleftrightarrow[]{\;\;\;\; \mathcal{S} \;\;\;\;} \quad 
    C^\rho_\sigma[SU(N);N_L] \,.
\end{equation}

\subsection{The partition function of circular theories}
\label{subsec:Crhosigma}

Another key ingredient for executing the dualization algorithm, which we will present in Section~\ref{sec:dualization_algorithm_circular}, is the choice of parameterization for the theory.  
A major advantage of the algorithm is that it enables an exact mapping of symmetries and parameters across the duality.  
This parameterization can be conveniently encoded in the $S^3_b$ partition function, which for a generic $C_\rho^\sigma[SU(N);N_L]$ we choose to be:  
\begin{align}\label{eq:parfun_circular}
	& \mathcal{Z}_{C_{\rho}^{\sigma}[SU(N);N_L]}\left(\vec{X};\vec{Y};B;E;m_A\right) = \nonumber \\
	& \quad = 
	e^{2 \pi i N_L Y_{1} B}
	\times
	\prod_{j=1}^{L}\prod_{k=j+1}^{L}e^{2\pi i Y_k \sum_{a=1}^{F_j}X_a^{(j)}}
	\nonumber\\
	& \qquad\times
	\int \prod_{j=1}^{L} \left(\mathrm{d}{\vec{Z}_{N_j}^{(j)}}\Delta_{N_j}\left(\vec{Z}^{(j)};m_A \right)\right)
	\prod_{j=1}^{L} e^{2 \pi i (Y_j-Y_{j+1}+\delta_{j,L}E)\sum_{a=1}^{N_j}Z_a^{(j)}}
	\nonumber\\
	& \qquad\quad \times
    \prod_{j=1}^{L}\prod_{a=1}^{N_j}\prod_{b=1}^{N_{j+1}} s_b\left( i\frac{Q}{2}-m_A\pm\left(Z_a^{(j)}-Z_b^{(j+1)}-\delta_{j,L}B\right) \right)
	\nonumber\\
	& \qquad\quad \times
	\prod_{j=1}^{L}\prod_{a=1}^{N_j}\prod_{b=1}^{F_{j}} s_b\left( i\frac{Q}{2}-m_A\pm\left(Z_a^{(j)}-X_b^{(j)}\right) \right) \,.
\end{align}
See Appendix~\ref{app:conventions_PF} for the convention on the partition function. 
The same parameterization is illustrated in Figure~\ref{fig:generic_Crhosigma}, where the generic $C_\rho^\sigma[SU(N);N_L]$ theory is depicted using the 3d $\mathcal{N}=4$ quiver notation.  

\begin{figure}[!ht]
\centering
    \includegraphics[width=.65\textwidth]{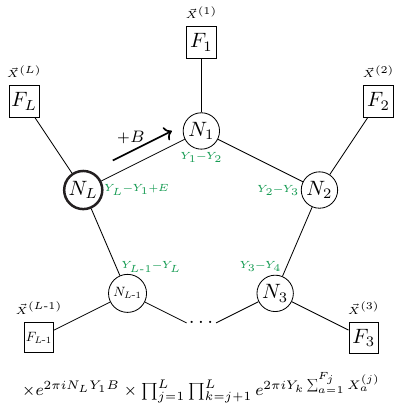}
    \caption{
    The quiver representation of the generic $C_{\rho}^{\sigma}[SU(N);N_L]$ theory. The FI parameter associated with each gauge node is shown in green. 
    The last node chosen according to the rules in Section~\ref{subsec:cutting} is highlighted in bold.
    An arrow from the last node to the first, with the label $+B$, indicates that the bifundamental chirals between $U(N_L)\times U(N_1)$ carries baryonic charges $\pm 1$ with the associated real mass $B$. The corresponding double sine factor is $\prod_{a,b}s_b(i{Q}/{2}-m_A\pm(Z_a^{(L)}-Z_b^{(1)}-B))$.
    At the bottom, we explicitly display the BF couplings in the $S^3_b$ partition function language.
    Throughout the paper, all quivers follow the conventions introduced here. When relevant, we indicate whether a hypermultiplet is twisted or not using the labels tw.h. or h., respectively.
    }
    \label{fig:generic_Crhosigma}
\end{figure}

The partition function depends on several parameters associated with the symmetries of the theory. First, there are $L$ parameters $Y_i$, with $Y_{L+1} = Y_1$, which parameterize the FI terms of the gauge nodes. In general, the $j$th gauge node has the FI parameter $Y_j-Y_{j+1}$, a choice that simplifies the mirror map. These are, however, not independent, since their sum vanishes. We therefore introduce an additional parameter $E$ so that the FI of the last node is $Y_L-Y_1+E$. With this extra parameter, we correctly obtain a total of $L$ independent parameters for the $L$ FI terms.

Similarly, there is another set of $L$ parameters $X_j$, associated with the flavor symmetries $S\left[\prod_{j=1}^L U(F_j)\right]$. Note that the diagonal $U(1)$ is gauged, which is consistent with the fact that the partition function depends only on differences of $X_j$ because the sum of $X_j$ can be set to zero by shifting the integration variables $Z^{(j)}_a$ simultaneously. In addition, the circular quiver admits an additional baryonic $U(1)_B$ symmetry under which a pair of bifundamental fields between the gauge nodes $U(N_L)$ and $U(N_1)$ have charges $\pm1$, respectively. We denote the corresponding parameter by $B$. 

Although we have introduced $E$ and $B$ for the last FI term and for the last bifundamental fields, respectively, these parameters could equivalently be assigned to other gauge groups, up to the redefinition of $\vec X$ and $\vec Y$ and background BF couplings. For definiteness, however, we stick to this choice of parameters and background BF couplings appearing in the second line of \eqref{eq:parfun_circular}.

Finally, we have the axial mass $m_A$ associated with the axial $U(1)_A$ symmetry, which is a non-R commutant of $U(1)_C \subset SU(2)_C$ in the $\mathcal N=4$ R-symmetry, $SU(2)_C \times SU(2)_H$. More precisely, we choose the $\mathcal N=2$ subgroup R-charge and the axial charge as
\begin{align}
    R &= C \,, \\
    A &= H-C \,,
\end{align}
where $C$ and $H$ are the Cartan charges of $SU(2)_C \times SU(2)_H$, respectively, normalized to be integers. Although this is not the canonical choice, the canonical assignment with R-charge 1/2 for hypermultiplets can be recovered by defining a new R-charge $R' = R+\frac{A}{2}=\frac{H+C}{2}$. This is equivalent to shifting $m_A \rightarrow m_A+i \frac{Q}{4}$ in the partition function. \\

As already stated, mirror symmetry has the effect of swapping the partitions $\rho$ and $\sigma$. The parameters $(\vec{X},B)$ and $(\vec{Y},E)$, related to the flavor and topological symmetries, respectively, are then swapped under mirror symmetry. Moreover, the two $SU(2)$ factors of the $\mathcal{N}=4$ R-symmetry are swapped, which is reflected in a redefinition of the axial mass $m_A \to iQ - m_A$.\footnote{One usually refers to hypermultiplets with this redefined axial mass as `twisted'. When needed, we distinguish between hypermultiplets and twisted hypermultiplets in the quiver diagrams using the labels h.~or tw.h.~respectively.} \
In general, however, it is not a trivial task to establish the precise map, nor to fix the correct background terms that are present in the duality. All of this can be tackled using the dualization algorithm, as already done for linear quivers in \cite{Hwang:2021ulb}.\footnote{See also \cite{Bottini:2021vms,Comi:2022aqo} for a comprehensive discussion of the dualization algorithm strategy and also \cite{Marino:2025uub,Giacomelli:2023zkk,Giacomelli:2024laq,Comi:2025zwu,Benvenuti:2023qtv} for applications and extensions.}

Although we operate with the 3d $\mathcal{N}=2^*$ partition function, the mass-deformed one with nonzero $m_A$, we will present the quivers in the 3d $\mathcal{N}=4$ language. In order to translate between the figures and the partition function contributions, see Table~\ref{tab:dictionary} in Appendix~\ref{app:conventions_PF}, where the dictionary between the two languages is provided.\\

\section{The dualization algorithm for good circular quivers} 
\label{sec:dualization_algorithm_circular}

The core idea of the dualization algorithm is to decompose a theory into its fundamental QFT components, referred to as \textit{QFT blocks}, which consist of hypermultiplets in various representations of the gauge groups.
To each QFT block, we apply a \textit{local basic duality move}, which can be derived assuming only Aharony duality \cite{Aharony:1997gp}, the 3d counterpart of the original Seiberg duality \cite{Seiberg:1994pq}. After gluing the dualized blocks back together and Higgsing gauge groups via a duality equivalent to the Hanany--Witten transition, the resulting theory is mirror dual to the original one.

This algorithm was initially developed for linear quivers in \cite{Bottini:2021vms,Hwang:2021ulb,Comi:2022aqo}. The QFT blocks and the basic duality moves introduced there are reviewed in Appendix~\ref{app:dualization_algorithm_ingredients}. In this section, instead, we introduce additional ingredients to extend the algorithm to circular theories. 
The main idea is that a circular quiver can be cut open and linearized; the resulting linear theory can then be dualized using the standard dualization algorithm, and lastly the quiver can be closed back into a circular one. This strategy requires a careful definition of the operations of opening and closing circular quivers, ensuring that no information is lost in the process.\footnote{A complementary approach to circular mirror symmetry was recently discussed in \cite{Jiang:2025kho}.}

To this end, we first define two additional blocks, which we call {\it baryonic} and {\it topological}, and provide their duality moves under the full $SL(2,\mathbb{Z})$ duality group. We then detail the steps of the dualization algorithm for circular theories, which involve linearizing a quiver, dualizing the resulting linear quiver, and subsequently closing it. Finally, we apply this extended algorithm to four examples: (1) a mirror pair of canonical circular quivers, for which the computation is carried out in full detail, (2) an $SL(2,\mathbb Z)$ triality arising from an ABJM-like theory, (3) an $S$-fold setup, and (4) $U(N)$ SQCD with one antisymmetric and one symmetric tensor.

\subsection{New ingredients}
\label{subsec:new_ingredients}

To perform the dualization algorithm on circular quivers, we first introduce two new blocks accounting for the baryonic mass and the diagonal FI term of the quiver. As discussed in Section~\ref{subsec:Crhosigma}, these are associated with the baryonic and (diagonal) topological symmetries, which need to be tracked throughout the dualization procedure. Therefore, we refer to them as the baryonic and topological blocks and define them as follows:
\begin{itemize}
    \item 
    The baryonic block $\mathcal{B}_{\text{bar}}$ is a shifted Identity-wall, as defined in \eqref{eq:shifted_Id_wall}, multiplied by a contact-term factor chosen for convenience in the dualization procedure.
    It is represented in the first line of Figure~\ref{fig:Bbar_Btop}, and its partition function contribution is
    \begin{equation}
        \mathcal{Z}_{\mathcal{B}_{\text{bar}}}^{(N,N)}(\vec{X};\vec{Y};m_A;B)
        =
        {}_{\vec{X}}\hat{\mathbb{I}}_{\vec{Y}+B} (m_A)
        \times
        e^{-\pi i N B^2}
        \,.
    \end{equation}
    Notice that it satisfies the property
    \begin{equation}
        \mathcal{Z}_{\mathcal{B}_{\text{bar}}}^{(N,N)}(\vec{X};\vec{Y};m_A;B)
        =
        \mathcal{Z}_{\mathcal{B}_{\text{bar}}}^{(N,N)}(\vec{Y};\vec{X};m_A;-B)
        \,.
    \end{equation}
    \item 
    The topological block $\mathcal{B}_{\text{top}}$ is an Identity-wall with an additional FI term, again multiplied by a contact-term factor for convenience in the dualization procedure.
    It is represented in the second line of Figure~\ref{fig:Bbar_Btop}, and its partition function contribution is
    \begin{equation}
        \mathcal{Z}_{\mathcal{B}_{\text{top}}}^{(N,N)}(\vec{X};\vec{Y};m_A;E)
        =
        {}_{\vec{X}}\hat{\mathbb{I}}_{\vec{Y}} (m_A)
        \times
        e^{2\pi i E \sum_{j=1}^{N} Y_j}
        \times
        e^{-\pi i N E^2}
        \,.
    \end{equation}
    Notice that it satisfies the property
    \begin{equation}
        \mathcal{Z}_{\mathcal{B}_{\text{top}}}^{(N,N)}(\vec{X};\vec{Y};m_A;E)
        =
        \mathcal{Z}_{\mathcal{B}_{\text{top}}}^{(N,N)}(\vec{Y};\vec{X};m_A;E)
        \,.
    \end{equation}
\end{itemize}
\begin{figure}[!ht]
    \centering    \includegraphics[width=.75\textwidth]{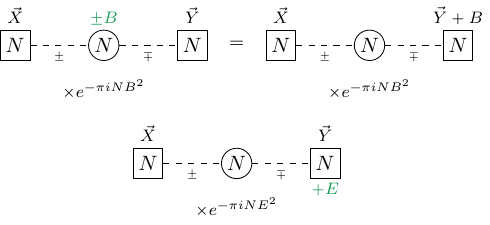}
    \caption{
    The baryonic block $\mathcal{B}_{\text{bar}}$ (top) and the topological block $\mathcal{B}_{\text{top}}$ (bottom). Each block can be defined using either of two equivalent choices of correlated signs, both resulting in the same block.}
    \label{fig:Bbar_Btop}
\end{figure}
The action of the full $SL(2,\mathbb{Z})$ duality group on $\mathcal{B}_{\text{bar}}$ and $\mathcal{B}_{\text{top}}$ is as follows.

\paragraph{$\mathcal{S}$-dualization.}
The $\mathcal{S}$ generator schematically acts on $\mathcal{B}_{\text{bar}}$ and $\mathcal{B}_{\text{top}}$ as
\begin{align}
    \mathcal{B}_{\text{bar}} &= \mathcal{S}\,\mathcal{B}_{\text{top}}\,\mathcal{S}^{-1} \,, \label{eq:baryonicSdual}\\
    \mathcal{B}_{\text{top}} &= \mathcal{S}\,\mathcal{B}_{\text{bar}}^{\vee}\,\mathcal{S}^{-1} \,,
\end{align}
where the superscript $\vee$ in $\mathcal{B}_{\text{bar}}$ indicates that the sign of baryonic mass (or the FI sign for $\mathcal{B}_{\text{top}}$), is flipped (see \eqref{eq:Btop_Sdualized}).
The identities in terms of the partition function are
\begin{align}
    &
    \mathcal{Z}_{\mathcal{B}_{\text{bar}}}^{(N,N)}(\vec{X};\vec{Y};m_A;B)
    =
    \int\udl{\vec{Z}_N}\Delta_N(\vec{Z};m_A) 
    \int\udl{\vec{W}_N}\Delta_N(\vec{W};m_A) 
    \times
    \label{eq:Bbar_Sdualized}\\
    &\qquad\qquad\qquad\qquad\qquad\quad\times
    \mathcal{Z}_{\mathcal{S}}^{(N,N)}(\vec{X};\vec{Z};m_A)
    \mathcal{Z}_{\mathcal{B}_{\text{top}}}^{(N,N)}(\vec{Z};\vec{W};m_A;B)
    \mathcal{Z}_{\mathcal{S}^{-1}}^{(N,N)}(\vec{W};\vec{Y};m_A) 
    \nonumber\,,
    \\[5pt]
    &
    \mathcal{Z}_{\mathcal{B}_{\text{top}}}^{(N,N)}(\vec{X};\vec{Y};m_A;E)
    =
    \int\udl{\vec{Z}_N}\Delta_N(\vec{Z};m_A) 
    \int\udl{\vec{W}_N}\Delta_N(\vec{W};m_A) 
    \times
    \label{eq:Btop_Sdualized}\\
    &\qquad\qquad\qquad\qquad\qquad\quad\times
    \mathcal{Z}_{\mathcal{S}}^{(N,N)}(\vec{X};\vec{Z};m_A)
    \mathcal{Z}_{\mathcal{B}_{\text{bar}}}^{(N,N)}(\vec{Z};\vec{W};m_A;-E)
    \mathcal{Z}_{\mathcal{S}^{-1}}^{(N,N)}(\vec{W};\vec{Y};m_A) 
    \nonumber\,.
\end{align}
They are proved in Appendix~\ref{app:Sdual_Bbar_Btop_proofs} using the definition \eqref{eq:shifted_Id_wall} of the shifted Identity-wall.

\paragraph{$\mathcal{T}$-dualization.}
The $\mathcal{T}$ generator schematically acts on $\mathcal{B}_{\text{bar}}$ and $\mathcal{B}_{\text{top}}$ as
\begin{align}
\mathcal{B}_\text{bar} &= \mathcal{T} \, \mathcal{B}_\text{top} \, \mathcal{B}_\text{bar} \, \mathcal{T}^{-1} \,, \\
\mathcal{B}_\text{top} &= \mathcal{T} \, \mathcal{B}_\text{top} \, \mathcal{T}^{-1} \,.
\end{align}
The corresponding partition function identities are
\begin{align}
    \mathcal{Z}_{\mathcal{B}_{\text{bar}}}^{(N,N)}(\vec{X};\vec{Y};m_A;B)
    &= 
    \int\udl{\vec{Z}_N}\Delta_N(\vec{Z};m_A) 
    \int\udl{\vec{V}_N}\Delta_N(\vec{V};m_A) 
    \int\udl{\vec{W}_N}\Delta_N(\vec{W};m_A)
    \times\nonumber \\
    &\quad \times 
    \mathcal{Z}_{\mathcal{T}}^{(N,N)}(\vec{X};\vec{Z};m_A) 
    \mathcal{Z}_{\mathcal{B}_{\text{top}}}^{(N,N)}(\vec{Z};\vec{V};m_A;B)
    \times \label{eq:Bbar_Tdualized} \\
    &\quad \times 
    \mathcal{Z}_{\mathcal{B}_{\text{bar}}}^{(N,N)}(\vec{V};\vec{W};m_A;B)
    \mathcal{Z}_{\mathcal{T}^{-1}}^{(N,N)}(\vec{W};\vec{Y};m_A) 
    \,, \nonumber \\[5pt]
    \mathcal{Z}_{\mathcal{B}_{\text{top}}}^{(N,N)}(\vec{X};\vec{Y};m_A;E)
    &= 
    \int\udl{\vec{Z}_N}\Delta_N(\vec{Z};m_A) 
    \int\udl{\vec{W}_N}\Delta_N(\vec{W};m_A)
    \times \label{eq:Btop_Tdualized} \\
    &\quad \times 
    \mathcal{Z}_{\mathcal{T}}^{(N,N)}(\vec{X};\vec{Z};m_A) 
    \mathcal{Z}_{\mathcal{B}_{\text{top}}}^{(N,N)}(\vec{Z};\vec{W};m_A;E)
    \mathcal{Z}_{\mathcal{T}^{-1}}^{(N,N)}(\vec{W};\vec{Y};m_A) 
    \,.\nonumber
\end{align}
It is straightforward to prove these identities by inserting the $\mathcal{T}$ operator defined in \eqref{eq:T_operator_PF}.

\paragraph{$\mathcal{T}^T$-dualization.}
The $\mathcal{T}^T$ generator schematically acts on $\mathcal{B}_{\text{bar}}$ and $\mathcal{B}_{\text{top}}$ as
\begin{align}
\mathcal{B}_\text{bar} &= \mathcal{T}^T \, \mathcal{B}_\text{bar} \, (\mathcal{T}^{-1})^T \,, \\
\mathcal{B}_\text{top} &= \mathcal{T}^T \, \mathcal{B}_\text{top} \, \mathcal{B}_\text{bar} \, (\mathcal{T}^{-1})^T \,.
\end{align}
The corresponding partition function identities are
\begin{align}
    \mathcal{Z}_{\mathcal{B}_{\text{bar}}}^{(N,N)}(\vec{X};\vec{Y};m_A;B)
    &= 
    \int\udl{\vec{Z}_N}\Delta_N(\vec{Z};m_A) 
    \int\udl{\vec{W}_N}\Delta_N(\vec{W};m_A)
    \times\label{eq:Bbar_Ttdualized} \\
    &\quad \times 
    \mathcal{Z}_{\mathcal{T}^T}^{(N,N)}(\vec{X};\vec{Z};m_A) 
    \mathcal{Z}_{\mathcal{B}_{\text{bar}}}^{(N,N)}(\vec{Z};\vec{W};m_A;B)
    \mathcal{Z}_{(\mathcal{T}^T)^{-1}}^{(N,N)}(\vec{W};\vec{Y};m_A) 
    \,, \nonumber\\[5pt]
    \mathcal{Z}_{\mathcal{B}_{\text{top}}}^{(N,N)}(\vec{X};\vec{Y};m_A;E)
    &= 
    \int\udl{\vec{Z}_N}\Delta_N(\vec{Z};m_A) 
    \int\udl{\vec{V}_N}\Delta_N(\vec{V};m_A) 
    \int\udl{\vec{W}_N}\Delta_N(\vec{W};m_A)
    \times\nonumber \\
    &\quad \times 
    \mathcal{Z}_{\mathcal{T}^T}^{(N,N)}(\vec{X};\vec{Z};m_A) 
    \mathcal{Z}_{\mathcal{B}_{\text{top}}}^{(N,N)}(\vec{Z};\vec{V};m_A;E)
    \times\nonumber \\
    &\quad \times 
    \mathcal{Z}_{\mathcal{B}_{\text{bar}}}^{(N,N)}(\vec{V};\vec{W};m_A;E)
    \mathcal{Z}_{(\mathcal{T}^T)^{-1}}^{(N,N)}(\vec{W};\vec{Y};m_A) 
    \,. \label{eq:Btop_Ttdualized}
\end{align}
These identities can be straightforwardly verified by inserting the $\mathcal{T}^T$ operator defined in \eqref{eq:Tt_operator_PF} and using the $\mathcal S$- and $\mathcal T$-dualizations given in \eqref{eq:Bbar_Sdualized}, \eqref{eq:Btop_Sdualized}, \eqref{eq:Bbar_Tdualized}, \eqref{eq:Btop_Tdualized}.

\subsection{The algorithm}
\label{subsec:algorithm_circular}

We now provide the algorithm to dualize good circular quivers. 
The steps are as follows.
\begin{enumerate}
\setcounter{enumi}{-1}
    \item Cut open the quiver according to the rules given in Section~\ref{subsec:cutting}.
    \item Decompose the resulting linear quiver into QFT blocks, including the baryonic and topological blocks that account for the $U(1)_B$ and $U(1)_E$ global symmetry factors. 
    \item Dualize the QFT blocks using the basic duality moves (see Section~\ref{subsec:new_ingredients} and Appendix~\ref{subsec:basic_moves}). 
    \item Glue back the dualized blocks.
    \item Since asymmetric $\mathcal S$-walls in the dualized quiver signal the presence of operator VEVs, perform the Hanany--Witten move (see Appendix~\ref{subsec:basic_moves}) repeatedly until no asymmetric $\mathcal{S}$-walls remain, i.e., until all VEVs are extinguished by Higgsing.
    \item Close the linear quiver back into a circular one to obtain the dual of the original circular theory.
\end{enumerate}

\subsection{Example 1: A mirror pair}

We now demonstrate the dualization algorithm at work on a simple example.
Consider the mirror pair in Figure~\ref{fig:mirror_pair}. Starting from the theory on the left, we fix the parametrization by selecting the special node (highlighted in bold) and the orientation (clockwise in this case) of the quiver, following the criteria of Section~\ref{subsec:cutting}. With this choice, the theory can be written as $C_{[3,3]}^{[2,2,1,1]}[SU(6);1]$, where the partitions are assigned as explained in Appendix~\ref{app:partitions}.\\

\begin{figure}[!ht]
\centering
    \includegraphics[width=1\textwidth]{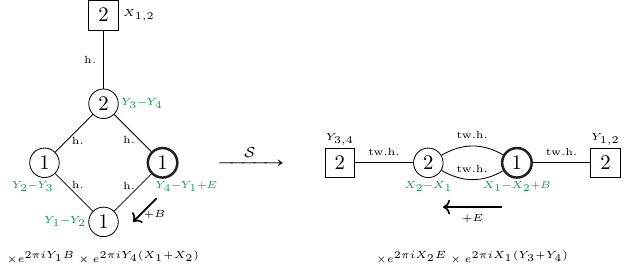}
    \caption{
    A circular mirror pair. The choice of the special node on the electric quiver on the left is such that the theory can be written as $C_{\rho}^{\sigma}[SU(6);1]$ with $\rho=[2,2,1,1]$ and $\sigma=[3,3]$, while its $\mathcal{S}$-dual on the right can be written as $C^{\rho}_{\sigma}[SU(6);1]$. The notation and conventions adopted are those showcased and explained in Figure~\ref{fig:generic_Crhosigma}. Moreover, close to each link, we wrote h.~for hyper in the electric theory and tw.h.~for twisted hyper in the magnetic theory.}
    \label{fig:mirror_pair}
\end{figure}

\noindent
The algorithm, illustrated in Figure~\ref{fig:mirror_pair_dualization}, proceeds through the following steps.
\begin{enumerate}
\setcounter{enumi}{-1}
	\item Cut open the quiver at the chosen special node. 
	\item Decompose the resulting linear quiver into blocks, as shown in the second line of Figure~\ref{fig:mirror_pair_dualization}. 
    Since the last FI parameter is shifted by $+E$ and the last bifundamental carries baryonic mass $+B$, there must be corresponding topological and baryonic blocks satisfying
	\begin{align}
        &
        \int\udl{\vec{V}_1}\Delta_1(\vec{V};m_A) \,
        \mathcal{Z}_{\mathcal{B}_{\text{top}}}^{(1,1)}(\vec{Z}^{(4)};\vec{V};m_A;E)
        \mathcal{Z}_{\mathcal{B}_{\text{bar}}}^{(1,1)}(\vec{V};\vec{Z}^{(0)};m_A;B)
		=\nonumber\\
        &\quad=
        \int\udl{\vec{V}_1}\Delta_1(\vec{V};m_A) \,
		e^{2 \pi i E Z^{(4)}} 
		{}^{\phantom{}}_{Z^{(4)}}\hat{\mathbb{I}}^{\phantom{}}_{V}(m_A) \;
		{}^{\phantom{}}_{V}\hat{\mathbb{I}}^{\phantom{}}_{Z^{(0)}+B}(m_A) \;
		e^{- \pi i N(B^2+E^2)} \nonumber\\
		&\quad=
		e^{2 \pi i E Z^{(4)}}
		{}^{\phantom{}}_{Z^{(4)}}\hat{\mathbb{I}}^{\phantom{}}_{Z^{(0)}+B} \;
		e^{-\pi i N(B^2+E^2)} \,,
	\end{align} 
	where the Identity-wall enforces the condition $Z^{(0)}=Z^{(4)}-B$. 
	\item Dualize the blocks using the basic duality moves. 
    \item Glue back the dualized blocks, as shown in the third line of Figure~\ref{fig:mirror_pair_dualization}.
 	\item Perform the Hanany--Witten move implementing the Higgs mechanism triggered by nonzero operator VEVs, as indicated by the arrows beneath the quiver in the third line of Figure~\ref{fig:mirror_pair_dualization}.
	\item Close the linear quiver back into a circular one as shown in the first diagram in the fourth line of Figure~\ref{fig:mirror_pair_dualization}. Then sequentially implement the Identity-walls. The effect of the Identity-wall at the bottom is to shift the FI of the leftmost gauge node from $+X_1$ to $+X_1+B$, as shown in the second diagram in the fourth line. The effect of the second Identity-wall involving the FI parameter $+E$ is to identify $\vec W^{(0)}$ with $\vec W^{(2)}-E$. Since the rightmost gauge node has the FI parameter $-X_2$, imposing $\vec W^{(0)} = \vec W^{(2)}-E$ leads to an additional BF coupling $e^{2 \pi i X_2 E}$, as shown in the last diagram.
\end{enumerate}

The final theory can be written as $C_{[3,3]}^{[2,2,1,1]}[SU(6);1]$, which is the correct mirror dual of the initial one.
Notice that, in the magnetic theory, once the special node, and hence the last node, is chosen consistently, the last bifundamental carries the baryonic mass $E$. At first sight, this seems inconsistent with \eqref{eq:Btop_Sdualized}, where the dualized parameter acquires an additional sign. However, this sign is compensated by the reversed orientation of the dual quiver.\footnote{Indeed, in Figure~\ref{fig:mirror_pair_dualization}, we have labeled the gauge Cartans after the dualization so that the special node (highlighted in bold) carries the ``last'' Cartan, in accordance with our conventions.}

The mirror dual of any good circular quiver can be constructed using the same procedure. The resulting mirror theory can be generated by the accompanying Mathematica code. \\

\begin{figure}[!ht]
\centering
    \includegraphics[width=\textwidth,center]{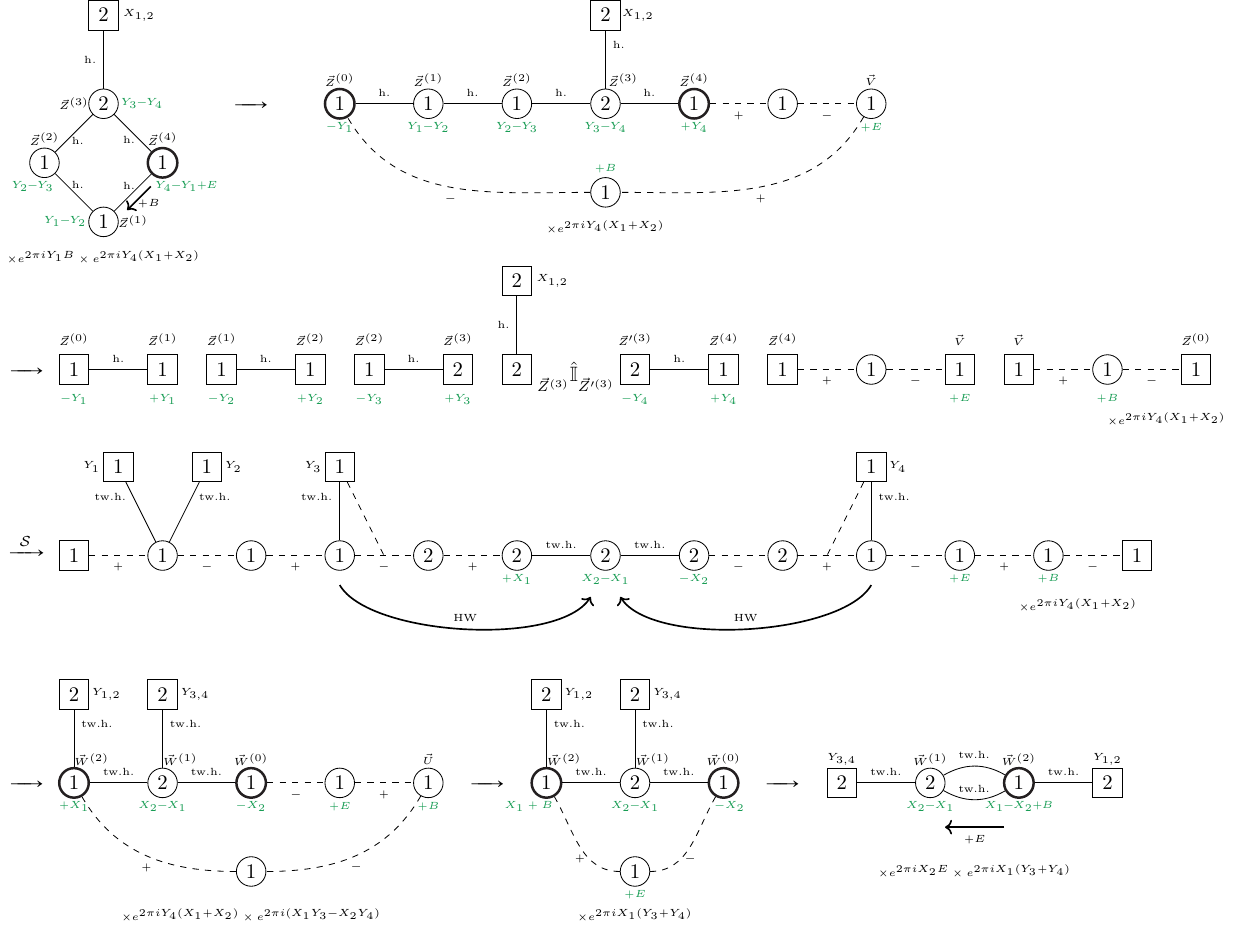}
    \caption{Application of the dualization algorithm to a circular mirror pair.
    }
    \label{fig:mirror_pair_dualization}
\end{figure}

\paragraph{Global symmetries and operator map}
The global symmetry of the dual pair is
\begin{equation}
    SU(2)_{X} \times
    SU(2)_{Y_{1,2}} \times
    SU(2)_{Y_{3,4}} \times
    U(1)_{B} \times
    SU(2)_{E} \,.
\end{equation}
Notice that the $U(1)_{E}$ diagonal topological symmetry on the electric side is enhanced to $SU(2)_{E}$ in the IR, which manifests as the $SU(2)$ symmetry acting on the pair of bifudnamental twisted hypermultiplets on the magnetic side.
Since the symmetry parameters can be tracked at every step of the algorithm, the algorithm naturally provides a map of global symmetries and gauge-invariant operators across the duality between the two theories shown in Figure~\ref{fig:mirror_pair}. 
We present several operator maps in Tables~\ref{tab:operators_mirror_pair_1} and~\ref{tab:operators_mirror_pair_2}, using the field notation introduced in Figure~\ref{fig:mirror_pair_fields} and the symmetry charges summarized in Table~\ref{tab:fields_charges}. Monopole operators are denoted by $\mathfrak M^{(\cdot)}$, where $(\cdot)$ indicates the corresponding magnetic flux. For example, given a theory with gauge group $U(N_1)\times U(N_2)\times\dots$, the monopole operator with $U(N_1)$ flux $\{0,0,\dots,0\}$, $U(N_2)$ flux $\{+1,0,\dots,0\}$, and so on, is denoted by $\mathfrak{M}^{(\{0,0,\dots\},\{+,0,\dots\},\dots)}$. In addition, a monopole operator dressed with an adjoint field $A_j$ is denoted by $\mathfrak{M}^{(\cdot)}_{A_j}$.

\vspace{15pt}

\begin{figure}[H]
\centering
    \includegraphics[width=\textwidth]{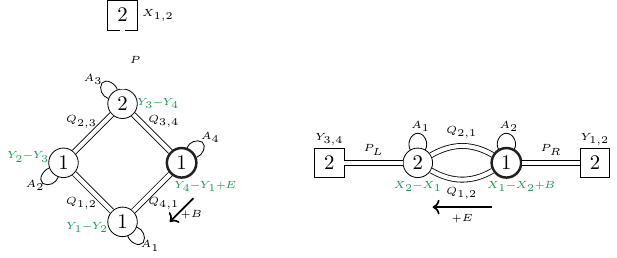}
    \caption{The circular mirror pair in the $\mathcal{N}=2$ notation. The pair of lines connecting two nodes $i$ and $j$ correspond to the $\mathcal N=2$ chiral fields $Q_{i,j}$ and $\tilde Q_{i,j}$ in the representations $\Box_i \times \Boxbar_j$ and $\Boxbar_i \times \Box_j$, respectively, composing an $\mathcal{N}=4$ hypermultiplet. For simplicity, only the labels $Q_{i,j}$ for the fundamental chirals are explicitly written in the figure.
    The arc $A_i$ represents the $\mathcal{N}=2$ adjoint chiral sitting inside the $\mathcal{N}=4$ vector multiplet of the gauge node $i$.}
    \label{fig:mirror_pair_fields}
\end{figure}

\vspace{15pt}

\begin{table}[!ht]
\centering
\input{Tables/fields_charges}
\caption{Charges conventions for the $\mathcal{N}=2$ fields in the mirror circular pair shown in Figure~\ref{fig:mirror_pair_fields}. The baryonic charge is not included, since it is assigned only to a specific link, as detailed in the figure.}
\label{tab:fields_charges}
\end{table}

\afterpage{%
\begin{landscape}
\setlength\extrarowheight{0pt}
\renewcommand{\arraystretch}{1.5}
\vspace*{\fill}
\begin{table}[H]
\input{Tables/operators_mirror_pair_el_mesons}
\caption{Operators' spectrum (part 1) for the mirror circular pair in Figure~\ref{fig:mirror_pair_fields}.}
\label{tab:operators_mirror_pair_1}
\end{table}
\vspace*{\fill}
\end{landscape}
}

\afterpage{%
\begin{landscape}
\setlength\extrarowheight{0pt}
\renewcommand{\arraystretch}{1.5}
\begin{table}
\input{Tables/operators_mirror_pair_mag_mesons}
\caption{Operators' spectrum (part 2) for the mirror circular pair in Figure~\ref{fig:mirror_pair_fields}.}
\label{tab:operators_mirror_pair_2}
\end{table}
\end{landscape}
}

\clearpage
\subsection{\texorpdfstring{Example 2: $SL(2,\mathbb Z)$ triality for an ABJM-like theory}{}}
\label{subsec:ABJM_triality}
Next, we present an example of the $SL(2,\mathbb{Z})$ generalization. Specifically, we consider a triality among the three circular quiver theories shown in Figure~\ref{fig:triality}. The first line exhibits a self-mirror pair of flavored circular quivers with $L$ gauge nodes, while the second line displays a circular quiver with $2 L$ gauge nodes and alternating Chern--Simons levels $\pm1$, which generalizes the ABJM model \cite{Aharony:2008ug}. These theories are known to describe the low-energy dynamics of a stack of $N$ M2-branes probing $\mathbb C^2/\mathbb Z_L \times \mathbb C^2/\mathbb Z_L$ in M-theory.

\begin{figure}[!ht]
	\centering
	\includegraphics[width=\textwidth]{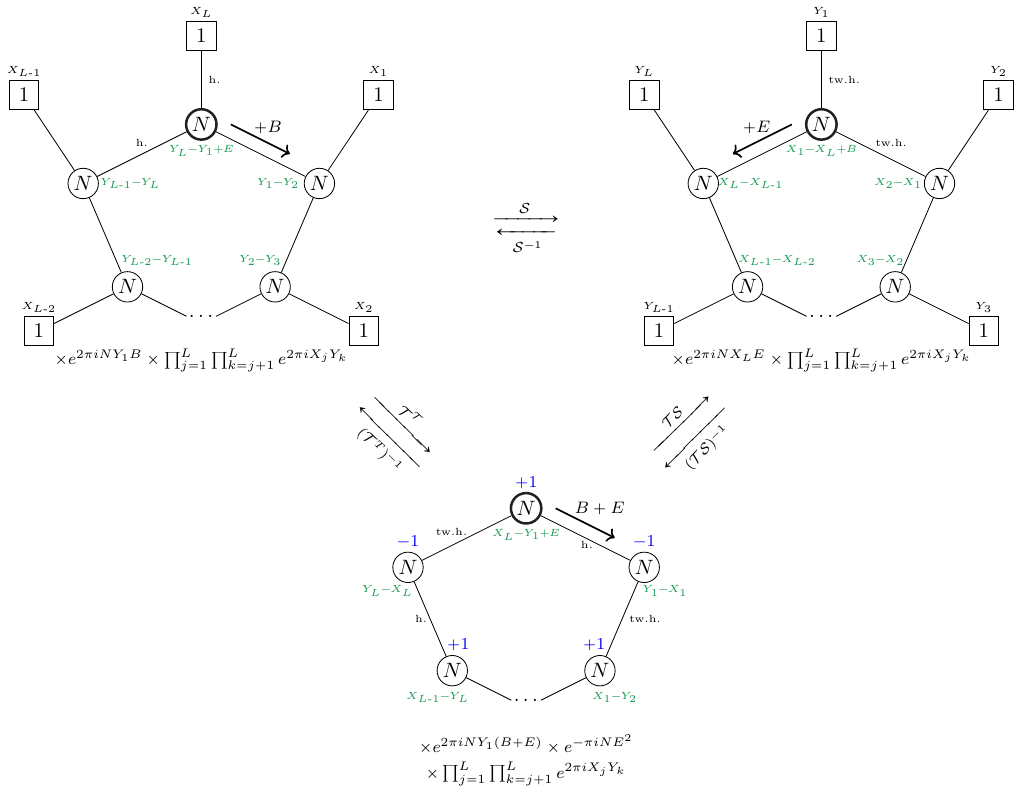}
	\caption{Triality for an ABJM-like theory. The starting theory in the upper-left corner has $L$ gauge nodes, and all its hypermultiplets are untwisted. An $\mathcal{S}$ transformation leads to the dual in the upper-right corner, which has $L$ gauge nodes and only twisted hypermultiplets. Acting with $\mathcal{T}^T$ on the first theory, or with $(\mathcal{T}\mathcal{S})^{-1}$ on the second, one obtains the dual theory shown at the bottom of the figure, which has $2L$ gauge nodes with alternating CS levels $\pm1$ and hypermultiplets that are alternatively twisted and untwisted.
    }
    \label{fig:triality}
\end{figure}

Let us begin with the theory in the upper-left corner of Figure~\ref{fig:triality}. The theory is self-dual under the $\mathcal S$-dualization, upon exchanging the topological symmetry with the flavor symmetry, including the baryonic one. To see this, we first decompose the circular quiver into blocks, including $\mathcal{B}_\text{top}$ and $\mathcal{B}_\text{bar}$, which will be glued back together after performing the $\mathcal S$-dualization. See Figure~\ref{fig:triality_Sdualization}, where the cutting and gluing steps are implemented implicitly.

\begin{figure}[!ht]
	\centering
	\includegraphics[width=\textwidth]{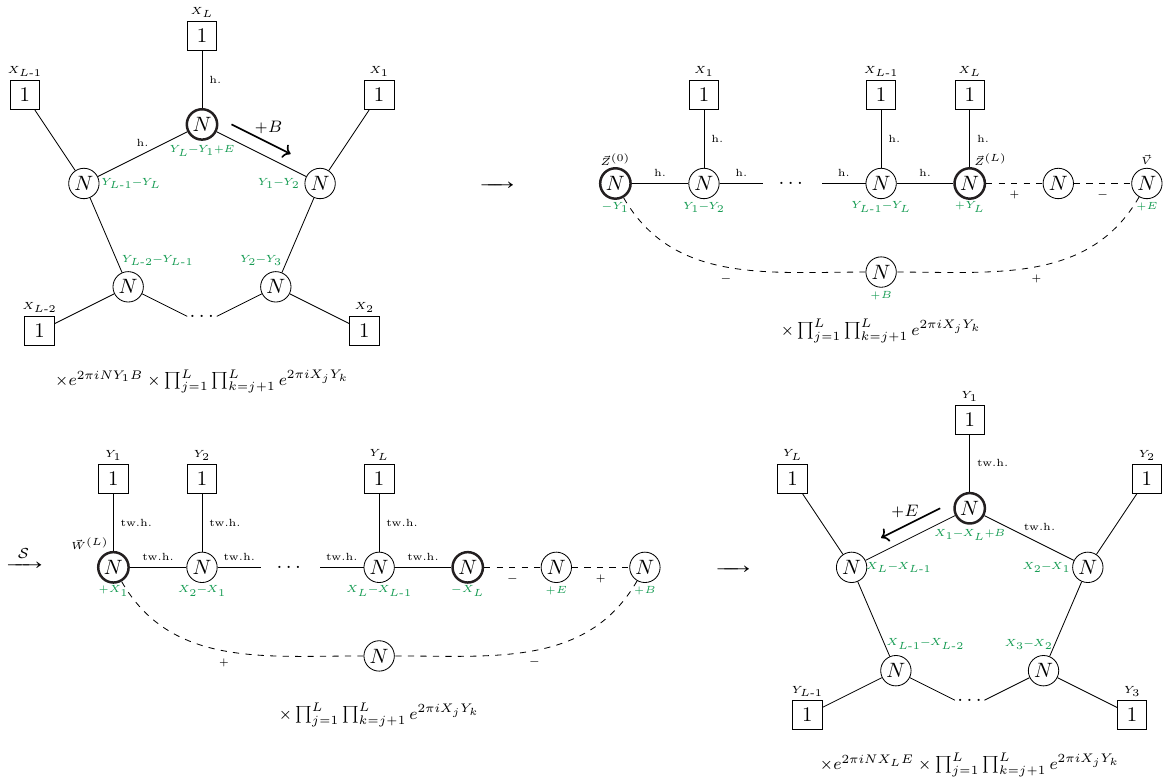}
	\caption{Steps in the $\mathcal S$-dualization procedure.}
    \label{fig:triality_Sdualization}
\end{figure}

As in the previous example, one can readily track how the symmetry parameters are mapped at each step of the dualization; under the $\mathcal S$-dualization, the global symmetries of the theory are mapped as follows:
\begin{equation}
\begin{split}
    SU(2)_C \quad &\longleftrightarrow \quad SU(2)_H \,, \\
    U(1)_B \quad &\longleftrightarrow \quad U(1)_E \,, \\
    S\left[\prod_{i = 1}^L U(1)_{X_i}\right] \quad &\longleftrightarrow \quad S\left[\prod_{i = 1}^L U(1)_{Y_{L-i+1}}\right] \,,
\end{split}    
\end{equation}
where $SU(2)_C \times SU(2)_H$ is the $\mathcal N=4$ $R$-symmetry, $U(1)_B$ is the baryonic symmetry, $S\left[\prod_{i = 1}^L U(1)_{X_i}\right]$ is the flavor symmetry, and $U(1)_E \times S\left[\prod_{i = 1}^{L} U(1)_{Y_{i}}\right]$ is the topological symmetry.
Notice that, for $L = 1$, the theory reduces to $\mathcal N=4$ $U(N)$ SQCD with one fundamental and one adjoint hypermultiplet, namely the ADHM quiver, which flows to an $\mathcal N=8$ SCFT in the IR \cite{Kapustin:2010xq}. In this case, some of these symmetries combine and enhance to the $SO(8)_R$ R-symmetry:
\begin{align}
    SU(2)_C \times SU(2)_H \times U(1)_B \times U(1)_E \quad \longrightarrow \quad SO(8)_R \,.
\end{align}

Next, let us consider the $\mathcal{T}^T$-dualization of the upper-left theory. The dual theory is a circular quiver with $2L$ $U(N)$ gauge nodes and alternating CS levels $\pm 1$, as depicted in the bottom line of Figure~\ref{fig:triality}.

As in the $\mathcal{S}$-dualization case, we follow the cutting prescription to decompose the circular quiver into $\mathcal{B}_\text{top}$, $\mathcal{B}_\text{bar}$ and a linear quiver, each of which can be $\mathcal{T}^T$-dualized using the basic duality moves (see Section~\ref{subsec:new_ingredients} and Appendix~\ref{app:dualization_algorithm_ingredients}). The schematic dualization procedure and the result are shown in Figure~\ref{fig:triality_Ttdualization}.
The $\mathcal{T}^T$-dual theory has $2L$ $U(N)$ gauge nodes, whose topological symmetries are given by
\begin{align}
    U(1)_E \times S\left[\prod_{i = 1}^L U(1)_{X_i}\right] \times S\left[\prod_{i = 1}^L U(1)_{Y_i}\right] \,,
\end{align}
whose rank is $2L-1$ rather than $2L$ because the anti-diagonal combination of $\prod_{i = 1}^L U(1)_{X_i}$ and $\prod_{i = 1}^L U(1)_{Y_i}$ is gauged due to the CS couplings. Also, the baryonic symmetry is now $U(1)_{B+E}$.

Notice that the $L=1$ case corresponds to the ABJM model with CS levels $\pm1$, which indeed exhibits enhanced $\mathcal{N}=8$ superconformal symmetry \cite{Aharony:2008ug}, in agreement with the supersymmetry enhancement in the upper frames.

\begin{figure}[!ht]
	\centering
	\includegraphics[width=\textwidth]{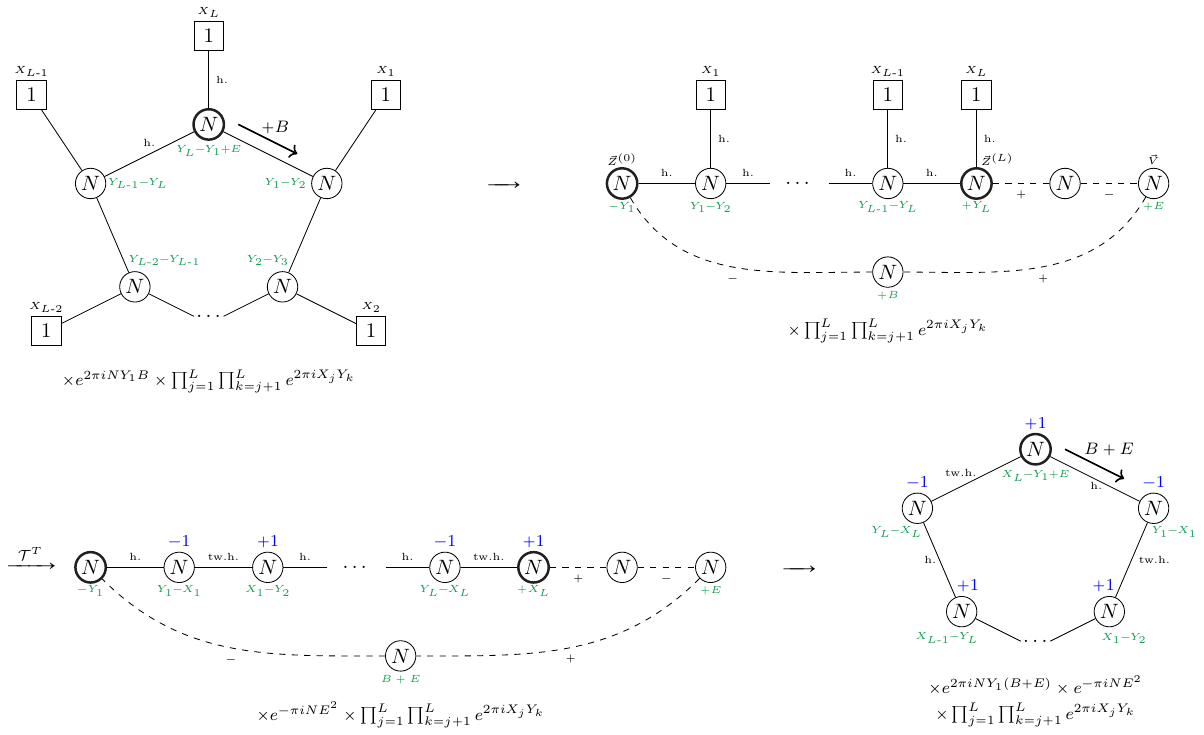}
	\caption{Steps in the $\mathcal T^T$-dualization procedure.}
    \label{fig:triality_Ttdualization}
\end{figure}

\subsection{Example 3: $\mathcal{S}$-fold theories}
The algorithm can be used also to recover the duals between 3d $\mathcal{S}$-fold theories (see e.g.~\cite{Terashima:2011qi,Ganor:2014pha,Gang:2015wya,Gang:2018wek,Gang:2018huc,Assel:2018vtq,Garozzo:2018kra,Garozzo:2019hbf,Garozzo:2019ejm,Beratto:2020qyk}). These models can be realized with a Type IIB brane setup consisting of compact D3-branes, with the peculiar feature of having an $\mathcal{S}$-wall inserted at some position along the compactified direction. For simplicity, we will restrict to an $\mathcal{N}=4$ theory with only one $\mathcal{S}$-wall, but our analysis can be extended to more general theories with CS levels and with multiple $\mathcal{S}$-walls.

Consider as an example the brane setup with $N$ compact D3-branes, one $\mathcal{S}$-wall and $n-1$ NS5-branes. The corresponding field theory is a circular quiver with $n$ $U(N)$ gauge nodes all connected by bifundamental hypermultiplets, except for one link which is replaced by an $\mathcal{S}$-wall theory; see Figure~\ref{fig:Sfold_th}.
The UV global symmetry of the theory consists only of the topological symmetries and we are parameterizing it as\footnote{Notice that this circular theory does not possess a baryonic symmetry. One way to observe this fact is that any baryonic symmetry could be reabsorbed in the definition of the parameter $E$ by a suitable gauge redefinition and using the property \eqref{eq:shifted_Swall} of the $\mathcal{S}$-wall.}
\begin{equation}\label{eq:symmSfold}
    U(1)_E\times S\left[\prod_{i=1}^nU(1)_{Y_i}\right]\,.
\end{equation}

\begin{figure}[!ht]
    \centering
    \includegraphics[width=0.9\linewidth]{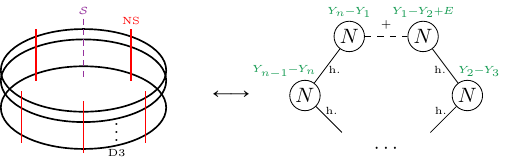}
    \caption{An example of $\mathcal{S}$-fold theory with the corresponding brane realization.}
    \label{fig:Sfold_th}
\end{figure}

This theory admits multiple duality frames, obtained by successively moving NS5-branes across the $\mathcal{S}$-wall, where each NS5-brane is turned into a D5-brane. In a generic frame, we have $n-k$ $U(N)$ gauge nodes all connected by bifundamental hypermultiplets, except for one link which is replaced by an $\mathcal{S}$-wall, and $k$ fundamental flavor attached to one of the nodes connected to the $\mathcal{S}$-wall, where $k$ runs from $0$ to $n-1$. This corresponds to a brane setup with $N$ compact D3s, one $\mathcal{S}$-wall, $n-k$ NS5s and $k$ D5s. For $k=n-1$ we obtain a brane setup consisting only of one $\mathcal{S}$-wall and $n-1$ D5-branes. The associated field theory contains a single $U(N)$ gauge node self-linked by an $\mathcal{S}$-wall and coupled to $n-1$ fundamental flavors, as shown in Figure~\ref{fig:Sfold_dualization_withbranes}.
\begin{figure}[!ht]
    \centering
    \includegraphics[width=\linewidth]{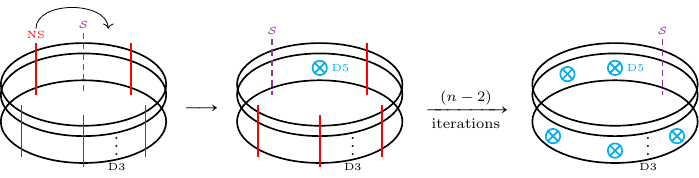}
    \caption{
    Multiple duality frames of the $\mathcal{S}$-fold example theory. 
    We start from $(n-1)$ NS5-branes and an $\mathcal{S}$-wall. In the first step, one NS5-brane is moved through the $\mathcal{S}$-wall, turning it into a D5-brane. Iterating this procedure $(n-2)$-more times, we obtain a setup with $(n-1)$ D5-branes and one $\mathcal{S}$-wall.}
    \label{fig:Sfold_dualization_withbranes}
\end{figure}

This duality was verified at the level of the superconformal index for low gauge rank in \cite{Garozzo:2019ejm}, where the map of the global symmetry parameters was worked out. Here we show how the algorithm can be used to derive this duality straightforwardly for generic rank, automatically producing the corresponding symmetry map. Although the quiver is circular, this analysis involves no subtlety in the choice of where to cut it open. Nevertheless, a convenient choice is to insert the $E$ parameter in the FI parameter of the node adjacent to the $\mathcal{S}$-wall, on the side opposite to the one chosen as the starting point for the dualization. 

Starting from the quiver in Figure~\ref{fig:Sfold_th}, we first $\mathcal{S}$-dualize the bifundamental on the left of the $\mathcal{S}$-wall using the basic duality move \eqref{eq:basic_duality_move_S_B10}. We also use twice the property \eqref{eq:shifted_Swall} to move the FI parameters around in a convenient way. Finally, we collapse the consecutive $\mathcal{S}$-wall and $\mathcal{S}^{-1}$-wall into an Identity-wall. This procedure is summarized in Figure~\ref{fig:Sfold_dualization}.
The resulting theory is the 3d $\mathcal{N}=4$ theory engineered by the second setup in Figure~\ref{fig:Sfold_dualization_withbranes}.
With these simple steps, we trade a bifundamental hypermultiplet for a fundamental one, whose mass parameter $Y_n$ is related to the FI of the node we lose in the process. This dualization is the field-theoretic analogue of the brane manipulation consisting of moving one NS5-brane across the $\mathcal{S}$-wall, thereby turning it into a D5-brane.
\begin{figure}[!ht]
    \centering
    \includegraphics[width=\textwidth]{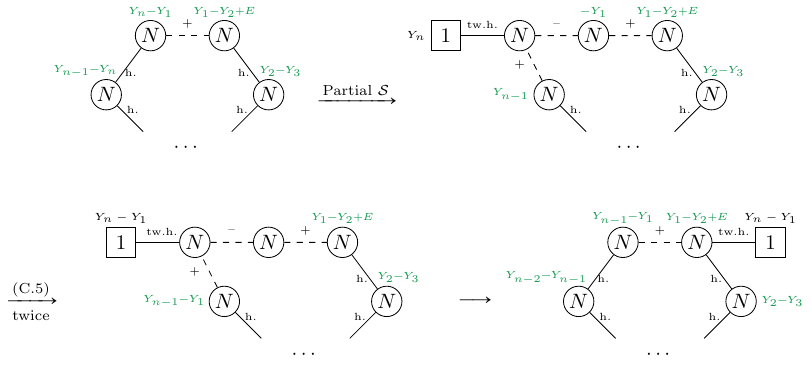}
    \caption{Field-theoretic counterpart of moving an NS5-brane across the $\mathcal{S}$-wall.}
    \label{fig:Sfold_dualization}
\end{figure}

A notable distinction exists between the dualization of $\mathcal{S}$-fold theories and that of ordinary quivers. In the case of ordinary quivers, dualizing a single QFT block is typically avoided, as it would introduce two $\mathcal{S}$-walls into the theory. Conversely, in the $\mathcal{S}$-fold theories under consideration, one of the two $\mathcal{S}$-walls generated during the dualization annihilates with the pre-existing $\mathcal{S}$-wall. This leaves a single $\mathcal{S}$-wall consistent with the original $\mathcal{S}$-fold structure. At the algorithmic level, this process represents a partial $\mathcal{S}$-dualization, which is physically realized by moving a single five-brane across the $\mathcal{S}$-wall, hence the ``partial $\mathcal{S}$'' used in Figure~\ref{fig:Sfold_dualization}.

One can repeat this procedure $k$ times to obtain a generic dual frame. If $k=n-1$, one obtains the single node gauge theory in Figure~\ref{fig:Sfold_final}, which is engineered by the right-most brane setup shown in Figure~\ref{fig:Sfold_dualization_withbranes}. 
The manifest global symmetry in this frame is
\begin{equation}
    U(1)_E\times SU(n)_Y\,,
\end{equation}
since now only $U(1)_E$ is a topological symmetry, whereas $SU(n)_y$ acts as a flavor symmetry. This shows that the topological symmetry \eqref{eq:symmSfold} of the starting theory in Figure~\ref{fig:Sfold_th} actually enhances to $U(1)_E \times SU(n)_Y$.

\begin{figure}[!ht]
    \centering
    \includegraphics[width=0.7\linewidth]{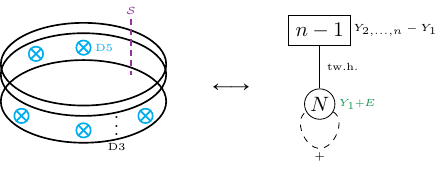}
    \caption{Result after $(n-1)$ dualizations. Notice that the SQCD has a redundant $Y_1$ parameter; indeed, already in the starting theory in Figure~\ref{fig:Sfold_th}, the $Y_1$ parameter can be reabsorbed by a redefinition of $E$.}
    \label{fig:Sfold_final}
\end{figure}

\subsection{Example 4: $U(N)$ SQCD with symmetric and antisymmetric hypers}
The dualization algorithm allows us to derive a mirror duality for $U(N)$ SQCD with $F$ fundamental flavors and two additional hypermultiplets, one in the symmetric and one in the antisymmetric representations.

The trick to derive the mirror dual of this theory is a straightforward generalization of the main strategy described in Section~\ref{subsec:algorithm_circular}, and it is based on the observation that a symmetric and an antisymmetric tensor can be obtained by taking a $U(N)\times U(N)$ bifundamental free-hypermultiplet and then performing the anti-diagonal gauging of the two $U(N)$ symmetries. This anti-identification can be implemented by a modified version of the $\mathcal{B}_\text{bar}$ QFT block:
\begin{equation} \label{eq:baryonic_block_twisted}
    \mathcal{Z}_{\bar{\mathcal{B}}_\text{bar}}^{(N,N)}(\vec{X},\vec{Y};m_A;B) = {}_{\vec{X}}\hat{\mathbb{I}}_{-\vec{Y}+B} (m_A) \times e^{-\pi i N B^2}
    \,.
\end{equation}
Its field-theoretical realization relies on building an Identity-wall gluing two $\mathcal{S}^{+1}$-walls as in Figure~\ref{fig:Bbar_modified}, instead of an $\mathcal{S}^{+1}$ and an $\mathcal{S}^{-1}$-wall as in Figure~\ref{fig:Bbar_Btop}.
\begin{figure}[!ht]
    \centering
    \includegraphics[width=0.7\linewidth]{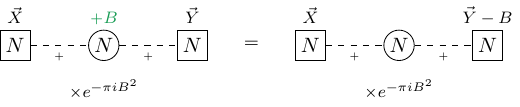}
    \caption{The modified baryonic block $\bar{\mathcal{B}}_\text{bar}$, obtained by sandwiching the FI parameter $B$ between two $\mathcal{S}^{+1}$-walls.}
    \label{fig:Bbar_modified}
\end{figure}

To see the action of this block \eqref{eq:baryonic_block_twisted}, we can consider a simple toy model where the two $U(N)$ global symmetries in Figure~\ref{fig:Bbar_modified} are both gauged, with the addition of a bifundamental hypermultiplet and $F$ flavors for one of the $U(N)$ nodes.
At the level of the $\mathbb{S}^3_b$ partition function we observe the following:
\begin{align}\label{eq:tensors_opening}
    & 
    \int d\vec{X}_N \Delta_N(\vec{X};m_A) \int d\vec{Y}_N \Delta_N(\vec{Y};m_A) 
    \mathcal{Z}_{\bar{\mathcal{B}}_{\text{bar}}}^{(N,N)}(\vec{X},\vec{Y};m_A;B) \times \nn\\
    & \quad \times 
    \prod_{j=1}^N \prod_{l=1}^F s_b\left( \frac{iQ}{2} - m_A \pm (X_j - W_l) \right)
    \prod_{j,k=1}^N s_b\left(\frac{iQ}{2} -m_A \pm (X_j-Y_k) \right) = \nn\\
    & \qquad = 
    \int d\vec{X}_N \Delta_N(\vec{X};m_A) \prod_{j=1}^N \prod_{l=1}^F s_b\left( \frac{iQ}{2} - m_A \pm (X_j - W_l) \right) 
    \times \\
    & \qquad\qquad \times 
    \prod_{j<k}^N s_b\left(\frac{iQ}{2} -m_A \pm (X_j+X_k-B) \right)
    \prod_{j\leq k}^N s_b\left(\frac{iQ}{2} -m_A \pm (X_j+X_k-B) \right)
    \nn\,.
\end{align}
The result is the partition function of a $U(N)$ SQCD with $F$ fundamentals plus two hypermultiplets, one symmetric and one antisymmetric. Notice that we can observe only the real mass $B$ associated to the diagonal combination of the two $U(1)$ global symmetries rotating independently the two tensors.

\begin{figure}[!ht]
    \centering
    \includegraphics[width=\linewidth]{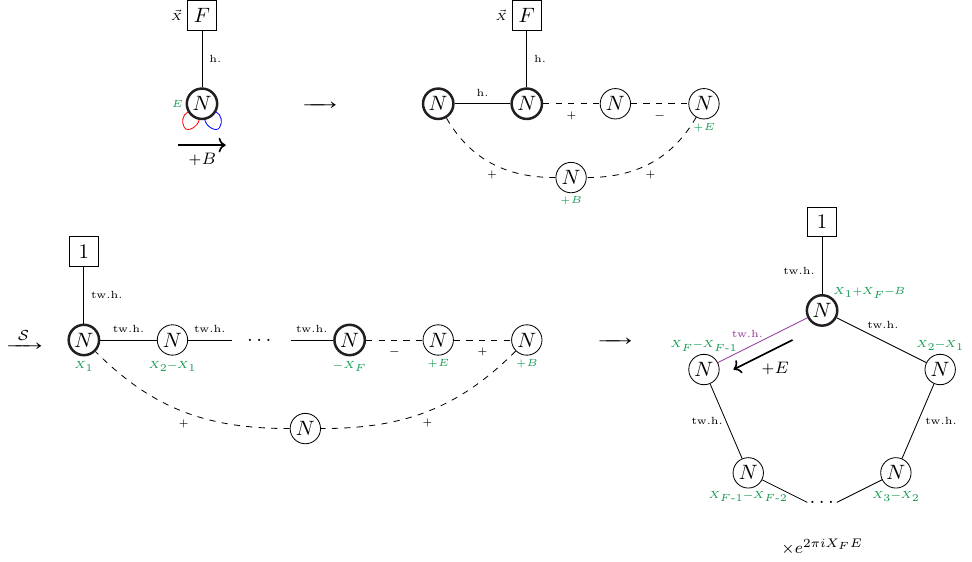}
    \caption{Mirror dualization of SQCD with symmetric and antisymmetric hypermultiplets, drawn respectively as red and blue arcs. In the first step, we open the quiver following the prescription in \eqref{eq:tensors_opening}, which leads to the unusual sequence of two $\mathcal{S}^{+1}$ operators. We then dualize each QFT block and implement Identity-walls, reaching the final theory in the bottom-right corner. The purple hypermultiplet indicates that it transforms in the $\Box \times \Box$ representation, rather than in the $\Box \times \Boxbar$ representation of the standard bifundamentals shown in black.}
    \label{fig:tensors_dualization}
\end{figure}

The procedure to derive the mirror duality is summarized in Figure~\ref{fig:tensors_dualization}.
Starting from a $U(N)$ SQCD with $F$ fundamental flavors, a symmetric tensor, and an antisymmetric tensor, we cut it open following the prescription outlined in Section~\ref{subsec:cutting}. The difference is that now we open it using the $\bar{\mathcal{B}}_\text{bar}$ block defined in \eqref{eq:baryonic_block_twisted}. Notice in particular that we now have a sequence of two $\mathcal{S}^{+1}$-walls instead of a standard sequence of alternating $\mathcal{S}^{+1}$ and $\mathcal{S}^{-1}$-walls.\footnote{Replacing $\mathcal{S}^2$ with $\mathcal{S}\mathcal{S}^{-1}=\hat{\mathbb{I}}$ leads to adjoint SQCD. This example will be considered later in Section~\ref{sec:local_badness_circular}.}
After opening the quiver, we perform the dualization of the linearized quiver as depicted in the second line of Figure~\ref{fig:tensors_dualization}, where the $F$ flavors are $\mathcal S$-dualized into bifundamentals, and the bifundamental into a single flavor. We then close back the dual quiver and obtain a circular theory with $F$ gauge nodes connected by bifundamentals, plus a single flavor. 
Notice that the bifundamental colored purple in Figure~\ref{fig:tensors_dualization} transforms in the representation $\Box \times \Box$ of the first two gauge nodes; the other bifundamentals, instead, transform as usual in the $\Box \times \Boxbar$ representation.
This rather unusual property implies that the quiver does not belong to the standard class discussed in Section~\ref{sec:good_circular}.\footnote{The Higgs branch Hilbert series of the SQCD with symmetric and antisymmetric tensors was studied in \cite{Dey:2013fea}. Therein, this theory was realized performing an orientifold operation on the Kronheimer--Nakajima quiver.}

\begin{figure}
    \centering
    \includegraphics[width=0.7\linewidth]{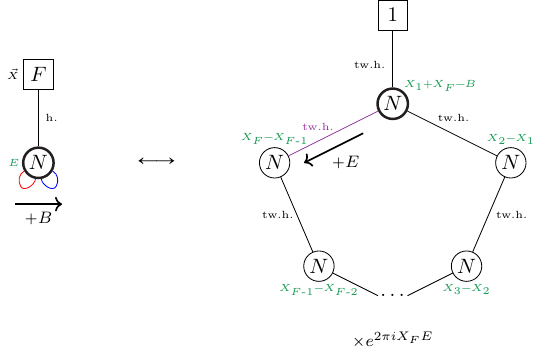}
    \caption{Mirror duality for QCD with $F$ fundamental, one antisymmetric, and one symmetric hypermultiplets. The red and blue arcs represent the two tensors. Since the theory has a single gauge node, the arrow indicating the baryonic symmetry simply denotes the presence of a real mass parameter $B$ for the two tensors. On the mirror side, the purple hypermultiplet indicates that it transforms in the $\Box \times \Box$ representation, rather than in the $\Box \times \Boxbar$ representation of the standard bifundamentals shown in black.}
    \label{fig:tensors_duality_finaly}
\end{figure}
The resulting duality, shown in Figure~\ref{fig:tensors_duality_finaly}, enjoys the following features. 
The global symmetry of the SQCD theory is
\begin{equation}
    SU(F)_X \times U(1)_A \times U(1)_S \times U(1)_E \,,
\end{equation}
where $SU(F)_X$ acts on the fundamental flavors, while the two abelian factors $U(1)_A$ and $U(1)_S$ rotate the antisymmetric and symmetric tensors, respectively. $U(1)_E$ corresponds to the topological symmetry. As already mentioned, the SQCD on the left hand side is considered at $A=S \equiv B$ because the parameter associated with the anti-diagonal subgroup is not captured by our construction. This is consistent with the fact that the anti-diagonal $U(1)$ is not a manifest symmetry of the dual-theory Lagrangian but instead emerges only in the IR.

On the other hand, in the dual theory the manifest global symmetry is
\begin{equation}
    S\left[\prod_{i=1}^F U(1)_{X_i}\right] \times U(1)_B \times U(1)_E \,,
\end{equation}
which enhances in the IR to 
\begin{equation}
    SU(F)_X \times U(1)_B \times U(1)_E \times U(1)_{\text{extra}}
    \,.
\end{equation}
In particular, the $U(1)^{F-1}$ topological symmetries are enhanced to $SU(F)$, while $U(1)_{\text{extra}}$ is purely emergent in the IR and not visible in the UV. 
$U(1)_{\text{extra}}$ should map across the duality to the anti-diagonal combination of $U(1)_A \times U(1)_S$. 
As we explain below, $U(1)_{\text{extra}}$ can effectively be regarded as an emergent topological symmetry, since it is expected to act on monopole operators of the dual theory. Thus, the duality maps flavor symmetries to enhanced topological symmetries and, as discussed later, also maps mesonic operators to topological operators, as expected for a mirror duality.
Although $U(1)_{\text{extra}}$ can be refined in the UV description of the SQCD, the same refinement cannot be performed in the dual theory. Therefore, the $\mathbb{S}^3_b$ partition function identity between the two dual theories, such as the one derived from the dualization algorithm, is meaningful only at zero real mass for this symmetry.\footnote{Similarly, we have performed a perturbative matching of the superconformal indices of the two theories after identifying the fugacities associated with $U(1)_A$ and $U(1)_S$. The anti-diagonal combination corresponds to $U(1)_{\text{extra}}$ on the mirror side, whose fugacity cannot be turned on at the Lagrangian level, since this symmetry emerges only in the IR.}\\

To better understand the emergence of this extra symmetry, let us detail the operator map. 
The moment map for $U(1)_{\text{extra}}$ can be constructed in the SQCD theory as a linear combination of quadratic operators built from the antisymmetric and symmetric tensors. We recall that in the SQCD one can construct three types of gauge invariant operators out of the hypermultiplets: bilinears of the fundamental, antisymmetric, and symmetric hypermultiplets. The $\mathcal{N}=4$ superpotential sets one combination of these operators to zero, while the two remaining linear combinations are the moment maps for the $U(1)_A \times U(1)_S$ symmetries, or equivalently for $U(1)_B \times U(1)_{\text{extra}}$ after a suitable change of basis.
In the mirror theory, the moment map for $U(1)_B$ is a linear combination of the traces of adjoint chiral multiplets, while the moment map for $U(1)_{\text{extra}}$ is a monopole operator with magnetic fluxes $(1,0,\ldots,0,-1)$ for each gauge group of the quiver. Therefore, this symmetry emerges only in the IR.  

In the SQCD theory, one can construct a tower of operators given by the meson matrix dressed with powers of the antisymmetric and/or symmetric tensors.
Consider, for simplicity, the case of a single dressing factor, either antisymmetric or symmetric. The meson matrix dressed once with the antisymmetric tensor maps to a collection of monopole operators forming an adjoint representation of $SU(F)$. These monopoles carry fluxes $\pm(1,0,\ldots,0)$ along a continuous sequence of gauge nodes, always including the node attached to the flavor, and zero magnetic flux for the rest of the gauge groups. 
Similarly, the meson matrix dressed once with the symmetric tensor maps to an almost identical collection of monopoles, but with the fluxes for the remaining gauge groups set to $(1,0,\ldots,0,-1)$ rather than zero. 
More precisely, the two collections of monopoles map to two linearly independent combinations of the two dressed mesons, since their degeneracy cannot be resolved without turning on the real mass parameter for the $U(1)_{\text{extra}}$ symmetry in the SQCD theory.\\

It would be interesting to further examine this new mirror duality in order to better understand its features, such as how the relation between the two tensors arises on the Coulomb branch of the mirror theory. We leave this analysis for future work.\\

\clearpage
\section{Local badness in circular quivers}
\label{sec:local_badness_circular}

We now turn to \emph{bad} or \emph{ugly} circular quivers. For circular quivers, one can distinguish two types of badness or ugliness: local and global. 

In this section, we focus on the locally bad or ugly cases, namely circular quivers with one or more bad or ugly nodes in the sense of \cite{Gaiotto:2008ak}. These are nodes whose rank $N$ satisfies $F_\text{eff} \leq 2 N -1$, where $F_\text{eff}$ denotes the number of fundamental hypermultiplets seen by the node.\footnote{If a bad node further satisfies $F_\text{eff}<N$, we refer to it as \emph{evil}.}
This contrasts with the case discussed in Section~\ref{sec:global_badness_circular}, where badness or ugliness is a global property of the quiver. In the following, for simplicity, we refer to both locally bad and locally ugly quivers simply as locally bad quivers.

Such local badness in circular quivers is no different from that in linear quivers. In \cite{Giacomelli:2024laq}, it was shown that a mirror description of a bad linear quiver can be obtained in two ways. The first is to apply the same dualization algorithm as in the good case, with additional care in implementing Higgsing, or equivalently Hanany--Witten moves. For bad theories, this procedure is called \emph{magnetic} dualization and leads in the IR to multiple mirror theories.

The second is the so-called \emph{electric} dualization, which we discuss shortly. This procedure produces multiple good IR theories, which can then be mirror-dualized in the standard manner and give rise to the same mirror theories obtained via magnetic dualization. Although both methods yield the same result, the electric dualization is generally more convenient for bad theories.

To implement the electric dualization algorithm for locally bad circular quivers, we first recall the corresponding partition function formula for bad SQCD \cite{Giacomelli:2023zkk}, which can be written schematically as
\begin{align}
    \raisebox{-1ex}{\includegraphics[height=12ex]{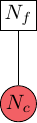}} \quad
    & =
    \sum_{n=0}^{\substack{\lfloor N_f/2 \rfloor\\-(N_f-N_c)-1}}
    \Bigg[\delta(\text{FI}\pm\dots)\times
    \raisebox{-1ex}{\includegraphics[height=12ex]{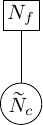}}
    \substack{\begin{array}{l}\times\text{\scriptsize sing.} \\ \times\text{\scriptsize BF}\end{array}}
    \Bigg]_{\widetilde{N}_c=\lfloor N_f/2 \rfloor-n}
    +\quad
    \raisebox{-1ex}{\includegraphics[height=12ex]{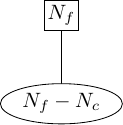}}
    \substack{\begin{array}{l}\times\text{\scriptsize sing.} \\ \times\text{\scriptsize BF}\end{array}}
    \label{eq:bad_SQCD}
\end{align}
where the details omitted in this schematic expression can be found in \cite{Giacomelli:2023zkk}.
Using this formula, the electric dualization algorithm for a locally bad circular quiver, schematically represented in Figure~\ref{fig:electric_circular_algorithm}, proceeds as follows.
\begin{enumerate}
    \item Given a bad circular theory, choose any ugly, bad, or evil node and carve it out from the rest of the quiver, thereby isolating a bad SQCD.
    \item Dualize the carved-out SQCD into a sum of good theories accompanied by free sectors, as dictated by the dualization formula for the bad SQCD \eqref{eq:bad_SQCD}.
    \item  Glue the dualized parts back together, obtaining a collection of circular quivers.
    \item If some of the quivers generated in the previous step still contain ugly or bad nodes, iterate the algorithm until all resulting frames contain only good gauge nodes.
\end{enumerate} 
As noted in \cite{Giacomelli:2024laq}, this procedure is subject to two consistency conditions that rule out \emph{unphysical} frames.\footnote{First, for generic values of $m$ and $Q$, there must exist a simultaneous solution to all delta function constraints in the frame. Second, all singlet contributions must remain non-singular when these constraints are imposed. A frame in which a singlet contribution develops a singularity is regarded as a singular limit of another physical frame and is therefore excluded.}
The result is therefore a collection of theories, written on their electric side, each equipped with its own singlets, BF couplings, and possibly delta functions specifying which VEVs must be turned on to reach the corresponding singular point in the moduli space.\\

\begin{figure}[!ht]
	\centering
	\includegraphics[width=\textwidth]{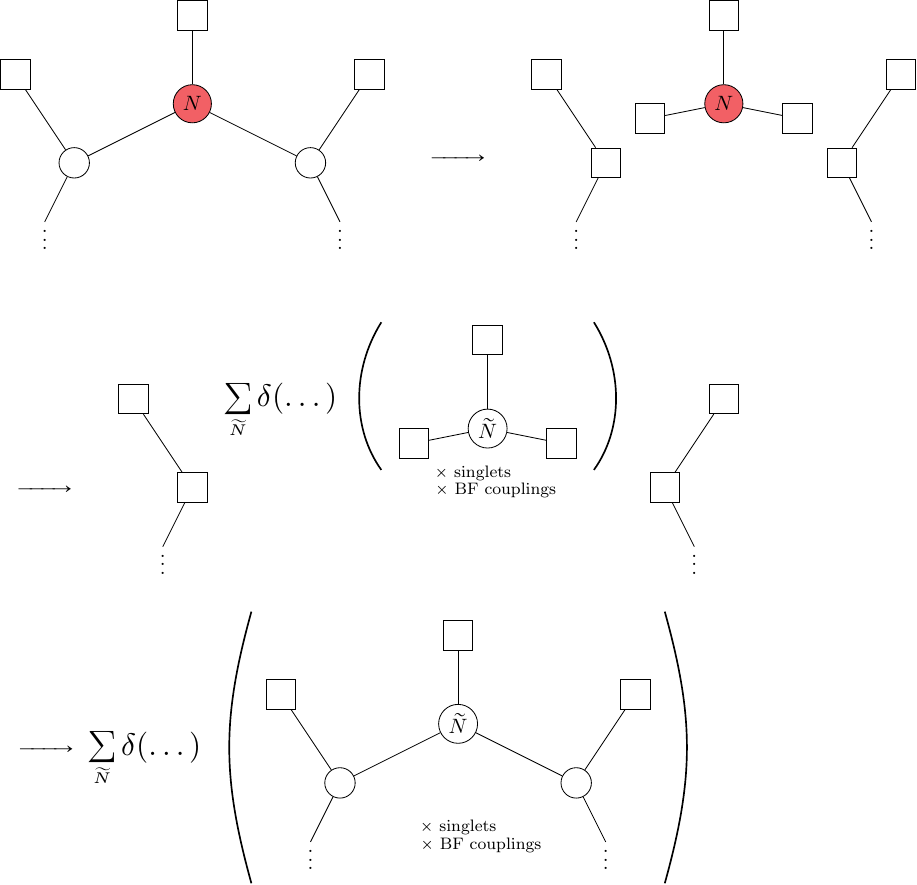}
	\caption{The electric dualization algorithm applied to a locally bad circular theory. For simplicity, we do not explicitly display the ranks, singlets, BF couplings, or delta function arguments. The red node is bad. In principle, one of the resulting frames may contain no delta, as in the last contribution in \eqref{eq:bad_SQCD}; this frame is omitted here, since the purpose of the figure is only to illustrate the procedure rather than its details.
    }
    \label{fig:electric_circular_algorithm}
\end{figure}

As an example, consider the locally bad circular quiver shown at the top of Figure~\ref{fig:locally_bad_example}. It contains one bad gauge node, shown in red, and one evil node, shown in magenta. Applying the algorithm described above, one can dualize it into the sum of frames depicted below.

\begin{figure}[!ht]
	\centering
	\includegraphics[width=\textwidth]{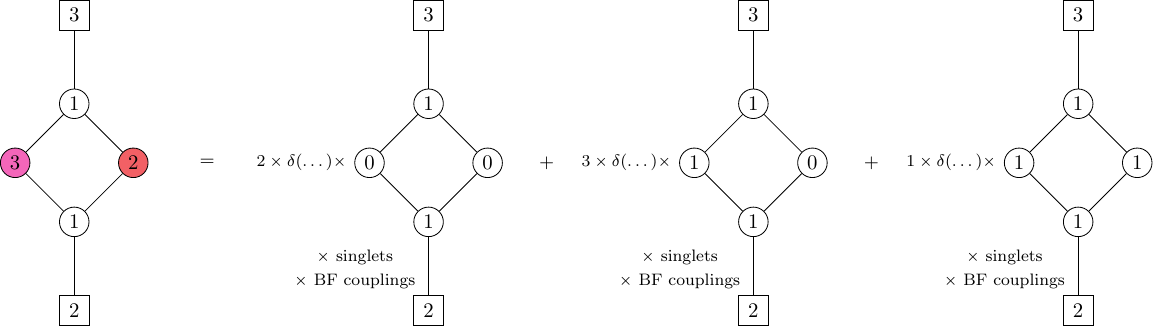}
	\caption{Example of a locally bad theory with a bad gauge node, shown in red, and an evil node, shown in magenta, together with its dual frames on the right. All links in the figure represent untwisted hypermultiplets. The numbers in front of each frame denote its multiplicity. The quivers presented here are shown in simplified form, omitting the singlet content, BF couplings, and delta functions that further distinguish dual frames with the same interacting sectors. These details can be found in the accompanying Mathematica code.
    }
    \label{fig:locally_bad_example}
\end{figure}

\section{Global badness in circular quivers}
\label{sec:global_badness_circular}
In the previous section, we discussed the case of circular quivers with bad or ugly nodes, which we called locally bad or ugly.
However, in circular theories, badness or ugliness can also arise in a different way: not from one (or more) underbalanced nodes, but as a property arising from the particular shape, or topology, of the quiver. We refer to this as \emph{global} badness or ugliness. This phenomenon can arise when the total number of flavors in the circular theory is at most one.
As in the local case, global badness (or ugliness) manifests itself through the presence of monopole operators whose scaling dimensions, computed with respect to the UV R-symmetry, fall below (or saturate) the unitarity bound. These monopoles carry non-zero magnetic flux under all gauge nodes\footnote{Notice that this phenomenon happens every time the quiver is shaped as an affine Dynkin diagram, with the circular quiver being the type-$A$ case. We will not discuss this phenomena in differently shaped quivers here, some related cases have been considered in \cite{Comi:2026gjx}.}.

The global ugliness and global badness for circular quivers can be defined as follows:
\begin{enumerate}
	\item \textit{Globally ugly}: a circular quiver with $L$ nodes of rank $N$ and a single flavor. 
    \item \textit{Globally bad}: a circular quiver with $L$ nodes of rank $N$ and no flavors at all.
\end{enumerate}
We begin by focusing on the globally ugly theories in Section~\ref{subsec:globally_ugly}. We then turn to the globally bad case in Section~\ref{subsec:globally_bad}, which proves to be highly non-trivial.
Finally, in Section~\ref{subsec:globally_and_locally_bad}, we comment on the case that exhibits both local and global badness.

\subsection{Globally ugly}
\label{subsec:globally_ugly}

Consider a circular theory with $L$ gauge nodes of rank $N$ and a single flavor on one gauge node, which is depicted in Figure~\ref{fig:globally_ugly_example}; see the left quiver. This is a variation of the good example in Section~\ref{subsec:ABJM_triality}, which now describes M2-branes probing $\mathbb C^2/\mathbb Z_L \times \mathbb C^2$ instead of $(\mathbb C^2/\mathbb Z_L)^2$ \cite{Porrati:1996xi}.
Its mirror dual, which can be easily obtained using the dualization algorithm described in Section~\ref{sec:dualization_algorithm_circular}, is the theory depicted on the right hand side of Figure~\ref{fig:globally_ugly_example}.
Notice that in the mirror theory the trace of the $U(N)$ adjoint hypermultiplet is free. This free sector is mapped to a pair of monopole operators in the original theory, carrying unit magnetic flux under each gauge node with opposite signs. As expected, the dimensions of these monopole operators saturate the unitarity bound, and they are thus free in the IR.
\begin{figure}[!ht]
\centering
    \includegraphics[width=0.8\textwidth]{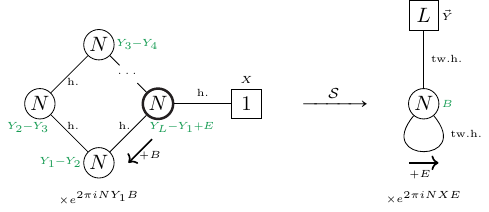}
    \caption{An example of a globally ugly circular theory, together with its mirror dual.}
    \label{fig:globally_ugly_example}
\end{figure}

\subsection{Globally bad}
\label{subsec:globally_bad}
Now we consider a quiver with $L$ gauge nodes of rank $N$, without flavors on any node. A straightforward inspection of monopole dimensions shows that any monopole carrying identical magnetic flux on all gauge nodes has UV dimension zero.
This observation allows us to derive a general expression for the dual partition function of a globally bad theory. As in the locally bad case, the result takes the form of a sum over delta function weighted frames, which are, interestingly, organized in a structure reminiscent of the permutation-group gauging.\\

The partition function of our quiver with $L$ gauge nodes of rank $N$ reads
\begin{align}\label{eq:globalbad_elecpf}
    \mathcal{Z}&=
    e^{2 \pi i N Y_{1} B}
	\int \prod_{j=1}^{L} \left[\mathrm{d}{\vec{Z}_{N}^{(j)}}\Delta_{N}\left(\vec{Z}^{(j)};m_A \right)\right]
    \nonumber\\
	& \qquad\qquad\qquad \times
    e^{2 \pi i (Y_L-Y_{1}+E)\sum_{a=1}^{N}Z_a^{(L)}}
    \,
	\prod_{j=1}^{L-1} e^{2 \pi i (Y_j-Y_{j+1})\sum_{a=1}^{N}Z_a^{(j)}}
	\nonumber\\
	& \qquad\qquad\qquad \times
	\prod_{j=1}^{L-1}\prod_{a=1}^{N}\prod_{b=1}^{N} s_b\left(i\frac{Q}{2}-m_A\pm\left(Z_a^{(j)}-Z_b^{(j+1)}\right) \right)
    \nonumber\\
	& \qquad\qquad\qquad \times
    \prod_{a=1}^{N}\prod_{b=1}^{N} s_b\left(i\frac{Q}{2}-m_A\pm\left(Z_a^{(L)}-Z_b^{(1)}-B\right) \right)
	\,.
\end{align}

As an example, the $L = 4$ case is depicted at the top-left corner of Figure~\ref{fig:globally_bad_dualization}.
Following the dualization algorithm in Section~\ref{subsec:algorithm_circular}, we first linearize the theory, as shown in the first row of Figure~\ref{fig:globally_bad_dualization}. 
Then we break the theory into fundamental QFT blocks and dualize each of them with the basic duality moves.
We can then glue the dualized blocks back together and implement the effect of the Identity-walls, leading to the last theory in Figure~\ref{fig:globally_bad_dualization}; the resulting partition function has the form
\begin{align}
    \mathcal{Z}^{\vee}
    &=
	\int \mathrm{d}{\vec{W}_{N}}\Delta_{N}(\vec{W};iQ/2-m_A)
    \times
    e^{2 \pi i B\sum_{a=1}^{N}W_a}
    \nonumber\\
	& \qquad \times
	{}_{\vec W}\hat{\mathbb{I}}_{\vec W-E} (iQ/2-m_A)
    \times
    \prod_{a=1}^{N}\prod_{b=1}^{L} s_b\left( m_A\pm\left(W_a-Y_b\right) \right)
	\,.
\end{align}
Now we can evaluate the Identity-wall in the integrand using its definition \eqref{eq:def_Id_wall} and that of the $U(N)$ integration measure given in \eqref{eq:U_integration_measure}. We then find the following expression for the partition function:
\begin{align}
    \mathcal{Z}^{\vee}
    &=
	\frac{1}{N!} \int \prod_{a=1}^{N}\udl{W_a}
    \,
    e^{2 \pi i B\sum_{a=1}^{N}W_a}
    \,
    \sum_{a=1}^{N}\sum_{\sigma\in S_N}\delta\left(W_a-W_{\sigma(a)}+E\right)
    \nonumber\\
	& \qquad\quad \times
    \prod_{a=1}^{N}\prod_{b=1}^{L} s_b\left( m_A\pm\left(W_a-Y_b\right) \right)
	\,.
\end{align}
As usual, the Dirac delta imposes a constraint on symmetry parameters, but now we also need to consider all possible permutations of these parameters. To better understand this, let us consider some explicit examples for fixed $N$.

\begin{figure}[!ht]
\centering
    \includegraphics[width=\textwidth]{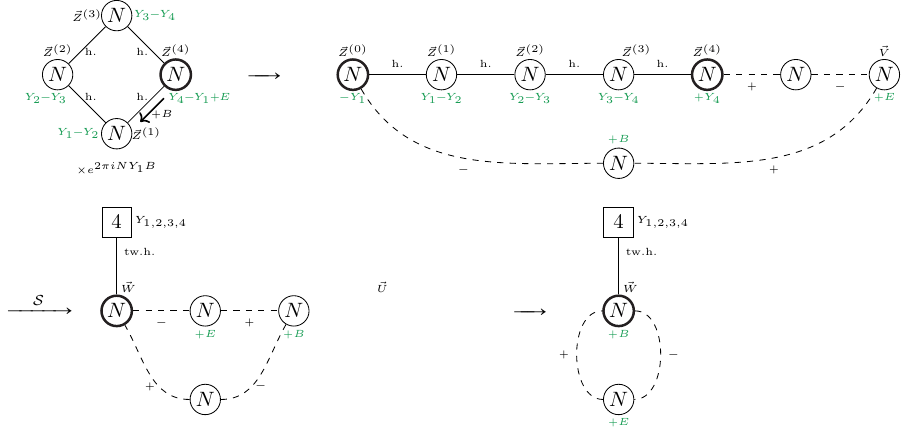}
    \caption{The algorithmic dualization of a globally bad circular theory. For simplicity, we consider the case with $4$ gauge nodes.}
    \label{fig:globally_bad_dualization}
\end{figure}

\subsubsection{Fixed $N$ examples}
\paragraph{$N=1$.} Setting $N=1$, one finds
\begin{align}
\label{eq:N=1}
    \mathcal{Z}^{\vee}
    &=
	\int \mathrm{d} W
    \,
    e^{2 \pi i B W}
    \,
    \delta (E)
    \,
    \prod_{b=1}^{L} s_b\left( m_A\pm\left(W-Y_b\right)\right) \,.
\end{align}
Since the resulting theory preserves $\mathcal N=4$ supersymmetry, the components of the partition function should also be consistent with it. In particular, we should recover the $\mathcal{N}=4$ vector multiplet, which consists of an $\mathcal N=2$ vector and an $\mathcal N=2$ adjoint chiral. The contribution of the former is trivial in the $U(1)$ case, whereas the latter reduces to a singlet for $U(1)$. Since such a singlet is absent in \eqref{eq:N=1}, we should introduce it by multiplying and dividing by the same factor, making use of the identity $s_b(x)s_b(-x)=1$ for the double-sine function. Hence, we obtain
\begin{align}\label{eq:globallybad_example_N1}
    \mathcal{Z}^{\vee}
    & = s_b \left(-i\frac{Q}{2} + 2m_A \right) \delta (E)\int \mathrm{d} W \, \Delta_1(W;iQ/2-m_A)
    \,
    e^{2 \pi i B W}
    \, 
    \prod_{b=1}^{L} s_b\left( m_A\pm\left(W-Y_b\right)\right) 
    \,,
\end{align}
where the first two factors correspond to  two gauge singlets decoupled from the interacting sector of the theory, with $\delta(E)$ being interpreted as a singlet of R-charge 0.

We illustrate the resulting dual pair in Figure~\ref{fig:globally_bad_example_N1}.
\begin{figure}[!ht]
    \centering
    \includegraphics[width=0.65\textwidth,center]{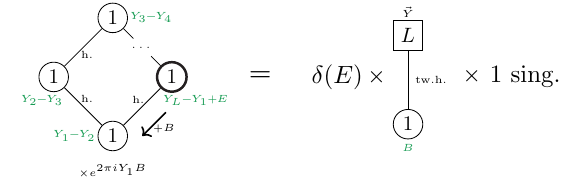}
    \caption{The sum of frames for the globally bad circular quiver of rank $1$ and length $L$; see \eqref{eq:globallybad_example_N1}.
    }
    \label{fig:globally_bad_example_N1}
\end{figure}
Notice that upon integrating over $E$, the presence of $\delta(E)$ yields an SQED with $L$ flavors on the right hand side. Instead, on the left hand side, integrating over $E$ corresponds to gauging the topological symmetry associated with the $L$-th node, which in turn ungauges that $U(1)$ gauge symmetry. The result is a good linear quiver, which is precisely the mirror dual of SQED with $L$ flavors. Note that this trick works only for $N=1$.

\paragraph{$N=2$.} Setting $N=2$, one finds
\begin{align}
    \mathcal{Z}^{\vee}
    &=
	\frac{1}{2} \int \mathrm{d}{W_{1}}\mathrm{d}{W_{2}}
    \,
    e^{2 \pi i B(W_1+W_2)}
    \,
    \prod_{a=1}^{2}\prod_{b=1}^{L} s_b\left( m_A\pm\left(W_a-Y_b\right)\right)
    \\
	& \qquad \times
    \Big[\delta(E)\delta(E)+\delta(W_1-W_2+E)\delta(W_2-W_1+E)\Big]
    \nn\\
    &=
	\frac{1}{2} \bigg[ \delta(E)^2
    \int \mathrm{d}{W_{1}}\mathrm{d}{W_{2}}
    \,
    e^{2 \pi i B(W_1+W_2)}
    \,
    \prod_{a=1}^{2}\prod_{b=1}^{L} s_b\left( m_A\pm\left(W_a-Y_b\right)\right)
    \label{eq:globallybad_example_N2} \\
    &\quad +
    \delta(2E) 
    \,
    e^{2 \pi i BE}
    \,
    \int \mathrm{d}{W_{1}}
    \,
    e^{2 \pi i B(2W_1)}
    \nonumber\\
	& \qquad\qquad\qquad \times
    \prod_{b=1}^{4} s_b\left( m_A\pm\left(W_1-Y_b\right)\right)s_b\left( m_A\pm\left(W_1-Y_b+E\right)\right) \bigg]
	\,.\nonumber
\end{align}
The result is a sum of two frames: the first contains two copies of SQED with $L$ flavors, while the second consists of a single copy of SQED with $2L$ flavors.
In particular, for the second frame, we perform the $W_2$-integration and impose the constraint from the first delta function:
\begin{align}
W_2 = W_1+E \,.
\end{align}
Consequently, the second delta function and the fundamental hyper contribution involving $W_2$ become
\begin{align}
\delta(W_2-W_1+E) \quad &\rightarrow \quad \delta(2 E) \,, \\
s_b\left(m_A\pm\left(W_2-Y_b\right)\right) \quad &\rightarrow \quad s_b\left(m_A\pm\left(W_1-Y_b+E\right)\right) .
\end{align}
If instead we impose the constraint from the second delta function first, namely $W_2 = W_1-E$, the first delta function again yields $\delta(2 E)$, while the sign of $E$ in the double sine is reversed: $s_b\left(m_A\pm\left(W_1-Y_b-E\right)\right)$. This is consistent, since the remaining delta function ultimately sets $E = 0$. In what follows, we adopt the first convention, in which $E$ enters the double-sine argument with a positive sign, for later convenience.
In both frames, the SQED factors lack the chiral adjoint contributions needed to complete the $\mathcal N=4$ multiplets.
This result is illustrated in Figure~\ref{fig:globally_bad_example_N2}. 

\begin{figure}[!ht]
    \centering
    \includegraphics[width=1\textwidth,center]{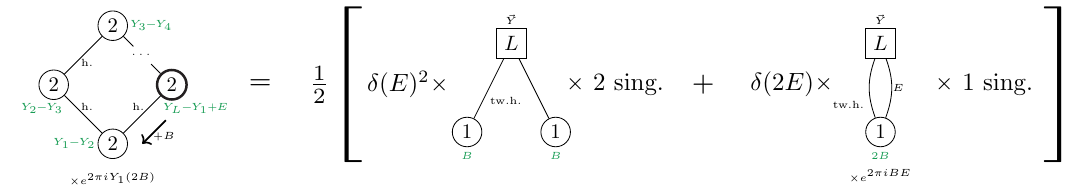}
    \caption{The sum of frames for the globally bad circular quiver of rank $2$ and length $L$; see \eqref{eq:globallybad_example_N2}.
    The parameter $E$ next to one of the edges in the second frame represents the baryonic shift for the corresponding hypermultiplet.
    }
    \label{fig:globally_bad_example_N2}
\end{figure}

\paragraph{$N=3$.} Setting $N=3$, one finds 
\begin{align}
    \mathcal{Z}^{\vee}
    &=
	\frac{1}{3!} \int \mathrm{d}{W_{1}}\mathrm{d}{W_{2}}\mathrm{d}{W_{3}}
    \,
    e^{2 \pi i B(W_1+W_2+W_3)}
    \,
    \prod_{a=1}^{3}\prod_{b=1}^{L} s_b\left( m_A\pm\left(W_a-Y_b\right)\right)
    \\
	& \qquad\quad \times
    \Big[
    \delta(E)\delta(E)\delta(E)
    +3\delta(W_1-W_2+E)\delta(W_2-W_1+E)\delta(E)
    \nonumber\\
	& \qquad\qquad
    +2\delta(W_1-W_2+E)\delta(W_2-W_3+E)\delta(W_3-W_1+E)
    \Big] = 
    \nn\\
    &=
	\frac{1}{3!} \bigg[ \delta(E)^3
    \int \mathrm{d}{W_{1}}\mathrm{d}{W_{2}}\mathrm{d}{W_{3}}
    \,
    e^{2 \pi i B(W_1+W_2+W_3)}
    \,
    \prod_{a=1}^{3}\prod_{b=1}^{L} 
    s_b\left( m_A\pm\left(W_a-Y_b\right)\right)
    \nonumber\\
    &\quad +
    3\delta(2E)\delta(E)
    \,
    e^{2 \pi i BE}
    \,
    \int \mathrm{d}{W_{1}}\mathrm{d}{W_{2}}
    \,
    e^{2 \pi i B(2W_1+W_2)}
    \nonumber\\
	& \qquad\qquad\qquad \times
    \prod_{b=1}^{L} 
    s_b\left( m_A\pm\left(W_1-Y_b\right)\right)
    s_b\left( m_A\pm\left(W_1-Y_b+E\right)\right)
    \nonumber\\
	& \qquad\qquad\qquad \times
    \prod_{b=1}^{L} 
    s_b\left( m_A\pm\left(W_2-Y_b\right)\right)
    \label{eq:globallybad_example_N3}\\
   &\quad +
    2\delta(3E)
    \,
    e^{2 \pi i (3BE)}
    \,
    \int \mathrm{d}{W_{1}}
    \,
    e^{2 \pi i B(3W_1)}
    \,
    \prod_{b=1}^{L} s_b\left( m_A\pm\left(W_1-Y_b\right)\right)
    \nonumber\\
	& \qquad\qquad\qquad \times
    \prod_{b=1}^{L} 
    s_b\left( m_A\pm\left(W_1-Y_b+E\right)\right)
    s_b\left( m_A\pm\left(W_1-Y_b+2E\right)\right) \bigg]
	\,.
\end{align}
The result is a sum of three frames.  
In the first frame, we obtain three copies of SQED with $L$ flavors.  
The second frame contains one copy of SQED with $L$ flavors and another copy with $2L$ flavors. As above, we perform the $W_2$-integration and impose the constraint from the first delta function, $W_2 = W_1+E$. We then relabel $W_3$ as $W_2$.
In the third frame, there is instead a single copy of SQED with $3L$ flavors. We perform the $W_2$- and $W_3$-integrations and impose the constraints from the first and second delta functions:
\begin{align}
W_2 &= W_1+E \,, \\
W_3 &= W_2+E = W_1+2 E \,.
\end{align}
Each frame appears with a different multiplicity, since in the sum over elements of $S_N$, only the cycle lengths of the permutations matter.  
As before, in order to interpret the result as the partition function of $\mathcal{N}=4$ SQED, the contributions of additional adjoint chirals must be included.
This result is illustrated graphically in Figure~\ref{fig:globally_bad_example_N3}.
\begin{figure}[!ht]
    \centering
    \includegraphics[width=1\textwidth,center]{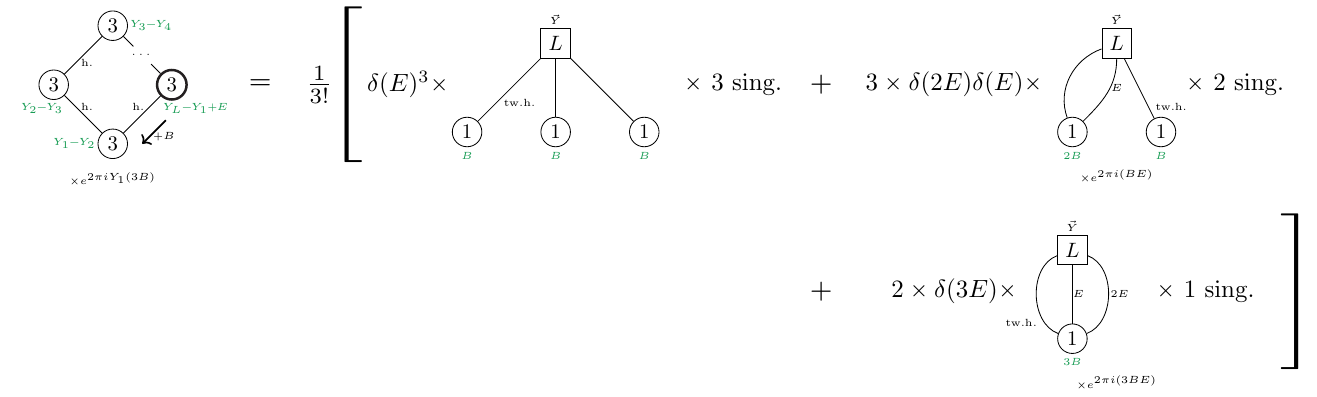}
    \caption{The sum of frames for the globally bad circular quiver of rank $3$ and length $L$; see \eqref{eq:globallybad_example_N3}.
    The $E$ or $2E$ close to some edges in the second and third frames represent the shifts in the hypermultiplet.
    }
    \label{fig:globally_bad_example_N3}
\end{figure}

\subsubsection{General result}
From the examples shown above, we can extrapolate the result for arbitrary $N$. It reads
\begin{align}\label{eq:globalbad_sumframes}
    \mathcal{Z}^{\vee}&=
    \frac{1}{N!}
    \sum_{\vec{k}}[\vec{k}]
    \Bigg\{
    \prod_{j=1}^{N}
    \Bigg[
    \delta\left(jE\right)
    \,
    e^{j (j-1) \pi i BE} 
    \\
    & \qquad \qquad \times
    \int\mathrm{d}{W}\,
    \,
    e^{2 \pi i \left(jB\right) W}
    \prod_{b=1}^{L}\prod_{c=1}^{j} 
    s_b\left( m_A\pm\left(W-Y_b+(c-1)E\right)\right)
    \Bigg]^{l_j}
    \Bigg\}
    \nn\,,
\end{align}
where the sum is taken over all partitions $\vec{k}=\{k_1,k_2,\dots,k_M\}=\{1^{l_1},2^{l_2},\dots,N^{l_N}\}$ of $N$, and $[\vec{k}]$ denotes the number of elements of $S_N$ in the conjugacy class corresponding to the partition $\vec{k}$:\footnote{The conjugacy class of $S_N$ labeled by the partition $\vec k$ consists of the elements whose cycle decomposition contains $l_i$ cycles of length $i$ for each $i$.}
\begin{align}
\label{eq:conjugacy class order}
    [\vec k] = \frac{N!}{\prod_{j = 1}^N j^{l_j} l_j!} \,.
\end{align}
The partition function $\mathcal{Z}^\vee$ is then dual to the partition function $\mathcal{Z}$ in \eqref{eq:globalbad_elecpf}.

The result can be stated as follows. The partition function of the circular theory with $L$ gauge nodes of rank $N$ and no flavors admit a decomposition into a sum over frames, which are in one-to-one correspondence with conjugacy classes of $S_N$, or equivalently with partitions of $N$. Each frame associated with a partition $\vec k$ appears with multiplicity $[\vec{k}]$ and consists of a product of SQED theories whose numbers of flavors are integer multiples of $L$, with the sum of these multiples equal to $N$. The explicit form of each dual frame can be obtained using the accompanying Mathematica code. \\

\subsubsection{Higgs branch index}
We now provide further evidence for the proposed mirror duality of the globally bad circular quiver by studying another observable, namely the superconformal index. Since a complete analysis of the superconformal index is more involved, we restrict attention to a particular limit of the index, usually referred to as the Higgs branch limit. This name arises because, for a good theory, this limit reproduces the Hilbert series of the Higgs branch of the moduli space. For a bad theory, on the other hand, the theory is not fully Higgsable, and the same limit does not exactly coincide with the Higgs branch Hilbert series. Nevertheless, it captures a subsector of the BPS states contributing to the superconformal index and therefore provides an efficient probe for testing the duality. By a slight abuse of terminology, we will refer to this quantity as the Higgs branch index.

\paragraph{Higgs branch index of the globally bad circular quiver}
We first examine the Higgs branch index on the circular quiver side. We will show that it exhibits the same sum-over-frames structure observed in the $S^3_b$ partition function above.\footnote{A computation of the geometric branch of the flavored circular quiver appeared in \cite{Comi:2026gjx} using an analogous technique which relies on results presented in this section.}

We start from the superconformal index \cite{Bhattacharya:2008zy} of the theory, which is obtained by the supersymmetric localization as follows \cite{Kim:2009wb,Imamura:2011su}:
\begin{align}
    \mathcal{I}_N
    =& 
    \sum_{\vec{m}^{(1)} \in \mathbb{Z}_{{N}/S_{{N}}}} 
    \sum_{\vec{m}^{(2)} \in \mathbb{Z}_{{N}/S_{{N}}}} 
    \dots
    \sum_{\vec{m}^{(L)} \in \mathbb{Z}_{{N}/S_{{N}}}}
    \oint \prod_{l=1}^{L} \left( \frac{\left( \frac{y_l}{y_{l+1}} \right)^{\sum_{j=1}^N m^{(l)}_j}}{\left| \mathcal{W}_{\vec{m}^{(l)}} \right|} \prod_{i=1}^{N}\frac{\mathrm{d}z_i^{(l)}}{2\pi i z_i^{(l)}}  \right) w^{\sum_{j=1}^N m^{(L)}_j} \nn \\
    & \times \prod_{l=1}^L \left[ \prod_{j<k}^N x^{-|m^{(l)}_j-m^{(l)}_k|} \left( 1 -  x^{|m^{(l)}_j-m^{(l)}_k|} \left( z^{(l)}_j/z^{(l)}_k \right)^\pm \right) 
    \right.\nn \\
    & \qquad\qquad\times\left.
    \prod_{j,k=1}^N a^{|m^{(l)}_j-m^{(l)}_k|} 
    \frac{\left( x^{1+|m^{(l)}_j-m^{(l)}_k|}a^{2} z^{(l)}_j/z^{(l)}_k ; x^2 \right)}{\left( x^{1+|m^{(l)}_j-m^{(l)}_k|}a^{-2} z^{(l)}_j/z^{(l)}_k ; x^2 \right)} \right] \nn \\
    & \times\prod_{l=1}^L \prod_{j,k=1}^N \left(\frac{x^\frac12}{a}\right)^{|m^{(l)}_j - m^{(l+1)}_k|}
    \frac{\left( x^{\frac32+|m^{(l)}_j-m^{(l+1)}_k|} a^{-1} \left(z^{(l)}_j/z^{(l+1)}_k\right)^\pm ; x^2 \right)}{\left( x^{\frac12+|m^{(l)}_j-m^{(l+1)}_k|} a \left(z^{(l)}_j/z^{(l+1)}_k\right)^\pm ; x^2 \right)}
\end{align}
where $y_{L+1} = y_1$, $\vec m^{(L+1)} = \vec m^{(1)}$, and $\vec{z}^{\,(L+1)} = \vec{z}^{\,(1)} b$ and $y_{L+1} = y_1$ have been assumed. $|\mathcal W_{\vec m^{(l)}}|$ is the Weyl group order of the $l$-th gauge node left unbroken by the flux $\vec m^{(l)}$. The real mass parameters in the $S^3_b$ partition function associated with the global symmetries are mapped to the holonomies in the index as follows:
\begin{align}
\begin{aligned}
\label{eq:parameter_map}
E \quad &\rightarrow \quad w \,, \\
B \quad &\rightarrow \quad b \,, \\
m_A \quad &\rightarrow \quad a \,, \\
Y_l \quad &\rightarrow \quad y_l \,.
\end{aligned}
\end{align}
These two sets of variables are identified via the exponential map. For instance, the holonomy associated with the diagonal topological symmetry is given by $w = e^{2 \pi i E}$, and similarly for the other variables.
Note that here we have recovered the superconformal R-charge by assigning R-charge $1/2$ to the hypermultiplets, so that the dimensions of the monopole operators become manifest. With this assignment, the Higgs branch index is obtained by introducing \cite{Razamat:2014pta}
\begin{align}
    h = x a^2\,,\qquad
    c = x / a^2\,,
\end{align} 
and then taking the well-defined limit $c \rightarrow 0$.\footnote{On the other hand, the Coulomb limit is obtained by taking $h \rightarrow 0$.} In this limit, only terms with vanishing power of $c$ survive. Each monopole flux sector carries a monomial with the following power of $c$
\begin{equation}
    \sum_{l=1}^{L} \left( \frac{1}{2} \sum_{i,\,j}^{N} \left| m_{i}^{(l)} - m_{j}^{(l+1)} \right| - \sum_{i < j} \left| m_{i}^{(l)} - m_{j}^{(l)} \right| \right) \,,
    \label{eq:c_power}
\end{equation}
which corresponds to the conformal dimension associated with the monopole configuration $\{\vec{m}^{(1)},\vec{m}^{(2)},\dots,\vec{m}^{(L)}\}$. Such configurations contribute to the Higgs branch index only when the associated conformal dimension vanishes, signaling the decoupling of the corresponding monopole operators in the IR. Namely, for the globally bad circular quiver, there are IR-free monopole operators contributing to the Higgs branch index, which is why this limit does not precisely coincide with the Higgs branch Hilbert series.

One can check that \eqref{eq:c_power} vanishes only when
\begin{equation}
    \{m_i^{(1)}\}=\{m_i^{(2)}\}=\dots=\{m_i^{(L)}\} \equiv \{m_i\} \quad \forall i=1,\dots,N \,,
\end{equation}
meaning that only these flux sectors contribute to the Higgs branch index. The index can be written as
\begin{align}
\label{eq:HBI}
    \mathcal{I}_N^{\text{HB}} 
    =&
    \sum_{\vec{m} \in \mathbb{Z}_{{N}/S_{{N}}}} 
    \frac{w^{\sum_{j=1}^N m_j}}{\left| \mathcal{W}_{\vec{m}} \right|} 
    \oint \left(\prod_{l=1}^{L}\sum_{i=1}^{N}\frac{\mathrm{d}z_i^{(l)}}{2\pi i z_i^{(l)}}\right)
    (1-h)^{NL} 
    \nn\\
    &\qquad\qquad\qquad\qquad\times
    \left( \prod_{l=1}^{L} \prod_{i \neq j}^{N} \left[ \left( 1 - z_{{i}}^{(l)} \left(z_{j}^{(l)}\right)^{-1} \right) \left(1 - z_{{i}}^{(l)} \left(z_{j}^{(l)}\right)^{-1} h \right) \right]^{\delta_{m_{i},m_j}} \right) 
    \nn\\
    &\qquad\qquad\qquad\qquad\times
    \left( \prod_{l=1}^{L} \prod_{i,\,j}^{N} \left(1 - \left(z_{{i}}^{(l)} /z_{j}^{(l+1)}\right)^{\pm1} 
    h^{1/2}\right)^{-\delta_{m_{i},m_j}} \right) 
    \,,
\end{align}
where the summation runs over the magnetic fluxes of the diagonal $U(N)$. Note that the integral appearing in the summand depends only on the residual gauge group preserved by the flux. Namely, for flux
\begin{align}
    \vec m =\{\underbracket{n_1,\dots,n_1}_{k_1},\underbracket{n_2,\dots,n_2}_{k_2},\dots,\underbracket{n_M,\dots,n_M}_{k_M}\} \,,
\end{align}
the diagonal $U(N)$ is further broken to $U(k_1) \times U(k_2) \times \dots \times U(k_M)$, where $\sum_{j=1}^M k_j = N$. Thus, the original flux sum can be decomposed as
\begin{align}
    \sum_{\vec m \in \mathbb Z_N/S_N} \quad \rightarrow \quad \sum_{\vec k} \sum_{n_1 \neq n_2 \neq \dots \neq n_M} \frac{1}{\prod_{i=1}^N l_i!}\,,
\end{align}
where the first sum runs over all partitions $\vec{k} = \{ k_1,\ldots,k_M \} = \{ 1^{l_1},\ldots,N^{l_N} \}$ of $N$, and the second runs over $M$ distinct integers. Here $M$ denotes the length of the partition $\vec k$.
The Higgs branch index \eqref{eq:HBI} can then be rewritten as
\begin{align}
    \mathcal{I}_N^{\text{HB}}(h,b,w=e^{2 \pi i E}) = \sum_{\vec{k}} \frac{1}{\prod_{i=1}^N l_i!} \left(\prod_{j=1}^M A_{k_j}(h,b)\right) \left(\sum_{n_1 \neq n_2 \neq \ldots \neq n_M} e^{2\pi i E \sum_{j=1}^M k_j n_j }\right)
    \,,
\end{align}
where $A_k$ is defined by 
\begin{align}
\label{eq:Molin}
    &A_k(h,b) =\nonumber \\
    &=
    \frac{1}{k!} \oint \left(\prod_{l=1}^{L}\sum_{i=1}^{k}\frac{\mathrm{d}z_i^{(l)}}{2\pi i z_i^{(l)}}\right)
    (1-h)^{kL}
    \left( \prod_{l=1}^{L} \prod_{i \neq j}^{k} \left( 1 - z_{{i}}^{(l)} \left(z_{j}^{(l)}\right)^{-1} \right) \left(1 - z_{{i}}^{(l)} \left(z_{j}^{(l)}\right)^{-1} h \right) \right) 
    \nonumber \\
    &\qquad\times
    \left( \prod_{l=1}^{L} \prod_{i,\,j}^{k} \left( 1 - \left(z_{{i}}^{(l)}/z_{j}^{(l+1)}\right)^{\pm1}
    h^{1/2}  \right)^{-1} \right) \Bigg|_{z_i^{(L+1)}=z_i^{(1)} b} \,,
\end{align}
which coincides with the standard Molien integral formula for the Hilbert series applied to the circular quiver with $L$ nodes whose ranks all equal to $M$, although it is not exactly the Higgs branch Hilbert series because the theory is not fully Higgsable. We have also used $w = e^{2 \pi i E}$ and
\begin{align}
    \frac{1}{|\mathcal W_{\vec m}|} = \frac{1}{\prod_{j = 1}^M k_j!} \,.
\end{align}

We are now left with two tasks: the evaluation of $A_k$ and the summation over $\vec n$. The latter is simpler and can be carried out using the standard Fourier-series identity, which yields a periodic Dirac delta:
\begin{align}
    \sum_{n=-\infty}^{+\infty}e^{2 \pi i E n}=\delta(E)
    \,.
\end{align}
Thus, the final expression for the Higgs branch index is given by a sum of products of the $A_{k_j}$, multiplied by Dirac delta distributions. Let us consider a concrete example. For $N=3$, the Higgs branch index is written as
\begin{align}
    \mathcal{I}_{N=3}^{\text{HB}} 
    &=
    \left(\sum_{n_1=-\infty}^{+\infty}e^{iE (3 n_1)} A_3\right)
    +\left(\sum_{n_1 \neq n_2=-\infty}^{+\infty}e^{iE (2 n_1+n_2)}A_2A_1\right)
    \nonumber\\
    &\quad
    +\frac{1}{6}\left(\sum_{n_1\neq n_2\neq n_3=-\infty}^{+\infty}e^{iE(n_1+n_2+n_3)}A_1^3\right)
    \nonumber\\
    &=
    \delta(3E)A_3
    +\Big[\delta(2E)\delta(E)-\delta(3E)\Big]A_2A_1
    +\frac{1}{6}\Big[\delta(E)^3-3\delta(2E)\delta(E)+2\delta(3 E)\Big]A_1^3
    \nonumber\\
    &= \delta(3E) \left(A_3-A_2 A_1+\frac13 A_1^3\right)
    +\delta(2 E) \delta(E) \left(A_2 A_1-\frac12 A_1^3\right)
    +\frac16 \delta(E)^3 A_1^3
    \,, \label{eq:HBI_N=3}
\end{align}
where, for brevity, we have suppressed the dependence on $h$ and $b$.

Now let us turn to the Molien integral \eqref{eq:Molin}. It can be evaluated by summing the residues of the poles inside the unit circle. For $|h| < 1$ and $|b| = 1$, the contour integrations can be carried out sequentially by picking up, for each integration variable, the residues at the poles involving positive powers of $h$, as well as the residue at the origin. For $k=L=1$, the integrand takes the simple form
\begin{align}
    \frac{1-h}{z^{(1)}_1 \left(1-b \sqrt{h}\right) \left(1-b^{-1} \sqrt{h}\right)} \,,
\end{align}
and the only pole is at $z^{(1)}_1 = 0$. Therefore, the integral for $k = L = 1$ is
\begin{align}
    A_1(h,b) = \frac{1-h}{\left(1-b \sqrt{h}\right) \left(1-b^{-1} \sqrt{h}\right)} \,.
\end{align}
For $k=1$ and $L=2$, the integrand is
\begin{align}
    \frac{z^{(1)}_1 z^{(2)}_1 (1-h)^2}{\left(z^{(1)}_1-\sqrt{h} z^{(2)}_1\right) \left(z^{(1)}_1-b^{-1} \sqrt{h} z^{(2)}_1\right) \left(\sqrt{h} z^{(1)}_1-z^{(2)}_1\right) \left(b \sqrt{h} z^{(1)}_1-z^{(2)}_1\right)} \,.
\end{align}
As a function of $z^{(1)}_1$, the poles involving positive powers of $h$ are located at $\sqrt h z^{(2)}_1$ and $b^{-1} \sqrt h z^{(2)}_1$. Summing the corresponding residues gives
\begin{align}
    \frac{1-h^2}{z^{(2)}_1 \left(1-b h\right) \left(1-b^{-1} h\right)} \,,
\end{align}
where has a single pole at $z^{(2)}_1 = 0$. The integral therefore evaluates to
\begin{align}
    \frac{1-h^2}{\left(1-b h\right) \left(1-b^{-1} h\right)} \,.
\end{align}
Analogously, we have evaluated the integral for $1 \leq k \leq 7$ and for several values of $L$:
\begin{align}
\begin{aligned}
    k &= 1: \qquad\qquad\; L = 1, \, \cdots, \, 10 \,, \\
    k &= 2: \qquad\qquad\; L = 1, \, \cdots, \, 5 \,, \\
    k &= 3: \qquad\qquad\; L = 1, \, \cdots, \, 4 \,, \\
    k &= 4: \qquad\qquad\; L = 1, \, 2, \\
    k &= 5, \, 6, \, 7: \qquad L = 1\,,
\end{aligned}
\end{align}
which leads to the general formula\footnote{In addition to the residue calculation, these results were also confirmed by expanding the integrand as a series in $h$.}
\begin{gather}
    A_k(h,b) = \frac{1}{k!} \sum_{\vec p} [\vec p] \prod_{j = 1}^k \left[f(h^j,b^j)\right]^{q_j} \,, \\
    f(h,b) = \frac{1-h^L}{\left(1-b h^\frac{L}{2}\right) \left(1-b^{-1} h^\frac{L}{2}\right)} \,,
\end{gather}
where the sum runs over all partitions of $k$, denoted by $\vec{p}=\{1^{q_1},2^{q_2},\dots,k^{q_k}\}$. As before, $[\vec p]$ is given by
\begin{align}
    [\vec p\,] = \frac{k!}{\prod_{j = 1}^k j^{q_j} q_j!} \,.
\end{align}
For example, $A_{k=1,\, 2,\, 3}$ appearing in \eqref{eq:HBI_N=3} for $N = 3$ can be evaluated as\footnote{Notice that these equations closely resemble the symmetric product considered in \cite{Hanany:2012dm}. It was also observed in \cite{Comi:2026gjx} that these relations are valid for the Geometric branch limit of any generic good circular quivers, with flavors and Chern--Simons interactions.}
\begin{align}
    A_1(h,b) =& f(h,b) \,,\\
    A_2(h,b) =& \frac{1}{2} \bigg[ f(h,b)^2 + f(h^2,b^2) \bigg] \,,\\
    A_3(h,b) =& \frac{1}{6} \bigg[ f(h,b)^3 + 3 f(h^2,b^2) f(h,b) + 2 f(h^3,b^3) \bigg] \,,
\end{align}
or equivalently
\begin{align}
    A_3-A_2 A_1+\frac13 A_1^3 &= \frac13 f(h^3,b^3) \\
    A_2 A_1-\frac12 A_1^3 &= \frac{1}{2} f(h^2,b^2) f(h,b) \\
    \frac16 A_1^3 &= \frac16 f(h,b)^3 \,.
\end{align}
Substituting these expressions into \eqref{eq:HBI_N=3}, we obtain the following formula for the Higgs branch index for $N=3$:
\begin{align}
    &\mathcal{I}_{N=3}^{\text{HB}}(h,b,w = e^{2 \pi i E}) \nonumber \\
    &\quad = \frac{1}{6} \Big[ 2\delta(3E) f(h^3,b^3) + 3\delta(2E)\delta(E) f(h^2,b^2)f(h,b) + \delta(E)^3 f(h,b)^3 \Big] \,.
\end{align}
Generalizing this result, we are led to the following simple expression for arbitrary $N$:
\begin{align}
    \mathcal{I}_N^{\text{HB}}(h,b,w = e^{2 \pi i E}) = \frac{1}{N!} \sum_{\vec{k}} [\vec k] \left(\prod_{j = 1}^N \left[\delta\left(j E\right) f\left(h^{j},b^{j}\right)\right]^{l_j}\right) .
\end{align}
This exhibits exactly the same summation over partitions $\vec k$ of $N$, with multiplicities $[\vec k]$, that appears in the dual partition function \eqref{eq:globalbad_sumframes}. Indeed, we will derive exactly the identical formula independently on the dual side, starting from the index of the proposed mirror theory.

\paragraph{Dual Coulomb branch index, $S_N$-gauging, and the ADHM quiver}
We now consider the same limit on the dual side, where the theory is described by multiple frames, each consisting of several SQED sectors in general. Since mirror symmetry exchanges the Higgs and Coulomb branches, the Higgs branch limit we considered corresponds to the Coulomb branch limit from the dual perspective.s

As seen in \eqref{eq:globalbad_sumframes}, the dual $S^3_b$ partition function has a sum-over-frames structure and can be rewritten as
\begin{align}
\label{eq:dualZ}
    &\mathcal{Z}^{\vee}=
    \frac{1}{N!}
    \sum_{\vec{k}}[\vec{k}] \times\nonumber \\
    &\qquad
    \times \Bigg\{
    \prod_{j=1}^{N}
    \Bigg[
    \delta\left(jE\right)
    \,
    e^{j (j-1) \pi i B E} 
    \,
    s_b\left(i\frac{Q}{2}-2 j\left(i\frac{Q}{4}+m_A\right)\right)
    \,
    s_b\left(-i\frac{Q}{2}+2 j\left(i\frac{Q}{4}+m_A\right)\right)
    \nn \\
    & \qquad \qquad \times
    \int\mathrm{d}{Z}\,
    \,
    e^{2 \pi i \left(jB\right) Z}
    \prod_{n=1}^{L}\prod_{q=1}^{j} 
    s_b\left( i\frac{Q}{4}+m_A\pm\left(Z-Y_n+(q-1)E\right)\right)
    \Bigg]^{l_j}
    \Bigg\}
    \,,
\end{align}
where, as before, $\vec{k} = \{ 1^{l_1},\ldots,N^{l_N} \}$ runs over all partitions of $N$. Relative to \eqref{eq:globalbad_sumframes}, we have restored the superconformal R-charge by shifting $m_A \rightarrow m_A+iQ/4$, and we have also introduced extra singlets whose contributions are given by the last two factors in the second line. Although these factors cancel each other and hence contribute trivially to the partition function, we include them for later comparison with the ADHM quiver. As we will see, half of them will belong to the SQED sector, and the other half correspond to additional free hypermultiplets charged under accidental symmetries in the IR.

It is straightforward to translate \eqref{eq:dualZ} into the superconformal index: 
\begin{align}
&\hat{\mathcal I}(x,a,b,w=e^{2 \pi i E},\vec y) = \frac{1}{N!} \sum_{\vec k} [\vec k] \times\nonumber \\
&\qquad \times \left\{ \prod_{j = 1}^N \left[\delta(j E) \, 
\frac{\left(x^{j} a^{2 j};x^2\right)}{\left(x^{2-j} a^{-2 j};x^2\right)} \sum_{m = -\infty}^\infty \oint \frac{dz}{2 \pi i z} b^{j m} x^{\frac{j L |m|}{2}} a^{j L |m|}
\right.\right. \nonumber \\
&\qquad \qquad \left.\left.\times
\frac{\left(x^{2-j} a^{-2 j};x^2\right)}{\left(x^{j} a^{2 j} ;x^2\right)}
    \left(\prod_{n=1}^L \prod_{q = 1}^j \frac{\left(x^{\frac32+|m|} a \left(\frac{z w^{q-1}}{y_n}\right)^\pm ;x^2\right)}{\left(x^{\frac12+|m|} a^{-1} \left(\frac{z w^{q-1}}{y_n}\right)^\pm ;x^2\right)}\right) \right]^{l_j}\right\} \,,
\end{align}
where we have used the same parameter map as in \eqref{eq:parameter_map}.
The Coulomb branch index of the mirror theory, which must coincide with the Higgs branch index of the original theory, is obtained by taking the same limit $x\to 0\,, \, 1/a \to 0$ while keeping $h=xa^2$ fixed:
\begin{align}
\hat{\mathcal I}^{\text{CB}}(h,b,w=e^{2 \pi i E}) &= \lim_{\substack{x, \, 1/a \rightarrow 0, \\ h = x a^2 \text{ fixed}}} \hat{\mathcal I}(x,a,b,w=e^{2 \pi i E},\vec y) \\
&= \frac{1}{N!} \sum_{\vec k} [\vec k] \left\{ \prod_{j = 1}^N \left[\delta(j E) \left(1-h^j\right) \sum_{m = -\infty}^\infty \frac{b^{j m} h^{\frac{j L}{2} |m|}}{1-h^j}\right]^{l_j}\right\} \\
&= \frac{1}{N!} \sum_{\vec k} [\vec k] \left\{ \prod_{j = 1}^N \left[\delta(j E) \left(1-h^j\right) g(h^j,b^j)\right]^{l_j}\right\} \,, \label{eq:CBI}
\end{align}
where
\begin{align}
\label{eq:gf}
g(h,b) \equiv \sum_{m = -\infty}^\infty \frac{b^{m} h^{\frac{L}{2} |m|}}{1-h} = \frac{1-h^L}{(1-h) \left(1-b h^\frac{L}{2}\right) \left(1-b^{-1} h^\frac{L}{2}\right)} = \frac{f(h,b)}{1-h} \,.
\end{align}
In particular, the identity \eqref{eq:gf} relating $g(h,b)$ and $f(h,b)$ reflects mirror symmetry between SQED with $L$ flavors and the abelian circular quiver with $L$ nodes, up to the decoupled diagonal $U(1)$. We define $g(h,b)$ in this way to make the $\mathcal N=4$ SQED structure manifest, even though the factor of $1-h$ in the denominator of $g(h,b)$ is canceled by the prefactor multiplying $g(h,b)$. This prefactor, together with the delta function, can be interpreted as the contribution of decoupled singlets. 

Most crucially, this Coulomb branch index exactly matches the Higgs branch index of the original theory. This agreement between the two indices provides further support for the proposed mirror duality of the globally bad circular quiver and its sum-over-frames structure.\\

Furthermore, the sum-over-frames structure of the index reveals an interesting connection between the globally bad circular quiver and the so-called ADHM quiver. Recall that the globally bad theory with $L$ nodes can be engineered as a stack of D3-branes wrapped on a circle and intersecting $L$ NS5-branes. Under S-duality, this configuration is mapped to a stack of D3-branes intersecting $L$ D5-branes, whose low-energy dynamics can be captured by an SQCD theory with one adjoint and $L$ fundamental hypermultiplets, namely the ADHM quiver theory. Unlike the globally bad circular quiver, the ADHM theory is only ugly and contains hyper- and twisted hypermultiplets of conformal dimension $1/2$ decoupled in the IR. This brane picture suggests that the globally bad theory and the ADHM theory describe the same IR physics, up to subtleties primarily associated with decoupled operators and accidental symmetries.

Here we provide further evidence for this relation by comparing Coulomb branch indices of the ADHM quiver and the mirror of the globally bad circular quiver.\footnote{For the ADHM quiver, the Coulomb limit of the superconformal index indeed reproduces the Coulomb branch Hilbert series, since the theory is not bad.} The Coulomb branch of the $U(N)$ ADHM quiver with $L$ flavors matches the symmetric product of $N$ copies of the Coulomb branch of SQED with $L$ flavors \cite{deBoer:1996mp,Cremonesi:2013lqa}:
\begin{align}
    \mathcal{M_C} (\text{$U(N)$ ADHM quiver with $L$ flavors} ) = \text{Sym}^N [ \mathcal{M_C} ( \text{SQED with $L$ flavors} ) ]
    \,,
\end{align}
where $\mathcal M_C(\mathcal T)$ denotes the Coulomb branch moduli space of the theory $\mathcal T$. This relation is encoded in the Hilbert series as \cite{Hanany:2018cgo}:
\begin{align}
\label{eq:CBHS_ADHM}
\mathrm{HS}_{\mathcal M_C}(\text{$U(N)$ ADHM quiver with $L$ flavors}) = \frac{1}{N!} \sum_{\vec{k}} [\vec{k}] \prod_{j=1}^N \left[g(h^j,b^j)\right]^{l_j} \,,
\end{align}
where the sum again runs over all partitions of $N$, denoted by $\vec{k} = \{ 1^{l_1},2^{l_2},\ldots,N^{l_N} \}$. As before, $[ \vec{k} ]$ is the number of elements of $S_N$ in the conjugacy class corresponding to the partition $\vec{k}$, which is given by \eqref{eq:conjugacy class order}. As explained, $g(h,b)$, defined in \eqref{eq:gf}, corresponds to the Coulomb branch Hilbert series of $\mathcal N=4$ SQED with $L$ fundamental hypermultiplets \cite{Cremonesi:2013lqa}:
\begin{align}
    g(h,b) &= \text{HS}_{\mathcal M_C} (\text{SQED with $L$ flavors}) = \text{HS}(\mathbb{C}^2/\mathbb{Z}_L) \\
    &= \frac{1-h^L}{(1-h) \left(1-b h^\frac{L}{2}\right) \left(1-b^{-1} h^\frac{L}{2}\right)} \,, \label{eq:g}
\end{align}
which can be obtained either from the monopole formula \cite{Cremonesi:2013lqa} or as a limit of the superconformal index as we have done above.

We thus see that the Hilbert series of the ADHM quiver in \eqref{eq:CBHS_ADHM} agrees with the Coulomb branch index of the mirror of the globally bad circular quiver in \eqref{eq:CBI}, up to the additional factors associated with decoupled singlets. As explained above, such singlets are expected, since one theory is ugly whereas the other is bad. Nevertheless, this comparison lends further support to the relation between the two theories and sharpens it at the level of the Coulomb branch Hilbert series.\\

We have so far examined the Higgs branch index of the globally bad circular quiver, or equivalently the Coulomb branch index of its mirror theory, and its connection to the Coulomb branch Hilbert series of the ADHM quiver. Similarly, it would be interesting to consider the \emph{Coulomb branch} index of the bad circular quiver and its relation to the Higgs branch Hilbert series of the ADHM quiver.\footnote{It coincides with the Coulomb branch Hilbert series by mirror symmetry for the ADHM quiver with a single flavor.} While the latter admits several evaluation formulas derived in various contexts, including the instanton counting and the AdS/CFT correspondence---for example, see \cite{Benvenuti:2010pq,Hanany:2012dm,Rodriguez-Gomez:2013dpa,Barns-Graham:2017zpv,Crew:2020psc,Hwang:2025hfs}---the former has not been discussed in the literature, presumably due to the badness of the theory. We leave a detailed analysis of the Coulomb branch index of the the bad circular quiver, as well as its comparison with the Higgs branch of the ADHM quiver, to future work.

As a simple consistency check, nevertheless, let us comment on the Coulomb branch dimension of the globally bad circular quiver, which should coincide with the Higgs branch dimension of its mirror. Each gauge node of the circular quiver contributes an $N$-quaternionic-dimensional Coulomb branch, including the contributions from decoupled monopole operators. Hence the total Coulomb branch dimension is $LN$, where $L$ is the number of gauge nodes. This should match the Higgs branch dimension in each magnetic frame. To see this, recall that a $U(1)$ theory with $F$ flavors has a Higgs branch of quaternionic dimension $F-1$. In addition, half the total number of additional singlets and delta functions can be interpreted as the contribution of free hypermultiplets corresponding to the decoupled monopoles in the original theory and must be included. In \eqref{eq:dualZ}, a frame associated with a partition $\vec k$ consists of $l_j$ copies of $U(1)$ theories with $L \cdot j$ flavors, together with $l_j$ additional free hypermultiplets. Therefore, in this frame, the total Higgs branch dimension is
\begin{equation}
\sum_{j=1}^N \Big[ l_j (L \cdot j - 1) + l_j \Big] = LN \,,
\end{equation}
where we have used $\sum_{j=1}^N j l_j = N$. Although the decomposition into the SQED sector and free hypermultiplets depends on the partition $\vec k$, the total Higgs branch dimension is independent of the frame. Thus, the Higgs branch dimension of each magnetic frame precisely agrees with the Coulomb branch dimension of the globally bad circular quiver theory.

\subsubsection{Electric dualization of globally bad quivers} \label{subsec:electricdual_globallybad}

As explained at the beginning of this section, the global badness of a circular quiver is not determined by the presence of underbalanced nodes. Consequently, the electric dualization algorithm described in the previous section---in which an underbalanced node is replaced by a collection of (over)balanced nodes---does not apply to globally bad circular quivers. Nevertheless, having identified the magnetic dual frames of a globally bad quiver, one can obtain its electric dual frames by taking the mirror of each good magnetic frame. As discussed above, each magnetic dual frame consists of a product of SQED factors, together with delta functions encoding the contribution of monopole operators with nonzero VEVs. The mirror theories of these SQEDs are good abelian linear quivers with a single flavor attached to each end. Furthermore, as we will see, each linear quiver with the accompanying delta function, which is dual to a $U(1)$ vector multiplet coupled to an FI parameter, fits into an abelian circular quiver theory. Hence, the electric dual frames of the globally bad circular quiver are realized as a collection of abelian circular quiver theories, which, surprisingly, do not need to have the same length as the original quiver.

Let us now examine this procedure at the level of the $S^3_b$ partition function. First, recall the magnetic partition function \eqref{eq:globalbad_sumframes} of a globally bad theory with $L$ nodes of rank $N$, which we reproduce here for convenience:
\begin{align}
\label{eq:globally_bad_magnetic}
    \mathcal{Z}^{\vee} &=
    \frac{1}{N!}
    \sum_{\vec{k}}[\vec{k}]
    \left\{
    \prod_{j=1}^{N}
    \Bigg[
    \delta\left(jE\right)
    \,
    e^{j (j-1) \pi i BE} \int\mathrm{d}{Z} \, e^{2 \pi i \left(jB\right) Z}\right.
    \nonumber \\
    & \qquad \qquad \qquad \left.\times
    \prod_{n=1}^{L}\prod_{q=1}^{j} 
    s_b\left(m_A\pm\left(Z-Y_n+(q-1)E\right)\right)
    \Bigg]^{l_j}
    \right\}
    \,.
\end{align}
It consists of multiple frames, each given by a product of SQED theories. One should note that the standard $\mathcal N=4$ SQED contains a singlet coupled to the trace part of the Higgs branch operator matrix $\tilde Q Q$, which is absent in the expression above. Although this singlet can be restored by introducing a pair of additional singlets, as in the previous subsubsection where we compared with the ADHM quiver, we will keep the original expression without this singlet.

The mirror of such an SQED without the singlet is the \emph{flipped} abelian linear quiver shown in Figure~\ref{fig:N2_SQED}.
\begin{figure}[!ht]
    \centering
    \includegraphics[width=0.75\linewidth]{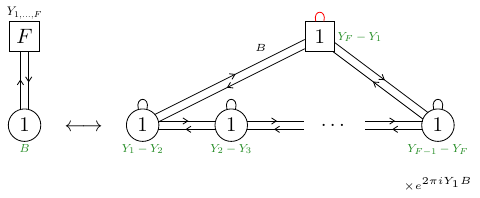}
    \caption{
    The $\mathcal N=2$ SQED with $F$ flavors, obtained by flipping the adjoint chiral of the $\mathcal N=4$ SQED, and its mirror dual: a flipped abelian linear quiver with a single flavor attached to each end. The flipping singlet on the dual side is denoted by a red arc.
    }
    \label{fig:N2_SQED}
\end{figure}
The partition function can then be written as 
\begin{align}
\label{eq:globally_bad_electric_linear}
    \mathcal{Z}^\vee_{\mathrm{elec}} &=
    \frac{1}{N!}
    \sum_{\vec{k}}[\vec{k}]
    \Bigg\{
    \prod_{j=1}^{N}
    \Bigg[
    \delta\left(jE\right)
    \,
    e^{j (j-1) \pi i BE} \, s_b\left(i \frac{Q}{2}-2 m_A\right) \, \mathrm L_L^j\left(m_A;B;E;\vec Y\right)
    \Bigg]^{l_j}
    \Bigg\}
    \,,
\end{align}
where $\mathrm L_L^j$ is the partition function of the linear quiver dual to the $\mathcal N=4$ SQED, which can be read from the quiver following the rule in Appendix~\ref{app:conventions_PF}, and $s_b(iQ/2-2 m_A)$ corresponds to the flipping singlet, colored red in Figure~\ref{fig:N2_SQED}. More interestingly, one can further dualize the delta function $\delta (jE)$ into an $\mathcal N=2$ $U(1)$ vector multiplet coupled to an FI parameter $j E$, which can be promoted to an $\mathcal N=4$ vector once combined with the flipping singlet. This $U(1)$ factor then combines with the linear quiver to form an abelian circular quiver with $L j$ nodes, where the $U(1)$ factor dual to the delta plays the role of the diagonal $U(1)$. Namely, the partition function can be rewritten as
\begin{align}
\label{eq:globally_bad_electric_circular}
    \mathcal{Z}^\vee_{\mathrm{elec}}&=
    \frac{1}{N!}
    \sum_{\vec{k}}[\vec{k}]
    \Bigg\{
    \prod_{j=1}^{N}
    \Bigg[
    e^{j (j-1) \pi i BE} \, \mathrm C_L^j\left(m_A;B;E;\vec Y\right)
    \Bigg]^{l_j}
    \Bigg\}
    \,,
\end{align}
where $\mathrm C_L^j$ is the partition function of the resulting $\mathcal N=4$ abelian circular quiver with $L j$ nodes, which again can be read from the quiver. Thus, we obtain two presentations of the electric dual of the globally bad circular quiver: one in terms of linear quivers, as in \eqref{eq:globally_bad_electric_linear}, and one in terms of circular quivers, as in \eqref{eq:globally_bad_electric_circular}. We will illustrate the detailed structure with an example shortly. In particular, we will see that the additional BF coupling between $B$ and $E$ can naturally be absorbed into the circular quiver theory by redistributing the baryonic charges among the bifundamental fields.

One quick comment is that, although the number of flavors appearing in \eqref{eq:globally_bad_magnetic} is $L j$, their associated parameters are only $Y_1, \, Y_2, \, \dots, \, Y_L$, repeated $j$ times. Since they are mapped to the FI parameters in $\mathrm L_L^j$ of the dual linear quiver, and subsequently to those in $\mathrm C_L^j$ of the dual circular quiver, the FI parameters in the electric frames are not completely generic but are fixed to particular combinations of the original parameters. Together with the FI parameter $j E$ for the diagonal $U(1)$, originating from the dualization of the delta function $\delta(jE)$, the FI parameters of the resulting abelian circular quiver are given by $j$ copies of $Y_1-Y_2, \, Y_2-Y_3, \, \dots, \, Y_L-Y_1+E$.\\

Most interestingly, this electric dualization reveals intriguing IR dynamics of the globally bad circular quiver theory. For concreteness, let us again consider the case with $N = 3$ and $L = 4$ as an example. Other cases can easily be obtained using the accompanying Mathematica code. The original globally bad circular quiver, together with its magnetic and electric duals, is shown in Figure~\ref{fig:elec_global_bad}.
\begin{figure}[!ht]
\centering
    \includegraphics[width=\textwidth]{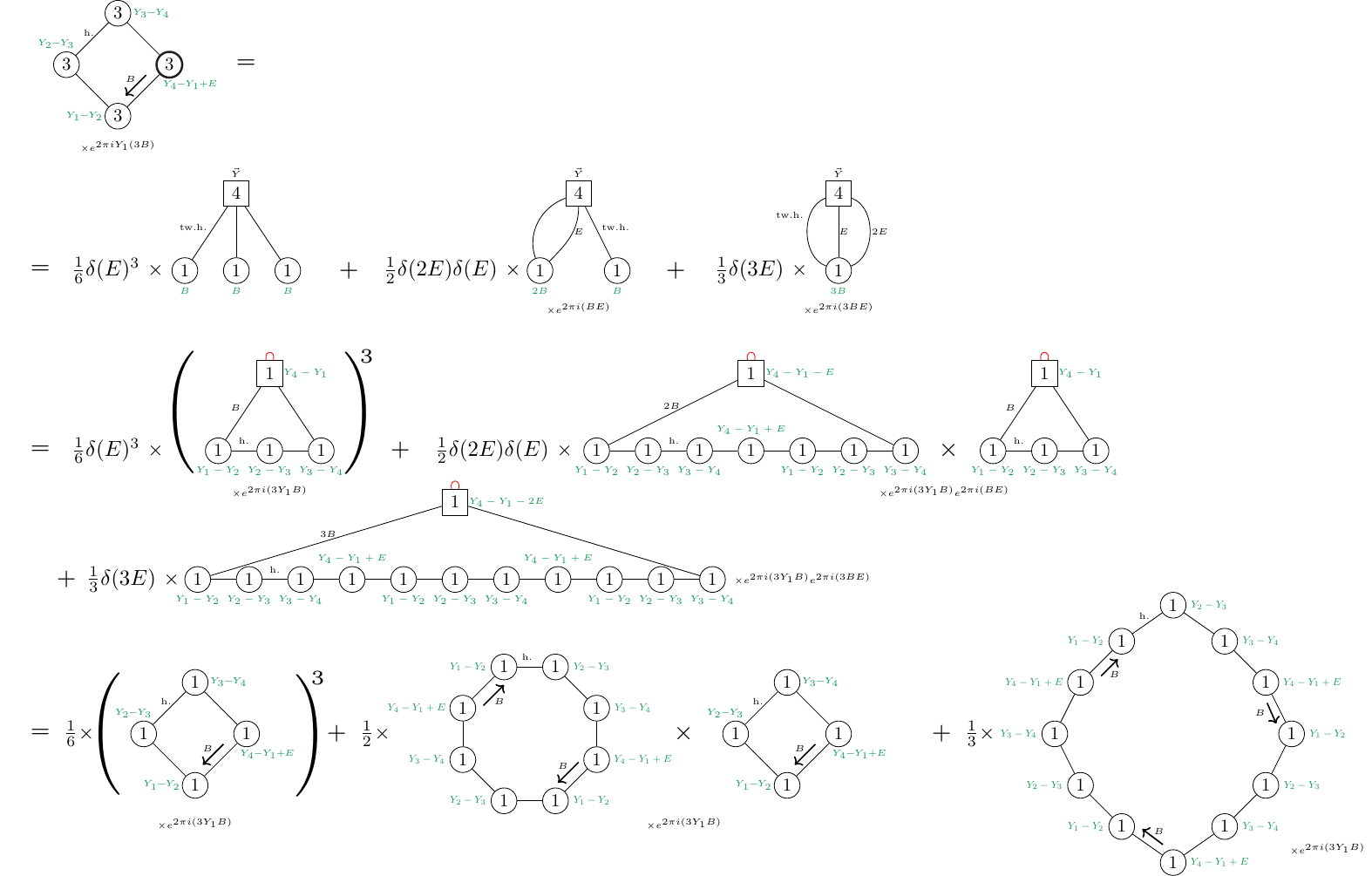}
    \caption{The magnetic and electric frames of the globally bad circular quiver with $N=3$ and $L=4$. For the electric frames realized as flipped linear quivers, red arcs denote the flipping $\mathcal N=2$ chiral singlets, while the remaining lines denote $\mathcal N=4$ hypermultiplets.
    }
    \label{fig:elec_global_bad}
\end{figure}
As explained above, the magnetic frames are described by different collections of SQED sectors, whereas the electric frames admit two descriptions: one in terms of abelian linear quivers and one in terms of abelian circular quivers.

We focus here on the latter description.
The first frame is described by three copies of an abelian quiver of length 4. The second frame consists of an abelian circular quiver of length 8 and another of length 4. Note that the FI parameters of the longer quiver are given by $Y_1-Y_2, \, Y_2-Y_3, \, Y_3-Y_4$ and $Y_4-Y_1+E$, repeated twice. This follows from the following identification of the mass parameters associated with the $8$ flavors of the magnetic QED:
\begin{gather}
Y_1 \,, \qquad Y_2 \,, \qquad Y_3 \,, \qquad Y_4 \,, \nonumber \\
Y_1-E \,, \quad Y_2-E \,, \quad Y_3-E \,, \quad Y_4-E \,.
\end{gather}
The last frame is given by an abelian circular quiver of length 12, whose FI terms are given by the same set of parameters repeated three times.

Notice that the circular quivers here are presented in a slightly different form from our convention so far. Once we follow the algorithm, a resulting circular quiver of length $L j$ has the BF term $j (j-1) \pi i B E$ and the nontrivial baryonic charge $j B$ only for the last bifundamental hyper between the last and first gauge nodes. One can redistribute this baryonic charge to other bifundamentals by shifting the gauge variables. Specifically, for a quiver of length $L j$, we shift the gauge variables as follows:
\begin{align}
Z^{(k L+1,\dots,L)} \quad &\rightarrow \quad Z^{(k L+1,\dots,L)}-(j-k-1) B
\end{align}
for $k = 0,\dots,j-1$
such that the bifundamental contributions become
\begin{align}
&s_b\left(iQ/2-m_A\pm\left(Z^{(i)}-Z^{(i+1)}-\delta_{i,Lj} B\right)\right) \nonumber \\
&\rightarrow \quad s_b\left(iQ/2-m_A\pm\left(Z^{(i)}-Z^{(i+1)}-\sum_{k=1}^j \delta_{i,L k} B\right)\right)
\end{align}
as shown in Figure~\ref{fig:elec_global_bad}.
Due to this shift, the original FI terms yield an additional BF term $-j(j-1) \pi i B E$, which exactly cancels out the original BF term resulting from the dualization.\\

These electric frames are associated with particular monopole VEVs, which manifest themselves as delta functions in the partition function. The electric frames in \eqref{eq:globally_bad_electric_linear}, presented in terms of linear quivers, contain the delta function factors $\delta(E)^3$, $\delta(2E) \delta (E)$, and $\delta(3E)$, respectively. These factors should correspond to monopole operators whose naive conformal dimensions vanish. From the Higgs branch index analysis in the previous subsubsection, we have seen that an arbitrary diagonal magnetic flux of the globally bad circular quiver, $\vec m = \vec m^{(1)} = \dots = \vec m^{(L)}$, has vanishing conformal dimension. These fluxes are generated by the following monopole operators:
\begin{align}
    \mathfrak M^{(\{\pm,0,0\},\{\pm,0,0\},\{\pm,0,0\},\{\pm,0,0\})} \quad &\longleftrightarrow \quad  \delta(E) \, s_b \left(i\tfrac{Q}{2} - 2m_A\right) , \\
    \mathfrak M^{(\{\pm,\pm,0\},\{\pm,\pm,0\},\{\pm,\pm,0\},\{\pm,\pm,0\})} \quad &\longleftrightarrow \quad  \delta(2 E) \, s_b \left(i\tfrac{Q}{2} - 2m_A\right) , \label{eq:double flux monopole} \\
    \mathfrak M^{(\{\pm,\pm,\pm\},\{\pm,\pm,\pm\},\{\pm,\pm,\pm\},\{\pm,\pm,\pm\})}  \quad &\longleftrightarrow \quad \delta(3 E) \, s_b \left(i\tfrac{Q}{2} - 2m_A\right) \,,\label{eq:triple flux monopole}
\end{align}
with all signs correlated. Here we have used the same notation as in Table~\ref{tab:operators_mirror_pair_1} to denote the associated magnetic flux. On the right hand side, we display the corresponding free hypermultiplet contribution to the partition function, realized as a delta function multiplied by $s_b (iQ/2 - 2m_A)$.

Thus, the first electric frame describes the effective theory on the Coulomb branch where three monopoles of the first type acquire nonzero VEVs. The second electric frame describes the branch where one first-type monopole and one second-type monopole acquire nonzero VEVs, while the last electric frame corresponds to the branch where a third-type monopole acquires a nonzero VEV.

The first electric frame is easy to understand: it is simply the effective theory at a generic point on the classical Coulomb branch of the diagonal subgroup $U(3) \subset U(3)^4$. This branch is partially parametrized by the vector multiplet scalar $\vec \sigma \equiv \vec \sigma^{(1)} = \dots = \vec \sigma^{(4)}$ with all components distinct, i.e., $\sigma_i \neq \sigma_j$ for $i \neq j$. If some components coincide, e.g., $\sigma_1 = \sigma_2 \neq \sigma_3$, two copies of the abelian quiver recombine into a nonabelian quiver with $U(2)$ gauge nodes, while the length of the quiver remains 4. On the other hand, the additional electric frames found above exhibit a rather different phenomenon: they remain abelian, but their lengths are enlarged to multiples of the original length, with the original parameters repeated periodically. This reflects a novel non-perturbative effect associated with monopole operators of non-unit flux, such as those given in \eqref{eq:double flux monopole} and \eqref{eq:triple flux monopole}.

\subsection{Badness of general circular quivers without flavors}
\label{subsec:globally_and_locally_bad}

We conclude this section by emphasizing that the classification of a quiver as ugly or bad, and as locally or globally bad, is frame-dependent. 
This distinction is important as a given theory, for example, can have many gauge nodes that are ugly and thus appear ugly, but after a sequence of electric dualizations we can reach a frame that has bad gauge nodes.
Similarly, a locally bad circular quiver may also have hidden global badness, which becomes manifest only after dualization. Thus, whenever we refer to the badness of a quiver, it should be understood as the badness in a given duality frame.

We recall that our definitions of the badness of a frame are the following:
\begin{itemize}
    \item A $U(N_C)$ gauge node is locally bad if the total number of fundamental hypermultiplets $F_{\text{eff}}$ satisfies $F_{\text{eff}} \leq 2N_C - 1$. We refer to ugliness when the inequality is saturated.

    \item A theory is globally bad if it is a circular theory with all ranks equal and total number of flavors satisfying $F_{\text{tot}} \leq 1$. We refer to ugliness when the inequality is saturated.
\end{itemize}
It follows from the definitions that a given frame can be either locally bad \textit{or} globally bad, but not both at the same time. Moreover, a theory that is locally bad can become globally bad after performing a sequence of local electric dualizations (see Section~\ref{sec:local_badness_circular}), but the opposite can not happen. Nonetheless, a theory that is locally bad can become globally bad only if the total number of flavors is one or zero, thus this possibility can be somehow predicted from the beginning.

This leads to the following prescription to find the frames describing a generic bad circular theory.
\begin{itemize}
    \item If the frame is locally bad we dualize it using the local electric algorithm described in Section~\ref{sec:local_badness_circular} and, if the resulting frames are still locally bad, we keep iterating this step.

    \item Once we obtain a set of frames that are locally good, if they are globally bad, we further dualize them using the results in Section~\ref{subsec:electricdual_globallybad}.
\end{itemize}

Consider, for example, the theory on the left hand side of Figure~\ref{fig:Globally_and_Locally_bad_example}. It contains a single bad node, colored red. Furthermore, since the quiver has zero flavors, it may become globally bad after iterated dualization, and this is indeed what occurs. If we implement the electric dualization algorithm as explained in Section~\ref{sec:local_badness_circular}, we obtain an electric dual description consisting of multiple frames with different ranks, deltas, singlets, and BF couplings, as shown on the right hand side of the figure.
\begin{figure}[!ht]
\centering
    \includegraphics[width=\textwidth]{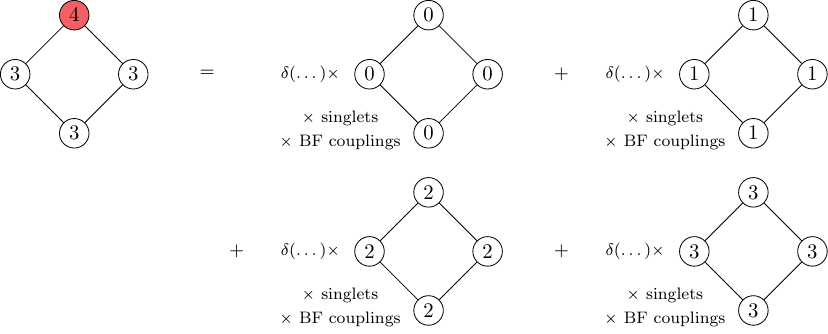}
    \caption{An example of hidden global badness. Although the original quiver exhibits only local badness due to a single bad node, colored red, it also has hidden global badness, which becomes manifest after dualization.}
    \label{fig:Globally_and_Locally_bad_example}
\end{figure}

The quivers appearing in these dual frames, except for the first one, are globally bad and can be described by the method discussed in the previous subsubsection. For the first frame, on the other hand, there is no interacting sector: the theory is described solely by decoupled singlets and therefore does not exhibit global badness. 

Running the electric dualization algorithm in many cases of this kind, one can formulate the following general statement for circular quivers with arbitrary ranks and no flavors.
If we call $C[L,\{N_i\}]$ a circular quiver with $L>2$ gauge nodes of rank $\{N_i\}=\{N_1,N_2,\dots,N_L\}$ and without any flavor, the set of dual frames is of the form
\begin{align}
    C[L,\{N_i\}] 
    =&
    \sum_{k=0}^{\text{min}\{N_i\}} (\text{copies of})\,\bigg[\delta(\text{FI}\pm\dots)\times C_{\text{const}}[L,k]
    \substack{\begin{array}{l}\times\text{singlets} \\ \times\text{BF couplings}\end{array}}
    \bigg]\,,
    \label{eq:generic_globally_locally_bad}
\end{align}
where we defined $C_{\text{const}}[L,k]$ as the circular quiver with $L$ gauge nodes all of rank $k$ and without any flavor. The phrase ``copies of'' indicates that there may be several contributions with the same interacting part but different $\delta$, singlets and BF couplings. As mentioned above, the dual frames exhibit global badness when $k \neq 0$.\\

\section*{Acknowledgments}
The authors would like to thank Hee-Cheol Kim and Chang Lei for useful discussions, and especially Simone Giacomelli, Sara Pasquetti, and Matteo Sacchi for collaboration at an early stage of this project and on closely related works.
RC is supported by the STFC grant ST/X000575/1.
CH is supported by the National Natural Science Foundation of China under Grant No.~12247103. The work of FM is supported by the Austrian Science Fund (FWF), START project ``Phases of quantum field theories: symmetries and vacua'' STA 73-N [grant DOI: 10.55776/STA73]. FM also acknowledges support from the Faculty of Physics, University of Vienna. 
Some results of this work have been presented prior publication at the ``Quivers, Symmetries and SCFTs'' workshop at ICTP Trieste (Sep 1-5, 2025).\\

\appendix

\section{Circular quivers and partitions}
\label{app:partitions}
In this appendix we review the parametrization of 3d $\mathcal{N}=4$ unitary circular good quivers in terms of an integer $N_L$ and of two partitions of an integer $N$. The derivation of the procedure to assign the two partitions to a given quiver exploits the Type IIB brane description of the theory and basically follows \cite{Assel:2012cj}, to which we add the derivation of the inverse formula expressing the quiver data in terms of $N_L$ and the two partitions, which we call $\rho$ and $\sigma$. A very similar discussion can also be found in \cite{Witten:2009xu}.

\subsection{From circular quivers to partitions}
Let us begin by describing how to write the partitions $\rho$ and $\sigma$ of the integer $N$ given a circular quiver theory of choice.
Once we have identified the $L$-th node and we have cut the quiver open, we have a linear brane system describing the resulting linear quiver. We draw the brane system by putting the node $U(N_1)$ to be the leftmost and the node $U(N_{L-1})$ to be the rightmost. Moreover, if $F_L\neq 0$ we put the corresponding D5-branes on the far right. For instance, considering the circular theory in Figure~\ref{fig:example_cutopen_1} we get the linear brane system in Figure~\ref{fig:cwisebrane}.

\begin{figure}[!ht]
    \centering
    \includegraphics[width=0.8\linewidth]{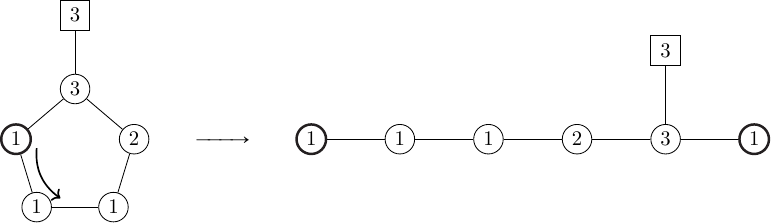}
    \caption{A first way of cutting open the chosen example of circular quiver. The node where we cut the quiver open has been chosen according to the prescription given in Section~\ref{subsec:cutting}. Notice that the first and last nodes in the linearized quivers on the right are identified, as they come from cutting open the bold node in the circular quivers on the left.}
    \label{fig:example_cutopen_1}
\end{figure}

\begin{figure}[!ht]
\centering
\includegraphics[width=.9\textwidth]{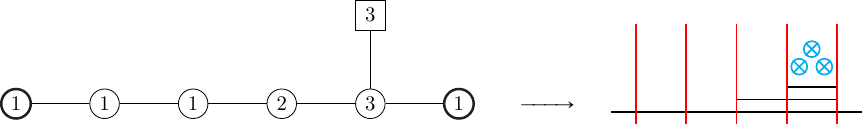}
\caption{The brane setup corresponding to the linear quiver in Figure~\ref{fig:example_cutopen_1}. The notation for the branes is explained in Figure~\ref{fig:circular_brane_setup}.}
\label{fig:cwisebrane}
\end{figure} 

In Figure~\ref{fig:cwisebrane} we have $F_L=0$ and, since in that case $L=1$ and $N_L=1$, we have one D3-brane which crosses the entire brane system from left to right. This is just a consequence of the $N_L$ D3-branes in the circular quiver which go all the way around the circle in Type IIB. 

For the purpose of writing down the corresponding partitions it is now convenient to momentarily remove from the brane system these $N_L$ D3-branes and bring all the D5-branes to the left taking into account Hanany--Witten transitions \cite{Hanany:1996ie}, in such a way that all the NS5-branes lie to the right of the D5-branes. Starting from the setup in Figure~\ref{fig:cwisebrane}, this operation results in the brane system depicted in Figure~\ref{fig:finalbrane}.

\begin{figure}[!ht]
\centering
\includegraphics[width=.9\textwidth]{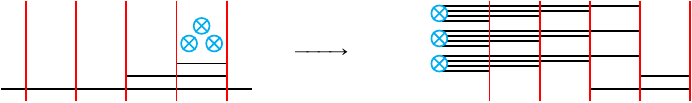}
\caption{The brane manipulations which allow us to read the partitions $\rho$ and $\sigma$ for the quiver in Figure~\ref{fig:cwisebrane}.}
\label{fig:finalbrane}
\end{figure} 

Once we have a brane system such as the one on the right of Figure~\ref{fig:finalbrane}, we can straightforwardly read the two corresponding partitions $\rho$ and $\sigma$. The procedure works as follows. For each D5-brane we have one element in the partition $\sigma$, which is simply equal to the number of D3-branes terminating on it from the right. Since this is equal to the number of NS5's the brane has crossed, we conclude that for each D5-brane contributing a flavor to the $i$-th gauge group in the initial configuration we have an element equal to $i$ in the partition $\sigma$. Overall, we find 
\be
\label{eq:sigma} 
\sigma=(L^{F_L},(L-1)^{F_{L-1}},\dots ,1^{F_1})\,.
\ee 
By construction $\sigma$ as defined in \eqref{eq:sigma} has at most $L$ rows (depending on whether $F_L$ is zero or not) and has exactly $\sum_i F_i$ columns, namely the total number of D5's in the brane system. It is always a partition of the number 
\be
\label{eq:integerN}
N=\sum_{i=1}^{L}iF_i\,.
\ee 
Let us now consider the partition $\rho$, whose elements are encoded in the number of D3-branes in between two consecutive NS5-branes in the final configuration, such as that on the right of Figure~\ref{fig:finalbrane}. Notice that, due to the removal of the $N_L$ D3-branes and the Hanany--Witten moves we have performed, the number of D3-branes in the $i$-th interval is not $N_i$ as in the initial configuration, but is rather 
\be
\label{eq:primedN} 
N_i'=N_i-N_L+\sum_{j>i}(j-i)F_j\,,\quad N_0'=N\,,\quad N_L'=0\,.
\ee 
After this preparation we can immediately state the result: the $i$-th element of the partition $\rho$ is simply equal to $N_{i-1}'-N_i'$. Using \eqref{eq:primedN} we can express this quantity in terms of the data of the circular quiver as 
\be
\label{eq:rho} 
\rho_i=N_{i-1}-N_i+ \sum_{j\geq i}F_j\,.
\ee 
Notice that from \eqref{eq:rho} we can observe that 
\be
\label{eq:rho2} 
\rho_{i}-\rho_{i+1}=N_{i-1}+N_{i+1}+F_i-2N_i\,,
\ee 
and the quantity on the r.h.s.~of \eqref{eq:rho2} is just equal to the excess number of the $i$-th gauge group of the circular quiver. This tells us that $\rho$ is actually a partition only if the initial circular quiver is good, otherwise it is just a sequence of integers whose sum is $N$. In any case, regardless of the goodness of the quiver, $\rho$ describes a sequence of $L$ integers which is the number of NS5-branes. We can also notice that the elements of $\sigma$ and $\rho$ as defined above match the linking numbers of D5 and NS5-branes respectively (see \cite{Witten:2009xu, Assel:2012cj}). From the Figure~\ref{fig:finalbrane} we can see that in our example we have $N=12$, $\sigma=(4,4,4)$ and $\rho=(3,3,2,2,2)$. If instead we go through the procedure explained above for the linear quiver in Figure~\ref{fig:example_cutopen_2}, which describes the same circular quiver but makes use of the other cutting choice, we find $N=9$, $\sigma=(3,3,3)$ and $\rho=(3,3,1,1,1)$. We therefore see that the quiver does not uniquely specify the partitions $\rho$ and $\sigma$ describing it. 

\begin{figure}[!ht]
    \centering
    \includegraphics[width=0.8\linewidth]{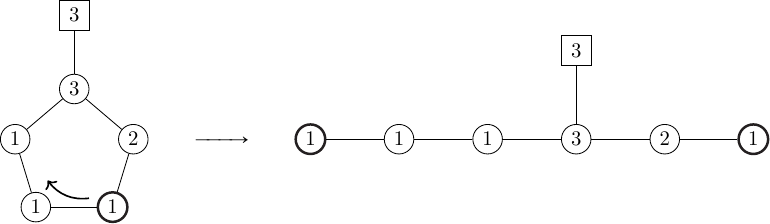}
    \caption{A second way of cutting open the chosen example of circular quiver. The node where we cut the quiver open has been chosen according to the prescription given in Section~\ref{subsec:cutting}. Notice that the first and last nodes in the linearized quivers on the right are identified, as they come from cutting open the bold node in the circular quivers on the left.}
    \label{fig:example_cutopen_2}
\end{figure}

Overall, we have found that we can assign to any circular quiver, possibly not in a unique way, an integer number $N_L$ (the minimum rank appearing in the quiver) and two partitions $\rho$ and $\sigma$ defined by \eqref{eq:rho} and \eqref{eq:sigma} respectively. These three objects actually fully encode the quiver and it is indeed possible to reconstruct the quiver data from them. This is given by an inversion formula we will discuss in the next Subsection. \\

Before moving on, let us use the insights gained so far to reconsider the idea behind our prescription for cutting the quiver open and numbering the nodes, which has been introduced in Section~\ref{subsec:cutting}. It is enough to illustrate this point by means of our example \eqref{fig:circular_brane_setup}. Say we cut at the central $U(1)$ node and proceed clockwise, as shown in Figure~\ref{fig:example_cutopen_3}.

\begin{figure}[!ht]
    \centering
    \includegraphics[width=0.8\linewidth]{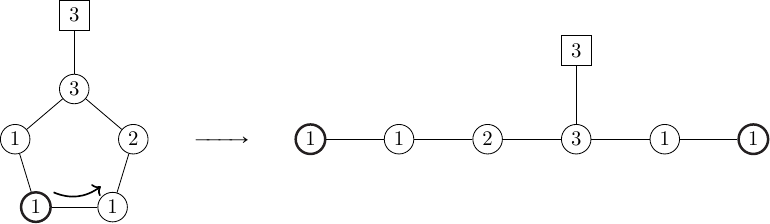}
    \caption{An alternative way of cutting open the chosen example of circular quiver, which however is not ideal to read the partitions $\rho$ and $\sigma$.}
    \label{fig:example_cutopen_3}
\end{figure}

After the Hanany--Witten moves we then land on the brane configuration depicted in the right side of Figure~\ref{fig:finalbrane2}.

\begin{figure}[!ht]
\centering
\includegraphics[width=.9\textwidth]{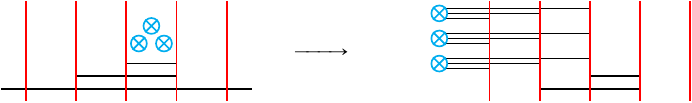}
\caption{The brane manipulations which allow us to read the partitions $\rho$ and $\sigma$ for the quiver in Figure~\ref{fig:example_cutopen_3}.}
\label{fig:finalbrane2}
\end{figure} 

As we see from Figure~\ref{fig:finalbrane2}, we find a NS5-brane completely disconnected from the rest of the quiver and when we try to read off the partitions from the brane system we get the same result we would find by getting rid of it completely. Its presence is however crucial for reconstructing the circular quiver we started from. We therefore find that, with this candidate cutting prescription, regardless of how we move the D5-branes, we end up with a decoupled NS5 whose existence is not encoded in the partitions $\rho$ and $\sigma$ and we would therefore need to provide extra data, besides the partitions, to keep track of its presence. This would lead to an overcomplication of the prescription we can avoid by sticking to the cutting prescription we have presented above. This is the main reason why our prescription is to cut at one end of the subquiver with nodes of constant rank equal to $N_L$. \\

Finally, let us comment on our prescription for numbering the nodes once we have chosen where to cut open the quiver. If in Figure~\ref{fig:example_cutopen_1} we proceed numbering the nodes counterclockwise we end up with the same linear brane system, just reflected. This in principle is fine and we can recover the same partitions $\rho$ and $\sigma$, but in order to do so we should move all the D5-branes to the right, not to the left, otherwise we run into the same problem above mentioned of decoupled NS5-branes which are not accounted for by the partitions (and also decoupled D5 if $F_L\neq 0$). We therefore see that if we allow both numbering options we should specify in each case how to perform Hanany--Witten transitions in order to properly read off the corresponding partitions $\rho$ and $\sigma$, instead of just stating to move all D5-branes to the left. We can avoid this complication by assigning a unique numbering convention from the start. \\

In conclusion, we have found that the circular quiver data are encoded by the integer $N_L$ and the partitions $\rho$ and $\sigma$. We denote the quiver in terms of these data as $C_{\rho}^{\sigma}[SU(N);N_L]$ (see Section~\ref{subsec:Crhosigma}). If we want to derive, in the case of good circular quivers, the mirror quiver we simply need to interchange $\rho$ and $\sigma$, keeping $N_L$ unchanged.

\subsection{From partitions to circular quivers}
Let us now come to the discussion of the inversion formulas which provide the quiver data once the integer $N_L$ and the partitions $\rho$ and $\sigma$ are known. 
They read
\begin{align}
    F_i&=(\sigma^T)_i-(\sigma^T)_{i+1} \,, \label{eq:invformula_flavors} \\
    N_i&=N_L+\sum_{j\leq i}(\sigma^T)_{j}-\sum_{j\leq i}\rho_j \,. \label{eq:invformula_color}
\end{align}
with $i=1,\dots,L$. They can be proved as follows.\\

From the definition \eqref{eq:sigma} of the partition $\sigma$ it is clear that the number of elements equal to $L$ is $F_L$, the number of elements equal to $L-1$ is $F_{L-1}$ and so on. The idea is simply to consider the transpose partition $\sigma^T$, whose elements are denoted as $(\sigma^T)_i$. From the definition \eqref{eq:sigma} we can easily see that the partition $\sigma^T$ has $L$ elements, like $\rho$, and the number of flavors at the $i$-th node $F_i$ is given by \eqref{eq:invformula_flavors} (where $(\sigma^T)_{L+1}=0$ has been set). 

The formula giving the rank of the $i$-th gauge node $N_i$ is a bit more involved and can be derived as follows. 
We start from the definition \eqref{eq:rho} of $\rho$, which we rewrite as 
\be
\label{eq:color1} 
N_i=N_{i-1}+(\sigma^T)_i-\rho_i\,.
\ee 
In the above equation we have used the fact that $(\sigma^T)_i=\sum_{j\geq i}F_j$. Indeed we also have the relation 
\be
\label{eq:color2} 
N_{i-1}=N_{i-2}+(\sigma^T)_{i-1}-\rho_{i-1}\,.
\ee 
If we now plug \eqref{eq:color2} into \eqref{eq:color1} we find 
\be
\label{eq:color3} 
N_i=N_{i-2}+(\sigma^T)_i+(\sigma^T)_{i-1}-\rho_i-\rho_{i-1}\,.
\ee 
We can now reiterate the procedure and trade $N_{i-2}$ for $N_{i-3}$ in \eqref{eq:color3}. Going on in this way we end up with 
\be
\label{eq:color4}
N_i=N_0+\sum_{j\leq i}(\sigma^T)_{j}-\sum_{j\leq i}\rho_j\,,
\ee 
and now we can simply impose that $N_0=N_L$, leading to \eqref{eq:invformula_color}.\\

If we apply these inversion formulas to our example in Figure~\ref{fig:example_cutopen_1} we have, as we have seen before, $N_L=1$, $\rho=(3,3,2,2,2)$ and $\sigma^T=(3,3,3,3)$. By applying \eqref{eq:invformula_flavors} we find that $F_4=3$ and all other $F_i$'s vanish. From \eqref{eq:invformula_color} we find $N_1=N_2=1$, $N_3=2$, $N_4=3$ and $N_5=N_0=1$. These are precisely the data describing our quiver. By doing the same exercise for the situation in Figure~\ref{fig:example_cutopen_2} we find that $F_3=3$ while all other $F_i$'s vanish. 
We also get $N_1=N_2=1$, $N_3=3$, $N_4=2$ and $N_5=N_0=1$. As expected, the ordering of nodes is different but overall the circular quiver is indeed the same. Notice that from \eqref{eq:invformula_color} it is clear that the inequality $\sigma^T\geq \rho$ is equivalent to stating that every node has rank at least $N_L$, namely that $N_L$ is the lowest rank appearing in the quiver.\\

\section{Conventions for the 3d partition function}
\label{app:conventions_PF}
In this appendix we introduce the notation for the 3d $\mathcal{N}=2^*$ $S^3_b$ partition function \cite{Kapustin:2009kz,Jafferis:2010un,Hama:2011ea}.
For a theory with gauge group $G$ and chiral  multiplets of R-charge $r$, in the representations $R_G$ and $R_F$ of the gauge and
flavor symmetry groups, the $S^3_b$ partition function is given by the following integral:
\begin{align}
	\mathcal{Z} (Y,k,\vec{X}) = \frac{1}{|W_G|} \int \prod_{j=1}^{\mathrm{rk} G} \mathrm{d} Z_j \times  & Z_{\text{cl}}(Y,k) 
	\frac{\prod_{\vec{w}_G \in R_G} \prod_{\vec{w}_F \in R_F} s_b \left( \frac{iQ}{2}(1-r) - \vec{w}_G(\vec{Z}) - \vec{w}_F(\vec{X}) \right)}{\prod_{\vec{\rho} \in G} s_b \big( \frac{iQ}{2} - \vec{\rho}(\vec{Z}) \big) }\,,
\end{align}
where $\vec{\rho}$ are the roots of $G$, $\vec{w}_G$ and $\vec{w}_F$ are the weights of the representations $R_G$ and $R_F$, $|W_G|$ is the dimension of the Weyl group of the gauge group $G$, $\vec{Z}$ and $\vec{X}$ are parameters in the Cartan subalgebra of the gauge and flavor groups, respectively, and $Q=b+b^{-1}$ with $b$ the squashing parameter of $S^3_b$. 
The classical term
\begin{align}
	Z_{\text{cl}}(Y,k) = \exp \left[ 2\pi i Y \sum_{j=1}^{\mathrm{rk} G} Z_j - \pi i k \sum_{j=1}^{\mathrm{rk} G} Z_j^2 \right] 
\end{align}
contains the contribution of the FI parameter $Y$ and of the Chern--Simons level $k\in\mathbb{Z}$. It is convenient to define the integration measure for the $U(N)$ group 
\begin{equation}
	\udl{\vec{Z}_{N}}=\frac{1}{N!}\prod_{j=1}^{N}\udl{Z_j} \,,
    \label{eq:U_integration_measure}
\end{equation}
and the combination
\begin{align}
\Delta_N(\vec{Z};m_A) =\Delta_N(\vec{Z}) A_N(\vec{Z};m_A)\,, 
\end{align}
containing the contributions from both the $\mathcal{N}=2$ vector $\Delta_N(\vec{Z}\,)$ and the $\mathcal{N}=2$ adjoint chiral multiplet $A_N(\vec{Z};m_A)$ sitting in the $\mathcal{N}=4$ vector multiplet:
\begin{equation}
	\Delta_N\big(\vec Z\big)=\frac{1}{\prod_{j<k}^{N}s_b\left(i\frac{Q}{2}\pm(Z_j-Z_k)\right)} \,,
	\,\,\, 
	A_N(\vec{Z};m_A)=\prod_{j,k=1}^N\sbfunc{-i\frac{Q}{2}+2m_A+(Z_j-Z_k)}\,.
\end{equation}
We also introduced $m_A$, namely the mass for the axial symmetry $U(1)_A$, which is the commutant of $U(1)_R$ in the $\mathcal{N}=4$ R-symmetry $SU(2) \times SU(2)$.

In the paper we draw the quivers in the 3d $\mathcal{N}=4$ notation. To translate between such a language and the 3d $\mathcal{N}=2^*$ partition function we employ for the dualization algorithm, see the dictionary in Table~\ref{tab:dictionary}.
\begin{table}[!ht]
\centering
\input{Tables/dictionary}
\caption{The dictionary for the quiver figures drawn in this paper. On the left the quiver elements are depicted in the 3d $\mathcal{N}=4$ language: hence we have single lines corresponding to hypermultiplets (with axial charge which can be twisted or not). On the right the quiver elements are depicted in the 3d $\mathcal{N}=2$ language: hence we have pairs of arrows corresponding to the chiral multiplets inside the $\mathcal{N}=4$ hypermultiplets. Moreover, the gauge nodes on the right have the $\mathcal{N}=2$ adjoint chiral contained in the $\mathcal{N}=4$ vector on the left. However, when a CS level (written in blue) is turned on, the corresponding gauge node has no adjoint chiral as it is integrated out. Close to the edges in the 3d $\mathcal{N}=2$ quivers the corresponding $S^3_b$ partition function charges has been written, with $m_A$ the axial mass and $Q=b+b^{-1}$ (see Section~\ref{subsec:Crhosigma} for more details on the conventions). The $\mathcal{N}=2$ adjoint chirals have the suitable charge to cubically couple to the adjacent hypermultiplets.}
\label{tab:dictionary}
\end{table} 
\\

\clearpage
\section{Ingredients for the dualization algorithm} 
\label{app:dualization_algorithm_ingredients}
In this appendix we briefly review the already known ingredients for the $SL(2,\mathbb{Z})$ dualization algorithm, providing the definitions of the $SL(2,\mathbb{Z})$ operators, the QFT blocks and the basic duality moves transforming them.

\subsection{\texorpdfstring{$SL(2,\mathbb{Z})$ operators}{}}
\label{subsec:operators}
We start by providing the partition function expressions of the $SL(2,\mathbb{Z})$ operators $\mathcal{S}$ and $\mathcal{T}$, their combination $\mathcal{T}^T$ and the Identity-wall.

\subsubsection{\texorpdfstring{The $\mathcal{S}$-wall}{}}
The $SL(2,\mathbb{Z})$ operator $\mathcal{S}$, also known as $\mathcal{S}$-wall in the field theory context, is defined as the $FT[U(N)]$ theory (see \cite{Comi:2022aqo} for the field theory details). In particular we denote its partition function as
\begin{equation}
    \mathcal{Z}_{\mathcal{S}}^{(N,N)}(\vec{X};\vec{Y};m_A) 
    = 
    \mathcal{Z}_{FT[U(N)]}(-\vec{X};\vec{Y};m_A) 
    = 
    \mathcal{Z}_{FT[U(N)]}(\vec{X};-\vec{Y};m_A) \,,
\end{equation}
where in the parenthesis we explicitly wrote the Cartan vectors $\vec{X}$ and $\vec{Y}$ of the emergent global symmetries $SU(N)_X \times SU(N)_Y$. We also wrote the mass $m_A$ for the axial symmetry $U(1)_A$.
Moreover we have that 
\begin{equation}
    \mathcal{Z}_{\mathcal{S}^{-1}}^{(N,N)}(\vec{X};\vec{Y};m_A) 
    = \mathcal{Z}_{\mathcal{S}}^{(N,N)}(-\vec{X};\vec{Y};m_A) 
    = \mathcal{Z}_{\mathcal{S}}^{(N,N)}(\vec{X};-\vec{Y};m_A) \,.
\end{equation}
The $\mathcal{S}$-wall is graphically represented in Figure~\ref{fig:S_operator}. There we also depicted the asymmetric $\mathcal{S}$-wall, where one of the $U(N)$ symmetries is broken to $U(M)\times U(1)$ with $M < N$. In the partition function notation we write
\begin{equation}
    \mathcal{Z}_{\mathcal{S}}^{(N,M<N)}(\vec{X};\vec{Y};V;m_A) 
    =
    \mathcal{Z}_{\mathcal{S}}^{(N,N)}(\vec{X};\vec{Y};m_A)\bigg|_{\subalign{&Y_{M+j}=\frac{N-M+1-2j}{2}(iQ-2m_A)+V \\ &\scalebox{.7}{$\footnotesize{(j=1,\dots,N-M)}$} }}
     \,,
\end{equation}
where the missing $N-M$ Cartans $Y_{M+j}$ (for $j=1,\dots,N-M$) are provided by a specialization of the parameter $V$. 
\begin{figure}[!ht]
    \centering
    \includegraphics[width=.7\textwidth]{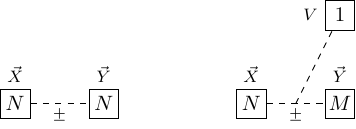}
    \caption{The symmetric $\mathcal{S}$-wall on the left and its asymmetric version on the right. The $\pm$ sign close to the dashed line indicates whether we have $\mathcal{S}$ or $\mathcal{S}^{-1}$.}
    \label{fig:S_operator}
\end{figure}

\noindent
Finally, by using the $FT[U(N)]$ property
\begin{equation}
    \mathcal{Z}_{FT[U(N)]}(\vec{X};\vec{Y};m_A) 
    e^{2\pi i B \sum_{j=1}^N X_j}
    =
    \mathcal{Z}_{FT[U(N)]}(\vec{X};\vec{Y}+B;m_A) \,,
\end{equation}
we can define the shifted $\mathcal{S}$-wall as
\begin{equation}
    \mathcal{Z}_{\mathcal{S}^{\pm1}}^{(N,N)}(\vec{X};\vec{Y}+ B;m_A) 
    = 
    \mathcal{Z}_{\mathcal{S}^{\pm1}}^{(N,N)}(\vec{X};\vec{Y};m_A) 
    e^{\mp 2\pi i B \sum_{j=1}^N X_j}
    \,.
    \label{eq:shifted_Swall}
\end{equation}

\subsubsection{\texorpdfstring{The Identity-wall}{}}
It is possible to glue together two $\mathcal{S}$-walls, where gluing means to consider two $U(N)$ flavour nodes and gauge a diagonal combination of them, with in addition an adjoint chiral field which couples in the superpotential with the moment map operators of the glued blocks.
When we perform such a gluing between two $\mathcal{S}$-walls of opposite sign, as a result we get what we call an Identity-wall, which identifies the non-gauged emergent symmetries of the two starting $\mathcal{S}$-walls (again, see \cite{Comi:2022aqo} for the field theory details). 
This is consistent with the $\mathcal{S}^2=-1$ (together with $\mathcal{S}\mathcal{S}^{-1}=1$) property of the $\mathcal{S}$ generator of the $SL(2,\mathbb{Z})$ group.
In the partition function notation this reads
\begin{align}
    {}_{\vec{X}}\hat{\mathbb{I}}_{\vec{Y}} (m_A)
    & =
    \int\udl{\vec{Z}_N}\Delta_N(\vec{Z};m_A) 
    \mathcal{Z}_{\mathcal{S}^{\pm1}}^{(N,N)}(\vec{X};\vec{Z};m_A)
    \mathcal{Z}_{\mathcal{S}^{\mp1}}^{(N,N)}(\vec{Z};\vec{Y};m_A)   
    \nonumber\\
    & = \frac{1}{\Delta_N(\vec{X};m_A)} \sum_{\sigma\in S_N}\prod_{j=1}^{N}\delta\left(X_j-Y_{\sigma(j)}\right) \,,
    \label{eq:def_Id_wall}
\end{align}
where $S_N$ is the $N$-elements permutations group and $\Delta_N(\vec{X},m_A)$ is the $\mathcal{N}=4$ vector contribution defined in Appendix~\ref{app:conventions_PF}.
In Figure~\ref{fig:Id_wall} we graphically represented this Identity-wall together with its asymmetric version, which comes from the gluing of a symmetric $\mathcal{S}$-wall and an asymmetric one.
In particular, considering $N>M$, it is defined as
\begin{align}
    {}_{\vec{X}}\hat{\mathbb{I}}_{\vec{Y},V} (m_A)
    & =
    \int\udl{\vec{Z}_N}\Delta_N(\vec{Z};m_A)
    \mathcal{Z}_{\mathcal{S}^{\pm1}}^{(N,N)}(\vec{X};\vec{Z};m_A)
    \mathcal{Z}_{\mathcal{S}^{\mp1}}^{(N,M)}(\vec{Z};\vec{Y};V;m_A) 
    \nonumber\\
    & = \frac{1}{\Delta_N(\vec{X};m_A)} \sum_{\sigma\in S_N}\prod_{j=1}^{N}\delta\left(X_j-Y_{\sigma(j)}\right) \bigg|_{\subalign{&Y_{M+k}=\frac{N-M+1-2k}{2}(iQ-2m_A)+V \\ &\scalebox{.7}{$\footnotesize{(k=1,\dots,N-M)}$} }}
    \,,
\end{align}
where $M$ of the $\vec{X}$ Cartans are identified with the $\vec{Y}$ Cartans, while the remaining $N-M$ Cartans $\vec{X}$ are identified with a specialization of the parameter $V$. 
\begin{figure}[!ht]
    \centering
    \includegraphics[width=.8\textwidth]{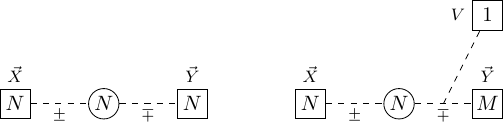}
    \caption{The symmetric Identity-wall on the left and its asymmetric version on the right.}
    \label{fig:Id_wall}
\end{figure}

\noindent
Finally, using the definition \eqref{eq:shifted_Swall} of the shifted $\mathcal{S}$-wall, we can define the shifted Identity-wall as follows:
\begin{align}
    {}_{\vec{X}}\hat{\mathbb{I}}_{\vec{Y}+B} (m_A)
    &=
    \int\udl{\vec{Z}_N}\Delta_N(\vec{Z};m_A) 
    \mathcal{Z}_{\mathcal{S}^{\pm1}}^{(N,N)}(\vec{X};\vec{Z};m_A)
    \mathcal{Z}_{\mathcal{S}^{\mp1}}^{(N,N)}(\vec{Z};\vec{Y}+B;m_A)
    \label{eq:shifted_Id_wall} \\
    & =
    \int\udl{\vec{Z}_N}\Delta_N(\vec{Z};m_A) 
    \mathcal{Z}_{\mathcal{S}^{\pm1}}^{(N,N)}(\vec{X};\vec{Z};m_A)
    e^{\pm 2\pi i B \sum_{j=1}^N Z_j}
    \mathcal{Z}_{\mathcal{S}^{\mp1}}^{(N,N)}(\vec{Z};\vec{Y};m_A) 
    \nonumber\\
    & =
    \int\udl{\vec{Z}_N}\Delta_N(\vec{Z};m_A) 
    \mathcal{Z}_{\mathcal{S}^{\pm1}}^{(N,N)}(\vec{X}-B;\vec{Z};m_A)
    \mathcal{Z}_{\mathcal{S}^{\mp1}}^{(N,N)}(\vec{Z};\vec{Y};m_A) 
    \nonumber\\
    & =
    {}_{\vec{X}-B}\hat{\mathbb{I}}_{\vec{Y}} (m_A)
    \nonumber\,.
\end{align}

\subsubsection{\texorpdfstring{The $\mathcal{T}$-wall}{}}
The $SL(2,\mathbb{Z})$ operator $\mathcal{T}$ is associated to the insertion of a CS level $+1$ (times some contact terms). Its partition function is
\begin{equation}
    \mathcal{Z}_{\mathcal{T}}^{(N,N)}(\vec{X};\vec{Y};m_A) 
    =
    e^{-i \pi \sum_{j=1}^N X_j^2}
    {}_{\vec{X}}\hat{\mathbb{I}}_{\vec{Y}} (m_A) \times
    e^{\frac{i \pi N}{24} \left( 8 m_A^2 (N-1) - 4 i m_A Q (N-1) + Q^2 \right)} 
    \,,
    \label{eq:T_operator_PF}
\end{equation}
while its graphical representation is provided in Figure~\ref{fig:T_operator}.
Obviously, one can build an asymmetric $\mathcal{T}$ operator by using an asymmetric Identity-wall in its definition.
Finally, to write its inverse we use the relation $(\mathcal{S}\mathcal{T})^3 = \bbone$ and hence we write $\mathcal{T}^{-1}=\mathcal{S}\mathcal{T}\mathcal{S}\mathcal{T}\mathcal{S}$.
\begin{figure}[!ht]
    \centering
    \includegraphics[width=.4\textwidth]{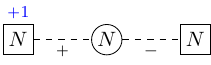}
    \caption{The $\mathcal{T}$-wall.}
    \label{fig:T_operator}
\end{figure}

\subsubsection{\texorpdfstring{The $\mathcal{T}^T$-wall}{}}
The $SL(2,\mathbb{Z})$ operator $\mathcal{T}^T$ is defined in terms of $\mathcal{S}$ and $\mathcal{T}$ as 
\begin{align}
    \mathcal{T}^{T} = \mathcal{T}\mathcal{S}^{-1}\mathcal{T} = \mathcal{S}^{\pm1}\mathcal{T}^{-1}\mathcal{S}^{\mp1} \,,
\end{align}
where the property $(\mathcal{S}\mathcal{T})^3 = \bbone$ has been used.
Its graphical representation is provided in Figure~\ref{fig:Tt_operator}, while its partition function is
\begin{align}
    \mathcal{Z}_{\mathcal{T}^T}^{(N,N)}(\vec{X};\vec{Y};m_A) 
    &=
    \int\udl{\vec{Z}_N}\Delta_N(\vec{Z};m_A) 
    \int\udl{\vec{W}_N}\Delta_N(\vec{W};m_A) 
    \times\label{eq:Tt_operator_PF}\\
    &\qquad\times
    \mathcal{Z}_{\mathcal{T}}^{(N,N)}(\vec{X};\vec{Z};m_A)
    \mathcal{Z}_{\mathcal{S}^{-1}}^{(N,N)}(\vec{Z};\vec{W};m_A)
    \mathcal{Z}_{\mathcal{T}}^{(N,N)}(\vec{W};\vec{Y};m_A) 
    \nonumber\,.
\end{align}
Obviously, one can build an asymmetric $\mathcal{T}^T$ operator by plugging an asymmetric $\mathcal{S}$-wall in its definition.
\begin{figure}[!ht]
    \centering
    \includegraphics[width=.75\textwidth]{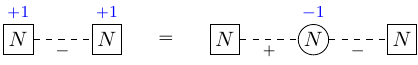}
    \caption{The $\mathcal{T}^T$-wall. It has been represented as $\mathcal{T}\mathcal{S}^{-1}\mathcal{T}$ on the left and as $\mathcal{S}^{\pm1}\mathcal{T}^{-1}\mathcal{S}^{\mp1}$ on the right.}
    \label{fig:Tt_operator}
\end{figure}

\subsection{QFT blocks}
In terms of fields, the content of our theories can be broken down into elementary pieces which we call fundamental QFT blocks. In particular, we define the following.
\begin{itemize}
    \item 
    The $\mathcal{B}_{(\alpha,0)}$ block (where $\alpha=\pm1$) is a hypermultiplet in the bifundamental representation of $U(N)_X \times U(M)_Y$ with FI $\alpha V$ on $U(N)_X$ and $-\alpha V$ on $U(M)_Y$. It is represented as the first quiver of Figure~\ref{fig:blocks} and its partition function expression is
    \begin{equation}
        \mathcal{Z}_{\mathcal{B}_{(\alpha,0)}}^{(N,M)}(\vec{X};\vec{Y};V;m_A) 
        =
        \prod_{j=1}^N\prod_{k=1}^M \sbfunc{\frac{i Q}{2}-m_A\pm(X_j-Y_k)}
        \times
        \mathrm{e}^{\alpha2\pi iV\left(\sum_{j=1}^NX_j-\sum_{k=1}^MY_k\right)} 
        \,.
    \end{equation}
    \item 
    The $\mathcal{B}_{(0,\alpha)}$ block (where $\alpha=\pm1$) is a hypermultiplet in the bifundamental representation of $U(M)_Y \times U(1)_{\alpha V}$, together with an Identity-wall glued on its side. It is represented as the second quiver of Figure~\ref{fig:blocks} and its partition function expression is
    \begin{equation}
        \mathcal{Z}_{\mathcal{B}_{(0,\alpha)}}^{(N,M)} (\vec{X};\vec{Y};V;\ell) 
        = \prod_{j=1}^M \sbfunc{\frac{iQ}{2}-\ell(N-M+1)\pm(Y_j-\alpha V)}{}^{\phantom{}}_{\vec X}\hat{\mathbb{I}}_{\vec Y,\alpha V}(m_A)\,,
    \end{equation}
    where $\ell$ can be either $m_A$ or $\frac{i Q}{2}-m_A$.
    On the l.h.s.~we do not list the argument $m_A$ of the 3d Identity-wall, as throughout the paper it will always be $m_A$.
    \item 
    The $\mathcal{B}_{(\alpha,q)}$ block (where $\alpha=\pm1$) is a $\mathcal{B}_{(\alpha,0)}$ block with the insertion of a CS level $\alpha q$ on the first node and $-\alpha q$ on the second.
    It is represented as the third quiver of Figure~\ref{fig:blocks} and its partition function expression is
    \begin{align}
        \mathcal{Z}_{\mathcal{B}_{(\alpha,q)}}^{(N,M)}(\vec{X};\vec{Y};V;m_A) &=e^{-\alpha \pi i q\left(\sum_{j=1}^NX_j^2-\sum_{k=1}^MY_k^2\right)}e^{\alpha2\pi iV\left(\sum_{j=1}^NX_j-\sum_{k=1}^MY_k\right)} \nonumber\\
        &\quad\times\prod_{j=1}^N\prod_{k=1}^M \sbfunc{\tfrac{iQ}{2}-m_A\pm(X_j-Y_k)}\,.
        \label{eq:B1q_def}
    \end{align}
\end{itemize}

\begin{figure}[!ht]
    \centering
    \includegraphics[width=\textwidth]{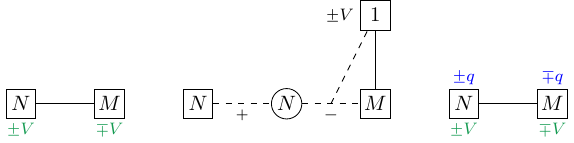}
    \caption{The basic QFT blocks $\mathcal{B}_{(\pm1,0)}$, $\mathcal{B}_{(0,\pm1)}$ and $\mathcal{B}_{(\pm1,q)}$ (from left to right).}
    \label{fig:blocks}
\end{figure}

\subsection{The basic duality moves}
\label{subsec:basic_moves}
With the ingredients introduced above we can write the basic duality moves for the $SL(2,\mathbb{Z})$ generators we are interested in.

\subsubsection{\texorpdfstring{$\mathcal{S}$-dualization}{}}
\label{subsubsec:Sdual_moves}
The basic $\mathcal{S}$-moves, represented in Figure~\ref{fig:S_basic_moves}, can be schematically written as
\begin{align}
    \mathcal{B}_{(-1,0)} &= \mathcal{S}\,\mathcal{B}_{(0,+1)}\,\mathcal{S}^{-1} \,,\label{eq:basic_duality_move_S_B10}\\
    \mathcal{B}_{(0,+1)} &= \mathcal{S}\,\mathcal{B}_{(+1,0)}\,\mathcal{S}^{-1} \,,\label{eq:basic_duality_move_S_B01}\\
    \mathcal{B}_{(1,+1)} &= \mathcal{S}\,\mathcal{B}_{(1,-1)}\,\mathcal{S}^{-1} \,.\label{eq:basic_duality_move_S_B11}
\end{align}

\begin{figure}[!ht]
    \centering
    \includegraphics[width=\textwidth]{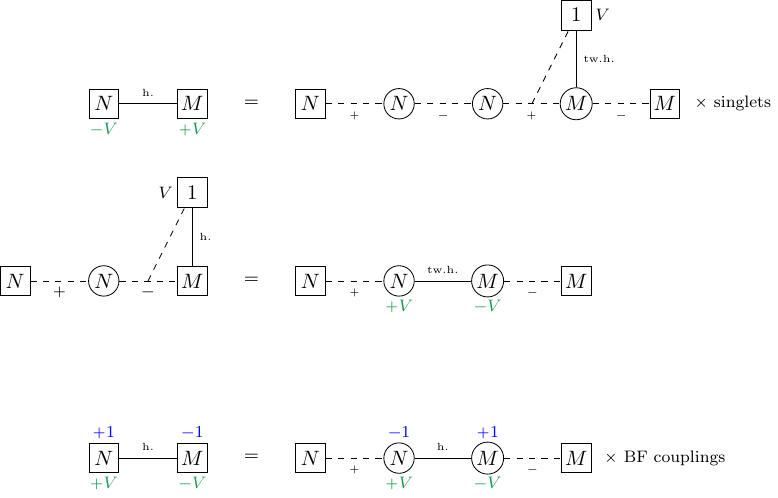}
    \caption{The basic $\mathcal{S}$-duality moves.}
    \label{fig:S_basic_moves}
\end{figure}

\noindent
The precise partition function statements are
\begin{align}
    &
    \mathcal{Z}_{\mathcal{B}_{(-1,0)}}^{(N,M)}(\vec{X};\vec{Y};V;m_A)
    = \\
    &\qquad =
    \prod_{j=1}^{N-M} s_b\left(i\frac{Q}{2}-j(iQ-2m_A) \right)
    \int\udl{\vec{Z}_N}\Delta_N(\vec{Z};m_A) 
    \int\udl{\vec{W}_M}\Delta_M(\vec{W};m_A) 
    \times\nonumber\\
    &\qquad\qquad\times
    \mathcal{Z}_{\mathcal{S}}^{(N,N)}(\vec{X};\vec{Z};m_A)
    \mathcal{Z}_{\mathcal{B}_{(0,+1)}}^{(N,M)} (\vec{Z};\vec{W};V;iQ/2-m_A)
    \mathcal{Z}_{\mathcal{S}^{-1}}^{(M,M)}(\vec{W};\vec{Y};m_A) 
    \nonumber\,,\\[5pt]
    &
    \mathcal{Z}_{\mathcal{B}_{(0,+1)}}^{(N,N)}(\vec{X};\vec{Y};V;m_A)
    = \\
    &\qquad =
    \int\udl{\vec{Z}_N}\Delta_N(\vec{Z}) 
    \int\udl{\vec{W}_N}\Delta_N(\vec{W}) 
    \times\nonumber\\
    &\qquad\qquad\times
    \mathcal{Z}_{\mathcal{S}}^{(N,N)}(\vec{X};\vec{Z};m_A)
    \mathcal{Z}_{\mathcal{B}_{(+1,0)}}^{(N,N)}(\vec{Z};\vec{W};V;iQ/2-m_A)
    \mathcal{Z}_{\mathcal{S}^{-1}}^{(N,N)}(\vec{W};\vec{Y};m_A) 
    \nonumber\,,\\[5pt]
    &
    \mathcal{Z}_{\mathcal{B}_{(1,+1)}}^{(N,M)}(\vec{X};\vec{Y};V;m_A)
    = \\
    &\qquad =
    e^{i\pi(N-M)V^2+i\pi\Theta}
    \int\udl{\vec{Z}_N}\Delta_N(\vec{Z};m_A) 
    \int\udl{\vec{W}_M}\Delta_M(\vec{W};m_A) 
    \times\nonumber\\
    &\qquad\qquad\times
    \mathcal{Z}_{\mathcal{S}}^{(N,N)}(\vec{X};\vec{Z};m_A)
    \mathcal{Z}_{\mathcal{B}_{(1,-1)}}^{(N,M)}(\vec{Z};\vec{W};V;m_A)
    \mathcal{Z}_{\mathcal{S}^{-1}}^{(M,M)}(\vec{W};\vec{Y};m_A) 
    \nonumber\,,
\end{align}
where $\Theta$ is a phase (non-vanishing only for $N \neq M$) which can be found in \cite{Comi:2022aqo}, together with the proofs of these identities.

\subsubsection{\texorpdfstring{$\mathcal{T}$-dualization}{}}
The basic $\mathcal{T}$-moves can be schematically written as
\begin{align}
    \mathcal{B}_{(+1,0)} &= \mathcal{T}\,\mathcal{B}_{(1,-1)}\,\mathcal{T}^{-1} \,,\\
    \mathcal{B}_{(0,+1)} &= \mathcal{T}\,\mathcal{B}_{(0,+1)}\,\mathcal{T}^{-1} \,,\\
    \mathcal{B}_{(1,+1)} &= \mathcal{T}\,\mathcal{B}_{(+1,0)}\,\mathcal{T}^{-1} \,,
\end{align}
where recall that $\mathcal{T}^{-1}=\mathcal{S}\mathcal{T}\mathcal{S}\mathcal{T}\mathcal{S}$.
The corresponding partition function statements (proved in \cite{Comi:2022aqo}) are
\begin{align}
    &
    \mathcal{Z}_{\mathcal{B}_{(1,0)}}^{(N,M)} (\vec{X};\vec{Y};V;m_A)
    =\\
    &\qquad=
    \int
    \udl{\vec{Z}^{(1)}_N}\Delta_N(\vec{Z}^{(1)};m_A)
    \left(\prod_{k=2}^6
    \udl{\vec{Z}^{(k)}_M}\Delta_M(\vec{Z}^{(k)};m_A)
    \right)
    \mathcal{Z}_{\mathcal{T}}^{(N,N)}(\vec{X};\vec{Z}^{(1)};m_A)
    \nn\\
    &\qquad\qquad\times
    \mathcal{Z}_{\mathcal{B}_{(1,-1)}}^{(N,M)}(\vec{Z}^{(1)};\vec{Z}^{(2)};V;m_A)
    \mathcal{Z}_{\mathcal{S}}^{(M,M)}(\vec{Z}^{(2)};\vec{Z}^{(3)};m_A)
    \mathcal{Z}_{\mathcal{T}}^{(M,M)}(\vec{Z}^{(3)};\vec{Z}^{(4)};m_A)
    \nn\\
    &\qquad\qquad\times
    \mathcal{Z}_{\mathcal{S}}^{(M,M)}(\vec{Z}^{(4)};\vec{Z}^{(5)};m_A)
    \mathcal{Z}_{\mathcal{T}}^{(M,M)}(\vec{Z}^{(5)};\vec{Z}^{(6)};m_A)
    \mathcal{Z}_{\mathcal{S}}^{(M,M)}(\vec{Z}^{(6)};\vec{Y};m_A)
    \,,\nn\\[10pt]
    &
    \mathcal{Z}_{\mathcal{B}_{(0,1)}}^{(N,N)} (\vec{X};\vec{Y};V;m_A)
    =\\
    &\qquad=
    \int\left(\prod_{k=1}^6
    \udl{\vec{Z}^{(k)}_N}\Delta_N(\vec{Z}^{(k)};m_A)
    \right)
    \mathcal{Z}_{\mathcal{T}}^{(N,N)}(\vec{X};\vec{Z}^{(1)};m_A)
    \nn\\
    &\qquad\qquad\times
    \mathcal{Z}_{\mathcal{B}_{(0,1)}}^{(N,N)}(\vec{Z}^{(1)};\vec{Z}^{(2)};V;m_A)
    \mathcal{Z}_{\mathcal{S}}^{(N,N)}(\vec{Z}^{(2)};\vec{Z}^{(3)};m_A)
    \mathcal{Z}_{\mathcal{T}}^{(N,N)}(\vec{Z}^{(3)};\vec{Z}^{(4)};m_A)
    \nn\\
    &\qquad\qquad\times
    \mathcal{Z}_{\mathcal{S}}^{(N,N)}(\vec{Z}^{(4)};\vec{Z}^{(5)};m_A)
    \mathcal{Z}_{\mathcal{T}}^{(N,N)}(\vec{Z}^{(5)};\vec{Z}^{(6)};m_A)
    \mathcal{Z}_{\mathcal{S}}^{(N,N)}(\vec{Z}^{(6)};\vec{Y};m_A)
    \,,\nn\\[10pt]
    &
    \mathcal{Z}_{\mathcal{B}_{(1,1)}}^{(N,M)}(\vec{X};\vec{Y};V;m_A)
    =\\
    &\qquad=
    \int\udl{\vec{Z}^{(1)}_N}
    \Delta_N(\vec{Z}^{(1)};m_A)
    \left(\prod_{k=2}^6\udl{\vec{Z}^{(k)}_M}
    \Delta_M(\vec{Z}^{(k)};m_A)\right)
    \mathcal{Z}_{\mathcal{T}}^{(N,N)}(\vec{X};\vec{Z}^{(1)};m_A)
    \nn\\
    &\qquad\qquad\times
    \mathcal{Z}_{\mathcal{B}_{(1,0)}}^{(N,M)}(\vec{Z}^{(1)};\vec{Z}^{(2)};V;m_A)
    \mathcal{Z}_{\mathcal{S}}^{(M,M)}(\vec{Z}^{(2)};\vec{Z}^{(3)};m_A)
    \mathcal{Z}_{\mathcal{T}}^{(M,M)}(\vec{Z}^{(3)};\vec{Z}^{(4)};m_A)
    \nn\\
    &\qquad\qquad\times
    \mathcal{Z}_{\mathcal{S}}^{(M,M)}(\vec{Z}^{(4)};\vec{Z}^{(5)};m_A)
    \mathcal{Z}_{\mathcal{T}}^{(M,M)}(\vec{Z}^{(5)};\vec{Z}^{(6)};m_A)
    \mathcal{Z}_{\mathcal{S}}^{(M,M)}(\vec{Z}^{(6)};\vec{Y};m_A)
    \,.\nn
\end{align}

\subsubsection{\texorpdfstring{$\mathcal{T}^T$-dualization}{}}
The basic $\mathcal{T}^T$-moves can be schematically written as
\begin{align}
    \mathcal{B}_{(+1,0)} &= \mathcal{T}^T\,\mathcal{B}_{(+1,0)}\,(\mathcal{T}^T)^{-1} \,,\\
    \mathcal{B}_{(0,-1)} &= \mathcal{T}^T\,\mathcal{B}_{(1,-1)}\,(\mathcal{T}^T)^{-1} \,,\\
    \mathcal{B}_{(1,+1)} &= \mathcal{T}^T\,\mathcal{B}_{(0,+1)}\,(\mathcal{T}^T)^{-1} \,.
\end{align}
The corresponding partition function statements (proved in \cite{Comi:2022aqo}) are
\begin{align}
    &
    \mathcal{Z}_{\mathcal{B}_{(1,1)}}^{(N,M)} (\vec{X};\vec{Y};V;m_A)
    =\nn\\
    &\qquad=
    e^{i\pi(N-M)V^2-i\pi\Theta}
    \prod_{j=1}^{N-M} s_b\left(iQ/2-j(iQ-2m_A)\right)
    \nn\\
    &\qquad\quad\times
    \int\left(\prod_{k=1}^3
    \udl{\vec{Z}^{(k)}_N}\Delta_N(\vec{Z}^{(k)};m_A)
    \right)
    \left(\prod_{k=4}^6
    \udl{\vec{Z}^{(k)}_M}\Delta_M(\vec{Z}^{(k)};m_A)
    \right)
    \mathcal{Z}_{\mathcal{T}}^{(N,N)}(-\vec{X};\vec{Z}^{(1)};m_A)
    \nn\\
    &\qquad\qquad\times 
    \mathcal{Z}_{\mathcal{S}}^{(N,N)}(\vec{Z}^{(1)};\vec{Z}^{(2)};m_A)
    \mathcal{Z}_{\mathcal{T}}^{(N,N)}(\vec{Z}^{(2)};\vec{Z}^{(3)};m_A)
    \mathcal{Z}_{\mathcal{B}_{(0,1)}}^{(N,M)}(\vec{Z}^{(3)};\vec{Z}^{(4)};V;iQ/2-m_A)
    \nn\\
    &\qquad\qquad\times
    \mathcal{Z}_{\mathcal{S}}^{(M,M)}(\vec{Z}^{(4)};\vec{Z}^{(5)};m_A)
    \mathcal{Z}_{\mathcal{T}}^{(M,M)}(\vec{Z}^{(5)};\vec{Z}^{(6)};m_A)
    \mathcal{Z}_{\mathcal{S}}^{(M,M)}(\vec{Z}^{(6)};-\vec{Y};m_A)
    \,,\nn\\ \\
    &
    \mathcal{Z}_{\mathcal{B}_{(1,0)}}^{(N,M)}(\vec{X};\vec{Y};V;m_A)
    =\nn\\
    &\qquad=
    e^{-i\pi(N-M)V^2-i\pi\Theta'}
    \int\left(\prod_{k=1}^3
    \udl{\vec{Z}^{(k)}_N}\Delta_N(\vec{Z}^{(k)};m_A)
    \right)
    \left(\prod_{k=4}^6
    \udl{\vec{Z}^{(k)}_M}\Delta_M(\vec{Z}^{(k)};m_A)
    \right)
    \nn\\
    &\qquad\qquad\times
    \mathcal{Z}_{\mathcal{T}}^{(N,N)}(-\vec{X};\vec{Z}^{(1)};m_A)
    \mathcal{Z}_{\mathcal{S}}^{(N,N)}(\vec{Z}^{(1)};\vec{Z}^{(2)};m_A)
    \mathcal{Z}_{\mathcal{T}}^{(N,N)}(\vec{Z}^{(2)};\vec{Z}^{(3)};m_A)
    \nn\\
    &\qquad\qquad\times
    \mathcal{Z}_{\mathcal{B}_{(1,0)}}^{(N,M)}(\vec{Z}^{(3)};\vec{Z}^{(4)};V;m_A)
    \mathcal{Z}_{\mathcal{S}}^{(M,M)}(\vec{Z}^{(4)};\vec{Z}^{(5)};m_A)
    \mathcal{Z}_{\mathcal{T}}^{(M,M)}(\vec{Z}^{(5)};\vec{Z}^{(6)};m_A)
    \nn\\
    &\qquad\qquad\times
    \mathcal{Z}_{\mathcal{S}}^{(M,M)}(\vec{Z}^{(6)};-\vec{Y};m_A)
    \,,\nn\\ \\
    &
    \mathcal{Z}_{\mathcal{B}_{(0,-1)}}^{(N,N)}(\vec{X};\vec{Y};V;m_A)
    =\nn\\
    &\qquad=
    \int\left(\prod_{k=1,2,5,6}
    \udl{\vec{Z}^{(k)}_N}\Delta_N(\vec{Z}^{(k)};m_A)
    \right)
    \left(\prod_{k=3,4}
    \udl{\vec{Z}^{(k)}_N}\Delta_N(\vec{Z}^{(k)})
    \right)
    \nn\\
    &\qquad\qquad\times 
    \mathcal{Z}_{\mathcal{T}}^{(N,N)}(-\vec{X};\vec{Z}^{(1)};m_A)
    \mathcal{Z}_{\mathcal{S}}^{(N,N)}(\vec{Z}^{(1)};\vec{Z}^{(2)};m_A)
    \nn\\
    &\qquad\qquad\times 
    \mathcal{Z}_{\mathcal{T}}^{(N,N)}(\vec{Z}^{(2)};\vec{Z}^{(3)};m_A)
	\mathcal{Z}_{\mathcal{B}_{(1,-1)}}^{(N,N)}(\vec{Z}^{(3)};\vec{Z}^{(4)};V;iQ/2-m_A) 
    \mathcal{Z}_{\mathcal{S}}^{(N,N)}(\vec{Z}^{(4)};\vec{Z}^{(5)};m_A)
    \nn\\
    &\qquad\qquad\times 
	\mathcal{Z}_{\mathcal{T}}^{(N,N)}(\vec{Z}^{(5)};\vec{Z}^{(6)};m_A)
	\mathcal{Z}_{\mathcal{S}}^{(N,N)}(\vec{Z}^{(6)};-\vec{Y};m_A)
    \,,
\end{align}
where $\Theta$ and $\Theta^\prime$ are phases (non-vanishing only for $N \neq M$) which can be found in \cite{Comi:2022aqo}, together with the proofs of these identities.

\subsubsection{The Hanany--Witten move}
The field theoretic realization of the Hanany--Witten move \cite{Hanany:1996ie}, represented in field theory in Figure~\ref{fig:HW_move}, can be schematically written as
\begin{align}
    \mathcal{B}_{(1,0)}\mathcal{B}_{(0,1)}\rightarrow\mathcal{B}_{(0,1)}\mathcal{B}_{(1,0)}  \,.
\end{align}
The precise partition function statement (proved in \cite{Comi:2022aqo}) is
\begin{align}
    &\int\udl{\vec{Z}_M}\Delta_M(\vec{Z})\,
    \mathcal{Z}_{\mathcal{B}_{(1,0)}}^{(N,M)}(\vec{X};\vec{Z};U;iQ/2-m_A)
    \mathcal{Z}_{\mathcal{B}_{(0,1)}}^{(M,L)}(\vec{Z};\vec{Y};V;iQ/2-m_A)
    \\
    &\quad 
    =
    e^{-2i\pi UV} s_b\left(-iQ/2+(N-\widetilde{M}+1)(iQ-2m_A) \right) 
    A_N(\vec{X};iQ-2m_A)
    A_L(\vec{Y};2m_A)
    \nonumber\\
    &\quad\quad\times 
    \int\udl{\vec{Z}_{\widetilde{M}}}\Delta_{\widetilde{M}}(\vec{Z}) \,
    \mathcal{Z}_{\mathcal{B}_{(0,1)}}^{(N,\widetilde{M})}(\vec{X};\vec{Z};V;iQ/2-m_A) \,
    \mathcal{Z}_{\mathcal{B}_{(1,0)}}^{(\widetilde{M},L)}(\vec{Z};\vec{Y};U;iQ/2-m_A)\nonumber\,,
\end{align}
where $\widetilde{M}=N+L-M+1$ and where the $\mathcal{N}=2$ adjoint contributions $A_{\text{rank}}(\text{Cartan};\text{charge})$ are defined in Appendix~\ref{app:conventions_PF}.

\begin{figure}[!ht]
    \centering
    \includegraphics[width=\textwidth]{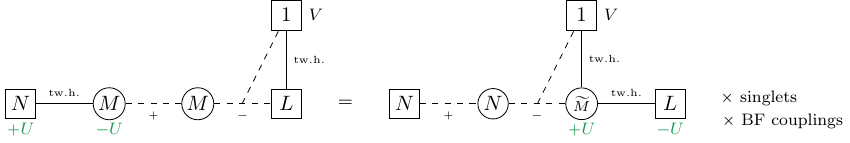}
    \caption{The Hanany--Witten move realized in field theory.}
    \label{fig:HW_move}
\end{figure}

\subsection{\texorpdfstring{Proofs of the $\mathcal{S}$-duality moves swapping $\mathcal{B}_{\text{bar}}$ and $\mathcal{B}_{\text{top}}$}{}}
\label{app:Sdual_Bbar_Btop_proofs}
In this final section we prove the $\mathcal{S}$-duality moves swapping $\mathcal{B}_{\text{bar}}$ and $\mathcal{B}_{\text{top}}$, which have been introduced in Section~\ref{subsec:new_ingredients}.\\

The proof of \eqref{eq:Bbar_Sdualized} goes as follows: 
\begin{align}
    &
    \mathcal{Z}_{\mathcal{B}_{\text{bar}}}^{(N,N)}(\vec{X};\vec{Y};m_A;B)
    = \\
    & \quad =
    {}_{\vec{X}}\hat{\mathbb{I}}_{\vec{Y}+B} (m_A) \times e^{-\pi i N B^2}
    =
    {}_{\vec{X}-B}\hat{\mathbb{I}}_{\vec{Y}} (m_A) \times e^{-\pi i N B^2}
    \nonumber\\
    & \quad =
    \int\udl{\vec{Z}_N}\Delta_N(\vec{Z};m_A) 
    \mathcal{Z}_{\mathcal{S}}^{(N,N)}(\vec{X}-B;\vec{Z};m_A)
    \mathcal{Z}_{\mathcal{S}^{-1}}^{(N,N)}(\vec{Z};\vec{Y};m_A)
    \times e^{-\pi i N B^2}
    \nonumber\\
    & \quad =
    \int\udl{\vec{Z}_N}\Delta_N(\vec{Z};m_A) 
    \int\udl{\vec{W}_N}\Delta_N(\vec{W};m_A) 
    \times\nonumber\\
    &\qquad\qquad\times
    \mathcal{Z}_{\mathcal{S}}^{(N,N)}(\vec{X}-B;\vec{Z};m_A)
    {}_{\vec{Z}}\hat{\mathbb{I}}_{\vec{W}} (m_A)
    \mathcal{Z}_{\mathcal{S}^{-1}}^{(N,N)}(\vec{W};\vec{Y};m_A)
    \times e^{-\pi i N B^2}
    \nonumber\\
    & \quad =
    \int\udl{\vec{Z}_N}\Delta_N(\vec{Z};m_A) 
    \int\udl{\vec{W}_N}\Delta_N(\vec{W};m_A) 
    \times\nonumber\\
    &\qquad\qquad\times
    \mathcal{Z}_{\mathcal{S}}^{(N,N)}(\vec{X};\vec{Z};m_A)
    \times e^{2\pi i B \sum_{j=1}^{N} Z_j} 
    \times e^{-\pi i B^2} \times
    {}_{\vec{Z}}\hat{\mathbb{I}}_{\vec{W}} (m_A) \times
    \mathcal{Z}_{\mathcal{S}^{-1}}^{(N,N)}(\vec{W};\vec{Y};m_A)
    \nonumber\\
    & \quad =
    \int\udl{\vec{Z}_N}\Delta_N(\vec{Z};m_A) 
    \int\udl{\vec{W}_N}\Delta_N(\vec{W};m_A) 
    \times\nonumber\\
    &\qquad\qquad\times
    \mathcal{Z}_{\mathcal{S}}^{(N,N)}(\vec{X};\vec{Z};m_A)
    \mathcal{Z}_{\mathcal{B}_{\text{top}}}^{(N,N)}(\vec{Z};\vec{W};m_A;B)
    \mathcal{Z}_{\mathcal{S}^{-1}}^{(N,N)}(\vec{W};\vec{Y};m_A) 
    \nonumber\,.
\end{align}
The proof of \eqref{eq:Btop_Sdualized} goes as follows: 
\begin{align}
    &
    \mathcal{Z}_{\mathcal{B}_{\text{top}}}^{(N,N)}(\vec{X};\vec{Y};m_A;E)
    = \\
    & \quad =
    {}_{\vec{X}}\hat{\mathbb{I}}_{\vec{Y}}(m_A) \times 
    e^{2\pi i E \sum_{j=1}^{N} Y_j} e^{-\pi i N E^2}
    \nonumber\\
    & \quad =
    \int\udl{\vec{V}_N}\Delta_N(\vec{V};m_A) 
    {}_{\vec{X}}\hat{\mathbb{I}}_{\vec{V}}(m_A) \times 
    e^{2\pi i E \sum_{j=1}^{N} V_j}  \times
    {}_{\vec{V}}\hat{\mathbb{I}}_{\vec{Y}}(m_A) \times
    e^{-\pi i N E^2}
    \nonumber\\
    & \quad =
    \int\udl{\vec{Z}_N}\Delta_N(\vec{Z};m_A) 
    \int\udl{\vec{V}_N}\Delta_N(\vec{V};m_A)
    \int\udl{\vec{W}_N}\Delta_N(\vec{W};m_A)
    \times\nonumber\\
    &\qquad\qquad\times
    \mathcal{Z}_{\mathcal{S}}^{(N,N)}(\vec{X};\vec{Z};m_A)
    \mathcal{Z}_{\mathcal{S}^{-1}}^{(N,N)}(\vec{Z};\vec{V};m_A) \times 
    e^{2\pi i E \sum_{j=1}^{N} V_j}  
    \times\nonumber\\
    &\qquad\qquad\times
    \mathcal{Z}_{\mathcal{S}}^{(N,N)}(\vec{V};\vec{W};m_A)
    \mathcal{Z}_{\mathcal{S}^{-1}}^{(N,N)}(\vec{W};\vec{Y};m_A) \times
    e^{-\pi i N E^2}
    \nonumber\\
    & \quad =
    \int\udl{\vec{Z}_N}\Delta_N(\vec{Z};m_A) 
    \int\udl{\vec{W}_N}\Delta_N(\vec{W};m_A)
    \times\nonumber\\
    &\qquad\qquad\times
    \mathcal{Z}_{\mathcal{S}}^{(N,N)}(\vec{X};\vec{Z};m_A)
    {}_{\vec{Z}}\hat{\mathbb{I}}_{\vec{W}-E} (m_A) \times e^{-\pi i N E^2} \times
    \mathcal{Z}_{\mathcal{S}^{-1}}^{(N,N)}(\vec{W};\vec{Y};m_A) 
    \nonumber\\
    & \quad =
    \int\udl{\vec{Z}_N}\Delta_N(\vec{Z};m_A) 
    \int\udl{\vec{W}_N}\Delta_N(\vec{W};m_A)
    \times\nonumber\\
    &\qquad\qquad\times
    \mathcal{Z}_{\mathcal{S}}^{(N,N)}(\vec{X};\vec{Z};m_A)
    \mathcal{Z}_{\mathcal{B}_{\text{bar}}}^{(N,N)}(\vec{Z};\vec{W};m_A;-E)
    \mathcal{Z}_{\mathcal{S}^{-1}}^{(N,N)}(\vec{W};\vec{Y};m_A) 
    \nonumber\,.
\end{align}
\\

\bibliographystyle{ytphys}
\bibliography{references}

\end{document}

%% file: Tables/fields_charges.tex
\begin{tabular}{c|cc|cc}
\toprule
& \multicolumn{2}{c|}{Electric Theory} 
& \multicolumn{2}{c}{Magnetic Theory}
\\[3pt] 
\hline    
Field
& $U(1)_R$	
& $U(1)_a$
& $U(1)_R$	
& $U(1)_a$
\\
\hline     
$Q_{i,i+1}$ or $P_j$
& $1/2$
& $+1$
& $1/2$
& $-1$
\\
\hline     
$A_i$
& $1$
& $-2$
& $1$
& $+2$
\\[3pt]
\bottomrule
\end{tabular}

%% file: Tables/operators_mirror_pair_el_mesons.tex
\begin{tabular}{ccccccccc}
\toprule
Electric Operators
& Magnetic Operators
& $U(1)_R$            
& $U(1)_a$ 	               
& $SU(2)_X$ 	
& $SU(2)_{Y_{1,2}}$
& $SU(2)_{Y_{3,4}}$
& $U(1)_B$ 	      
& $SU(2)_E$
\\[3pt] 
\hline    
$\Tr[{Q}_{1,2}\widetilde{Q}_{1,2}]$
& $\Tr[A_2]$	
& $1$
& $2$
& $\cdot$
& $\cdot$
& $\cdot$
& $0$
& $\cdot$
\\[3pt]
\hline 
$\Tr[\widetilde{P}P]$
& $\mathfrak{M}^{(\{+,0\},\{0\})},\mathfrak{M}^{(\{-,0\},\{0\})},\Tr[A_1]$
& $1$
& $2$
& adj
& $\cdot$
& $\cdot$
& $0$
& $\cdot$
\\[3pt]
\hline 
$\text{Tr}\Big[Q_{1,2}Q_{2,3}Q_{3,4}Q_{4,1}\Big]$
& $\mathfrak{M}^{(\{+\},\{+,0\})}$
& $2$   
& $4$  
& $\cdot$
& $\cdot$
& $\cdot$
& $1$
& $\cdot$
\\[3pt]
\hline 
$\Tr[\widetilde{Q}_{1,2}\widetilde{Q}_{2,3}\widetilde{Q}_{3,4}\widetilde{Q}_{4,1}]$
& $\mathfrak{M}^{(\{-,0\},\{-\})}$
& $2$   
& $4$  
& $\cdot$
& $\cdot$
& $\cdot$
& $1$
& $\cdot$
\\[3pt]
\hline 
$\Tr[\widetilde{P}Q_{3,4}Q_{4,1}Q_{1,2}Q_{2,3}P]$
& $
\mathfrak{M}^{(\{0,0\},\{+\})},
\mathfrak{M}^{(\{+,+\},\{+\})},
\mathfrak{M}^{(\{-,+\},\{+\})}$
& $3$		                  
& $6$
& adj
& $\cdot$
& $\cdot$
& $1$
& $\cdot$
\\[3pt]
\hline 
$\Tr[P\widetilde{Q}_{3,4}\widetilde{Q}_{4,1}\widetilde{Q}_{1,2}\widetilde{Q}_{2,3}\widetilde{P}]$
& $
\mathfrak{M}^{(\{0,0\},\{-\})},
\mathfrak{M}^{(\{-,-\},\{-\})},
\mathfrak{M}^{(\{+,-\},\{-\})}$
& $3$		                  
& $6$
& adj
& $\cdot$
& $\cdot$
& $1$
& $\cdot$
\\[3pt]
\hline     
$\Tr[A_4]$
& $\Tr[{Q}_{1,2}\widetilde{Q}_{1,2}]$
& $1$
& $-2$
& $\cdot$
& $\cdot$
& $\cdot$
& $0$
& $\cdot$
\\[3pt]
\bottomrule
\end{tabular}

%% file: Tables/operators_mirror_pair_mag_mesons.tex
\begin{tabular}{ccccccccc}
\toprule
Electric Operators
& Magnetic Operators
& $U(1)_R$            
& $U(1)_a$ 	               
& $SU(2)_X$ 	
& $SU(2)_{Y_{1,2}}$
& $SU(2)_{Y_{3,4}}$
& $U(1)_B$ 	      
& $SU(2)_E$
\\[3pt] 
\hline  
$\substack{\\[2pt]
\textstyle{\mathfrak{M}^{(\{+\},\{0\},\{0,0\},\{0\})}} \\[2pt]
\textstyle{\mathfrak{M}^{(\{-\},\{0\},\{0,0\},\{0\})}} \\[2pt]
\textstyle{\Tr[A_1]}
\\[2pt]}$
& $\Tr[\widetilde{P}_R P_R]$
& $1$
& $-2$
& $\cdot$
& adj
& $\cdot$
& $0$
& $\cdot$
\\[3pt]
\hline  
$\substack{\\[2pt]
\textstyle{\mathfrak{M}^{(\{0\},\{0\},\{+,0\},\{0\})}} \\[2pt]
\textstyle{\mathfrak{M}^{(\{0\},\{0\},\{-,0\},\{0\})}} \\[2pt]
\textstyle{\Tr[A_3]}
\\[2pt]}$
& $\Tr[\widetilde{P}_L P_L]$
& $1$
& $-2$
& $\cdot$
& $\cdot$
& adj
& $0$
& $\cdot$
\\
\hline 
$\substack{\\[2pt]
\textstyle{\mathfrak{M}^{(\{+\},\{+\},\{+,0\},\{+\})}} \\[2pt]
\textstyle{\mathfrak{M}^{(\{-\},\{-\},\{-,0\},\{-\})}} \\[2pt]
\textstyle{\Tr[A_2]}
\\[2pt]}$
&
$\substack{\\[2pt]
\textstyle{\text{Tr}\Big[{Q}_{1,2}{Q}_{2,1}\Big],\Tr[\widetilde{Q}_{1,2}\widetilde{Q}_{2,1}]} \\[2pt]
\textstyle{{Q}_{1,2}\widetilde{Q}_{1,2}+{Q}_{2,1}\widetilde{Q}_{2,1}}
\\[2pt]}$
& $1$
& $-2$
& $\cdot$
& $\cdot$
& $\cdot$
& $0$
& adj
\\[3pt]
\hline  
$\mathfrak{M}^{(\{0\},\{+\},\{0,0\},\{0\})}$
& $\Tr[\widetilde{P}_R Q_{2,1} P_L]$
& $3/2$
& $-3$
& $\cdot$
& $\Box$
& $\Boxbar$
& $0$
& $\cdot$
\\
\hline  
$\mathfrak{M}^{(\{0\},\{-\},\{0,0\},\{0\})}$
& $\Tr[P_R \widetilde{Q}_{2,1} \widetilde{P}_L]$
& $3/2$
& $-3$
& $\cdot$
& $\Boxbar$
& $\Box$
& $0$
& $\cdot$
\\
\hline  
$\mathfrak{M}^{(\{0\},\{0\},\{0,0\},\{+\})}$
& $\Tr[P_R \widetilde{Q}_{1,2} \widetilde{P}_L]$
& $3/2$
& $-3$
& $\cdot$
& $\Boxbar$
& $\Box$
& $0$
& $\Box$
\\
\hline  
$\mathfrak{M}^{(\{0\},\{0\},\{0,0\},\{-\})}$
& $\Tr[\widetilde{P}_R Q_{1,2} P_L]$
& $3/2$
& $-3$
& $\cdot$
& $\Box$
& $\Boxbar$
& $0$
& $\Boxbar$
\\
\hline  
$\substack{\\[2pt]
\textstyle{\mathfrak{M}^{(\{0\},\{+\},\{+,0\},\{+\})}} \\[2pt]
\textstyle{\mathfrak{M}^{(\{+2\},\{+\},\{+,0\},\{+\})}} \\[2pt]
\textstyle{\mathfrak{M}^{(\{+\},\{+\},\{+,0\},\{+\})}_{A_1}}
\\[2pt]}$
& $\Tr[\widetilde{P}_R Q_{1,2} {Q}_{2,1} P_R]$
& $2$
& $-4$
& $\cdot$
& adj
& $\cdot$
& $0$
& $\Box$
\\
\hline  
$\substack{\\[2pt]
\textstyle{\mathfrak{M}^{(\{0\},\{-\},\{-,0\},\{-\})}} \\[2pt]
\textstyle{\mathfrak{M}^{(\{-2\},\{-\},\{-,0\},\{-\})}} \\[2pt]
\textstyle{\mathfrak{M}^{(\{-\},\{-\},\{-,0\},\{-\})}_{A_2}}
\\[2pt]}$
& $\Tr[{P}_R \widetilde{Q}_{1,2} \widetilde{Q}_{2,1} \widetilde{P}_R]$
& $2$
& $-4$
& $\cdot$
& adj
& $\cdot$
& $0$
& $\Boxbar$
\\
\hline  
$\substack{\\[2pt]
\textstyle{\mathfrak{M}^{(\{+\},\{+\},\{0,0\},\{+\})}} \\[2pt]
\textstyle{\mathfrak{M}^{(\{+\},\{+\},\{+2,0\},\{+\})}} \\[2pt]
\textstyle{\mathfrak{M}^{(\{+\},\{+\},\{+,0\},\{+\})}_{A_3}}
\\[2pt]}$
& $\Tr[\widetilde{P}_L \widetilde{Q}_{1,2} \widetilde{Q}_{2,1} P_L]$
& $2$
& $-4$
& $\cdot$
& $\cdot$
& adj
& $0$
& $\Box$
\\
\hline  
$\substack{\\[2pt]
\textstyle{\mathfrak{M}^{(\{-\},\{-\},\{0,0\},\{-\})}} \\[2pt]
\textstyle{\mathfrak{M}^{(\{-\},\{-\},\{-2,0\},\{-\})}} \\[2pt]
\textstyle{\mathfrak{M}^{(\{-\},\{-\},\{-,0\},\{-\})}_{A_4}}
\\[2pt]}$
& $\Tr[{P}_L {Q}_{1,2} {Q}_{2,1} \widetilde{P}_L]$
& $2$
& $-4$
& $\cdot$
& $\cdot$
& adj
& $0$
& $\Boxbar$
\\
\bottomrule
\end{tabular}

%% file: Tables/dictionary.tex
\begin{tabular}{cc}
\toprule
3d $\mathcal{N}=4$ & 
3d $\mathcal{N}=2^*$
\\ 
\hline    
\raisebox{-4mm}{\includegraphics[valign=c,width=.25\textwidth]{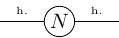}} & 
\includegraphics[valign=c,width=.4\textwidth]{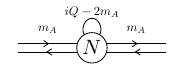} \\
\\[3pt]
\hline    
\raisebox{-4mm}{\includegraphics[valign=c,width=.25\textwidth]{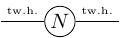}} & 
\includegraphics[valign=c,width=.4\textwidth]{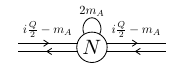} \\
\\[3pt]
\hline    
\raisebox{2mm}{\includegraphics[valign=c,width=.25\textwidth]{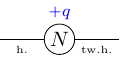}} & 
\includegraphics[valign=c,width=.4\textwidth]{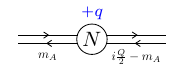} 
\\[3pt]
\bottomrule
\end{tabular}